\documentclass[pra,showpacs,showkeywords,graphics,epsfigure,superscriptaddress]{revtex4}

\usepackage{graphicx}
\usepackage{graphics}
\usepackage{bbm}
\usepackage{amsmath}
\usepackage{amsfonts}
\usepackage{amssymb}
\usepackage{latexsym}

\usepackage{braket}
\usepackage{amsbsy}
\usepackage{amsthm}
\usepackage{bm}
\usepackage{dsfont} 

\newcommand{\beq}{\begin{eqnarray}}
\newcommand{\eeq}{\end{eqnarray}}

\newcommand{\tr}{\textnormal{Tr}}

\newtheorem{theorem}{Theorem}

\newtheorem{corollary}{Corollary}
\newtheorem{definition}{Definition}
\newtheorem{proposition}{Proposition}

\newcommand{\A}{\mathcal{A}}
\newcommand{\B}{\mathcal{B}}
\newcommand{\Ha}{\mathcal{H}}

\newcommand{\V}{\mathcal{V}}

\begin{document}

\title{Bell's Universe: A Personal Recollection}
\author{Reinhold A. Bertlmann}
\affiliation{University of Vienna, Faculty of Physics, Boltzmanngasse 5, 1090 Vienna, Austria}
\email{Reinhold.Bertlmann@univie.ac.at}

\begin{abstract}

My collaboration and friendship with John Bell is recollected. I will explain his outstanding contributions in particle physics, in accelerator physics, and his joint work with Mary Bell. Mary's work in accelerator physics is also summarized. I recall our quantum debates, mention some personal reminiscences, and give my personal view on Bell's fundamental work on quantum theory, in particular, on the concept of contextuality and nonlocality of quantum physics. Finally, I describe the huge influence Bell had on my own work, in particular on entanglement and Bell inequalities in particle physics and their experimental verification, and on mathematical physics, where some geometric aspects of the quantum states are illustrated.
\end{abstract}


\maketitle


\vspace{0.5cm}
\begin{center}
\textbf{\emph{Dedicated to Mary Bell, John's lifelong companion.}}
\end{center}

\section{Collaboration with John Bell}

In April 1978 I moved from Vienna to Geneva to start with my Austrian Fellowship at CERN's Theory Division. Already in one of the first weeks, after one of the Theoretical Seminars, when all newcomers had a welcome tea in the Common Room, I got acquainted with John Stewart Bell. I remember when he approached me straightaway, \emph{``I'm John Bell, who are you?''} I answered a bit shy, \emph{``I am Reinhold Bertlmann from Vienna, Austria.''} \emph{``What are you working on?''} was his next question and I replied \emph{``Quarkonium...''}, which was already a \emph{magic} word since we immediately fell into a lively discussion about bound states of quark-antiquark systems, a very popular subject at that time, which continued in front of the blackboard in his office. A fruitful collaboration and warm friendship began.\\

\begin{figure}
\begin{center}
\includegraphics[width=0.65\textwidth]{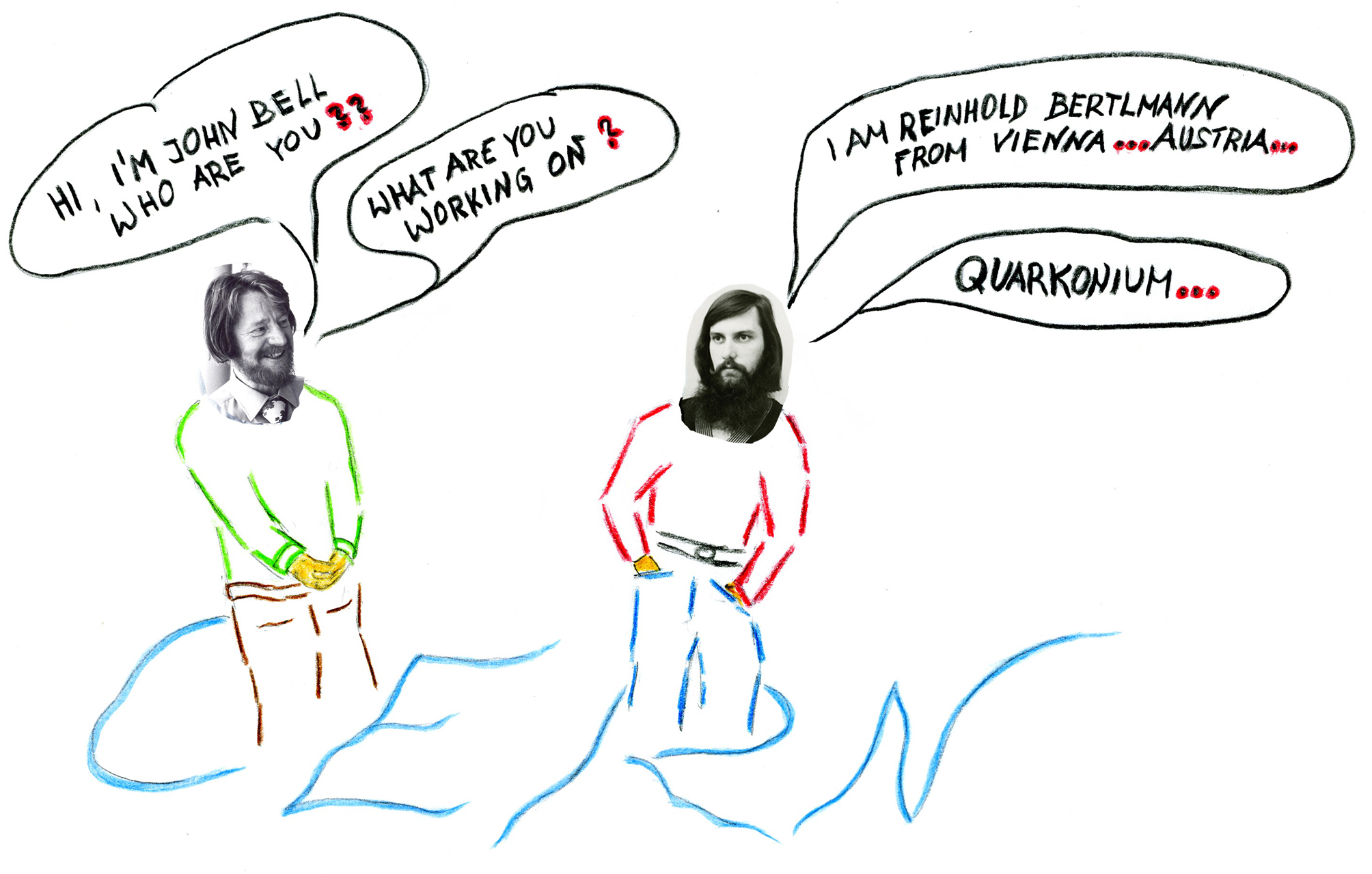}
\normalsize{
\caption{First encounter with John Bell at CERN in April 1978. Cartoon: \copyright Reinhold A. Bertlmann.}
\label{fig:first-encounter-John-Reinhold}}
\end{center}
\end{figure}

The first problem we attacked was how to understand the production of hadrons (strong interacting particles) in $e^+ e^-$ collisions. The experiments showed the following feature: At low energies there occurred pumps, the resonances, in the hadronic cross-section whereas at high energies the cross-section became quite flat or asymptotically smooth. Hadrons consist of quarks and antiquarks thus the $e^+ e^-$ collisions actually produce quark-antiquark ($q \bar{q}$) pairs. The idea was that at high energies, which corresponds to short distances, the $q \bar{q}$ pairs behave as quasi-free particles providing such the flat cross-section. However, at low energies, where the quarks can penetrate into larger distances they are confined and generate bound states -- called quarkonium -- which show up as resonances.

Our starting point was the idea of \emph{duality} that stated that smearing each resonance in energy already appropriately matches the corresponding result of the asymptotic cross-section determined by the short-distance interaction~\cite{BellBertlmannDual1980, Bertlmann-ActaPhysAustr1981}, an idea that can be traced back to a work of J.J. Sakurai \cite{Sakurai1973}. Theoretically, it can be understood in the following way. Allowing for an energy spread means---via the uncertainty relation---that we focus on short times. But for short times the corresponding wave does not spread far enough to feel the details of the long distances, the confining potential. So this part can be neglected and the wave function at the origin of the bound state, which determines the leptonic width or area of a resonance, matches the averaged quasi-free $q \bar{q}$ pair. However, if we want to push the idea of duality even further in order to become sensitive to the position of the bound state in the mass spectrum then we have to include into the wave function the contributions of larger distances, confinement.\\

\begin{figure}
\begin{center}
\includegraphics[width=0.59\textwidth]{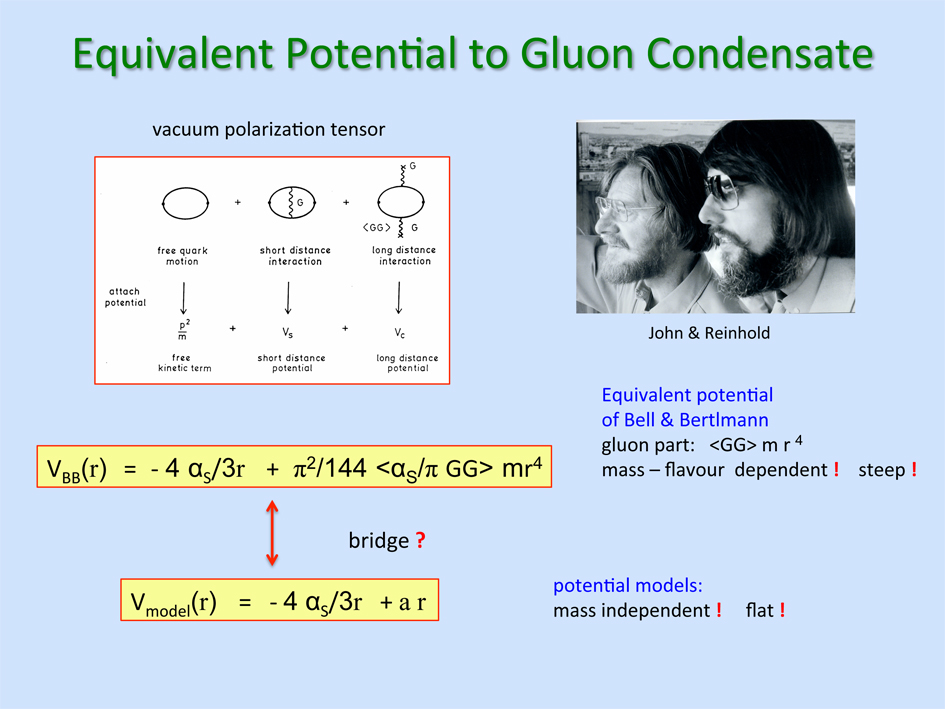}
\normalsize{
\caption{Perturbative expansion (Feynman diagrams) of the vacuum polarization tensor in QCD including the gluon condensate, and the \emph{equivalent potential of Bell and Bertlmann} to this expansion. Foto: \copyright Renate Bertlmann.}
\label{fig:BB-equivalent-potential}}
\end{center}
\end{figure}

How could we include confinement in our duality concept? Our starting point was the vacuum polarization tensor $\Pi(q^2)$, the vacuum expectation value of the time ordered product of two quark currents in quantum field theory, quantum chromo dynamics
\begin{eqnarray}\label{vacuum-polarisation-tensor}
i \int dx\; e^{iqx} \langle\Omega|T\;j_\mu(x)j_\nu(0)|\Omega\rangle  \;=\; \Pi(q^2)(q_\mu q_\nu \,-\,q^2 g_{\mu\nu})\;.
\end{eqnarray}
This quantity was proportional to the hadronic cross-section, where at low energy the bound states, the resonances, occurred. More precisely, the imaginary part of the vacuum polarization function (the forward scattering amplitude) was related to the total cross-section via the optical theorem
\begin{eqnarray}\label{optical-theorem}
{\rm{Im}}\,\Pi(E) \;\sim\; \sigma_{\rm{tot}}(E)\;,
\end{eqnarray}
and calculable within perturbation theory with help of Feynman diagrams, the loop diagrams depicted in Fig.~\ref{fig:BB-equivalent-potential}. At that time the Russian group, Shifman-Vainshtein-Zakharov (SVZ)~\cite{SVZ1979}, claimed that the so-called gluon condensate $\langle \frac{\alpha_{\rm{s}}}{\pi} G G \rangle$, the vacuum expectation value of two gluon fields, would be responsible for the influence of confinement (represented by the third loop diagram in Fig.~\ref{fig:BB-equivalent-potential}). This idea we wanted to examine further.\\

Approximating quantum field theory by potential theory, we could calculate both the perturbative \emph{and} the exact expression. For the energy smearing we had chosen an exponential as weight function, which was called a \emph{moment} by the mathematicians
\begin{eqnarray}\label{nonrelat-moment-general}
M(\tau)&\;=\;&\int dE\; e^{-E \tau}\; {\rm{Im}}\,\Pi(E) \;=\; \frac{3}{8m^2}\,\langle \vec{x} = 0|e^{-H \tau}|\vec{x} = 0 \rangle\;.
\end{eqnarray}
In this case, where we had rediscovered the $\tau$ dependent Green's function at $\vec{x} = 0\,$, the procedure corresponded to perturbation theory of a Hamiltonian $H$ with respect to an imaginary time $\tau\,$. This I found quite fascinating. The actual calculation provided the following result
\begin{eqnarray}\label{nonrelat-moment-explicit}
M(\tau)&\;=\;& \frac{3}{8m^2} \,4\pi\, (\frac{m}{4\pi\tau})^{\frac{3}{2}}\biggl\lbrace 1 \,+\, \frac{4}{3}\alpha_{\rm{s}} \sqrt{\pi m} \;\tau^{\frac{1}{2}} \,-\, \frac{4\pi^2}{288 m} \langle \frac{\alpha_{\rm{s}}}{\pi} GG \rangle \;\tau^3 \biggr\rbrace\;.
\end{eqnarray}
The leading term corresponded to the free motion of the quarks (first diagram in Fig.~\ref{fig:BB-equivalent-potential}); it was perturbed by the
$\alpha_{\rm{s}}$ term, representing the short distance interaction (second diagram), and by the gluon condensate $\langle \frac{\alpha_{\rm{s}}}{\pi} GG \rangle$ term, responsible for the longer distances (third diagram).\\

How did we get the levels of the bound states, the masses of the resonances? The ground state level we could extract by using the logarithmic derivative of a moment,
\begin{eqnarray}\label{ratio-of-moments}
R(\tau) &\;=\;& -\frac{d}{d\tau} \,\log \,M(\tau) \;=\; \frac{\int dE \;E \;e^{-E \tau} \,{\rm{Im}} \,\Pi(E)}{\int dE \;e^{-E \tau} \,{\rm{Im}} \,\Pi(E)} \quad\stackrel{\tau\longrightarrow\infty}{\longrightarrow}\quad E_1\;,
\end{eqnarray}
which approached the ground state energy for large (imaginary) times since the contributions of the higher levels were cut off. In this way we were able to predict the ground states of charmonium (the $J/\psi$ resonances) and of bottonium (the $\Upsilon$ resonances) to a high accuracy, quantitatively within $10 \%$ \cite{BellBertlmannMagic1981, BertlmannCharmonium-PhysLett1981, BertlmannCharmonium-NuclPhys1982, BellBertlmann-PhysLett1984}.\\

We observed a remarkable \emph{balance}, see Fig.~\ref{fig:ratio-moments}:

\begin{figure}
\begin{center}
\includegraphics[width=0.4\textwidth]{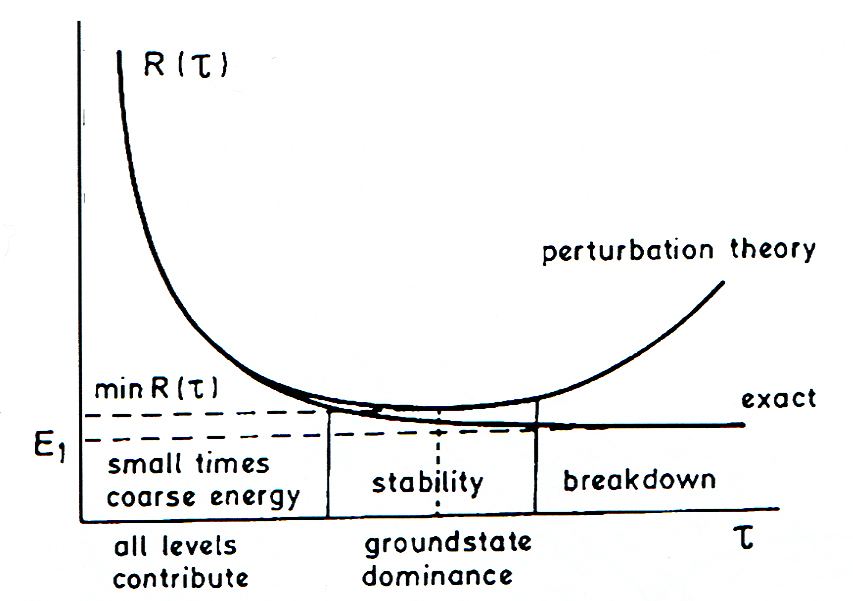}
\normalsize{
\caption{The ratio of moments $R(\tau)$ is depicted qualitatively. There occurs a balance between the short and long distance contributions in the ratio of moments. The stability of the expansion determined by the Feynman diagrams of Fig.~\ref{fig:BB-equivalent-potential} approximates quite accurately the exact ground state level.}
\label{fig:ratio-moments}}
\end{center}
\end{figure}

\emph{The energy average can be made coarse enough---involving small times---for the modified perturbation theory to work, while on the other hand fine enough for the individual levels to emerge clearly.}\\

This is a surprising feature, indeed, since intuitively we had expected that for a clearly emerging level the confinement force must be dominant and not just a small additional perturbation. The moments, however, forced us to re-educate our intuition, when modifying the perturbation with the gluon condensate term, levels do appear for {\em magical} reasons. Therefore we gave our paper the title \emph{``Magic Moments''} \cite{BellBertlmannMagic1981}.\\

\begin{figure}
\begin{center}
(a)\;\setlength{\fboxsep}{2pt}\setlength{\fboxrule}{0.8pt}\fbox{
\includegraphics[angle = 0, width = 61mm]{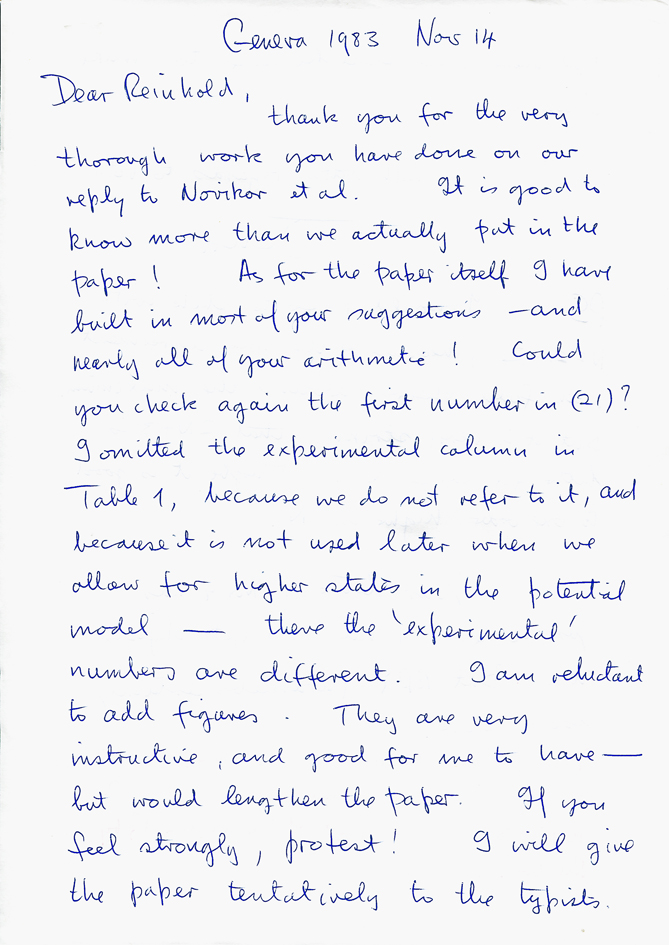}}
\hspace{5mm}
(b)\;\setlength{\fboxsep}{2pt}\setlength{\fboxrule}{0.8pt}\fbox{
\includegraphics[angle = 0, width = 61mm]{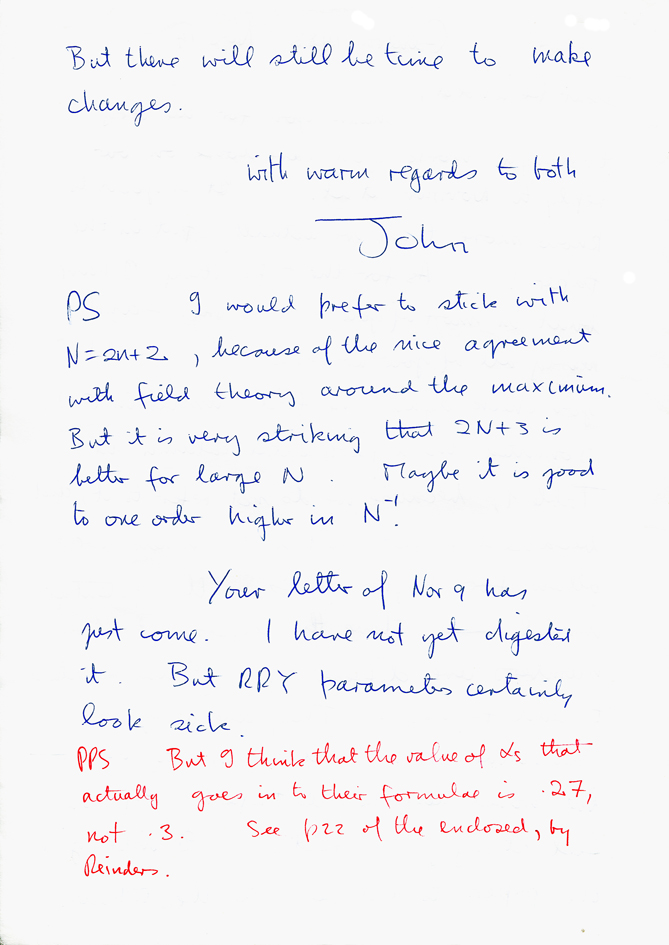}}
\normalsize{
\caption{(a) Page 1 of the handwritten letter of John Bell to Reinhold Bertlmann from 1983~\protect\cite{Bell-letter-Reinhold1983}, in connection with the preparation of their paper on the \emph{``SVZ moments for charmonium and potential model''}~\protect\cite{BellBertlmann-PhysLett1984}. (b) Page 2 of the letter.}
\label{fig:Letter-John-to-Reinhold}}
\end{center}
\end{figure}

At the time of our collaboration, the late 1970s and 1980s, there was no internet, no quick email exchange. The way we interacted, when I was absent from CERN, was via letters written by hand. Of course, this communication took some time, the writing itself, the search for stamps, the walk to the post office, etc. In retrospect these letters were beautiful documents expressing not only our scientific thoughts but also our personal attitudes, our characters, how we had investigated a physical subject and how we had presented our scientific work, what to include and what to leave out. As a typical example, I would like to show a letter of John (see Fig.~\ref{fig:Letter-John-to-Reinhold}), which he wrote to me in November 1983~\cite{Bell-letter-Reinhold1983}, while I was staying in Vienna. The letter was written in connection with the preparation of our paper on the \emph{``SVZ moments for charmonium and potential model''}~\cite{BellBertlmann-PhysLett1984}, which we were going to publish. It shows quite nicely our struggle for an accurate and clear presentation.\\

Since John and I were working within potential theory, which functioned remarkably well \cite{BellBertlmann-PhysLett1984}, it was quite natural for us to ask whether one can attach a potential to the occurrence of the gluon condensate. Indeed, we found ways to do this \cite{BellBertlmann-SVZ1981, BellBertlmann-LV1983}.

One way was to work within the moments, which regularize the divergence of the long-distance part of the gluon propagator, the gluon condensate contribution. In this case a static, nonrelativistic potential containing the gluon condensate can be extracted, which is called in the literature the \emph{equivalent potential of Bell and Bertlmann} \cite{BellBertlmann-SVZ1981}
\beq\label{BB-equival-potential}
V_{\rm{BB}}(r) &\;=\;& - \frac{4}{3} \frac{\alpha_{\rm{s}}}{r} \,+\, \frac{\pi^2}{144} \langle \frac{\alpha_{\rm{s}}}{\pi} GG \rangle \,m \;r^4 \;,
\eeq
where $\alpha_{\rm{s}}$ is the QCD coupling constant, $m$ the quark mass and $\langle \frac{\alpha_{\rm{s}}}{\pi} GG \rangle$ the gluon condensate.

The short-distance part was the well-known Coulomb potential, whereas the long-distance component, the gluon condensate contribution, emerged as a quartic potential $m\,r^4$ and is mass-, i.e., flavour-dependent. In this respect it differed considerably from the familiar mass-independent, rather flat potential models \cite{QuiggRosner1979, KrammerKrasemann1979, GrosseMartin1980, LuchaSchoeberlGromes1991}. However, for a final comparison with potential models one has to go further and take into account the higher order fluctuations \cite{Bertlmann-GGGG1984}.\\

\begin{figure}
\begin{center}
\includegraphics[width=1.02\textwidth]{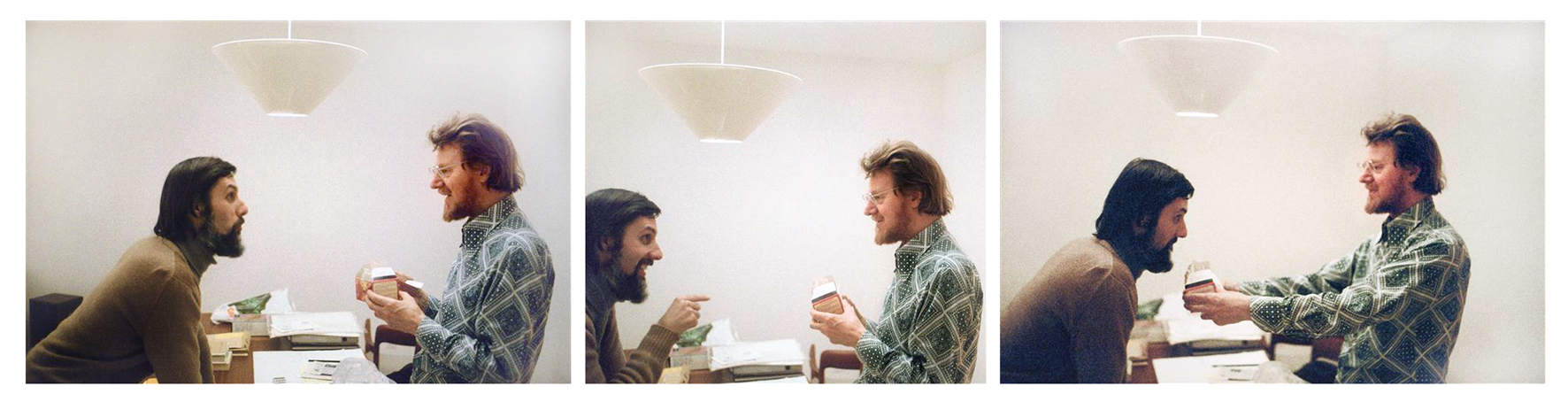}
\normalsize{
\caption{Bertlmann (left) and Bell (right) choosing the right sort of tea at Bell's home in 1980. Foto: \copyright Renate Bertlmann.}
\label{fig:Reinhold+John triptychon choosing tea}}
\end{center}
\end{figure}

I very well remember one of our afternoon rituals in our collaboration. John, a true Irishman, always had to drink a \emph{4~o'clock tea}\,, and this we regularly practiced in the CERN Cafeteria or at John's home, when we were working there. Choosing the right sort of tea was quite a ceremony, see Fig.~\ref{fig:Reinhold+John triptychon choosing tea}. Then, in this relaxed tea-atmosphere, we talked not only about physics but also about politics, philosophy, and when we were joined by my artist wife Renate, we three also had heated debates about modern art.\\

Next we studied very heavy quarkonium systems, which described e.g. bottonium (the $\Upsilon$ resonances). There John and I found another way to extract a potential from the gluon condensate effect~\cite{BellBertlmann-LV1983}. In that case the low-lying bound states, because of their small size, were dominated by the Coulomb potential and the condensate effect, an external colour field representing the gluon, could be added as a small perturbation. Leutwyler \cite{Leutwyler1981} and Voloshin \cite{Voloshin1979} had considered such a colour-electric Stark effect and calculated the energy spectrum for all quantum numbers $n$ and $l$. John and myself, on the other hand, were able to construct a gluonic potential, which by perturbing the Coulomb states provided the energy spectrum of Leutwyler-Voloshin
\beq\label{BB-gluon-potential-for-LV}
V_{\rm{gluon}}(r) \;=\;  \frac{4\pi^2}{81\beta} \langle \frac{\alpha_s}{\pi} GG \rangle \,\Big( r^3 \,-\, \frac{304}{81}\frac{r^2}{m\beta}
\,+\, \frac{53}{10}\frac{r}{m^2\beta^2} \,-\, \frac{113}{100}\frac{1}{m^3\beta^3} \Big) \;\quad \mbox{with} \;\quad \beta \,=\, \frac{4\alpha_s}{3}\;.
\eeq
The leading term in potential (\ref{BB-gluon-potential-for-LV}), the infinite mass limit, had a cubic $r^3$ dependence and was therefore mass-independent. But for finite masses there were further corrections necessary proportional to $\frac{r^2}{m}, \frac{r}{m^2}, \frac{constant}{m^3}$ such that the potential became flavour-dependent again.\\

While our equivalent potential (\ref{BB-equival-potential}) was constructed to reproduce the gluon condensate shifts in the moments \`{a} la Shifman, Vainshtein and Zakharov \cite{SVZ1979}, the potential (\ref{BB-gluon-potential-for-LV}) was designed to reproduce gluon condensate level shifts \`{a} la Leutwyler \cite{Leutwyler1981} and Voloshin \cite{Voloshin1979} in hydrogen-like heavy $q \bar{q}$ systems. The two potentials differed because they were fitted to different aspects of quantum field theory---and potential theory is, of course, not field theory.\\

In conclusion, whereas the moment procedure including the gluon condensate worked surprisingly well for predicting the ground state levels of quark-antiquark bound states, no adequate bridge was found between a field theory containing the gluon condensate, quantum chromodynamics, on one side, and popular potential models on the other. For an overview of this field I would like to refer to Ref.~\cite{BertlmannPotential1991}.

\section{John Bell -- the Particle Physicist}\label{sec:Bell-particle-physics}

Bell graduated with First-Class Honours in Experimental Physics in 1948 at Queen's University Belfast, where the senior staff members of the Physics Department were Karl Emeleus and Robert Sloane. He spent an additional year at the University and obtained a second degree in Mathematical Physics, where his teacher was the famous crystallographer Peter Paul Ewald. Subsequently, in 1949, he got a position the Atomic Energy Research Establishment (AERE) at Harwell, Oxfordshire, but was soon sent to the Telecommunications Research Establishment (TRE) at Malvern, Worcestershire. There he began to work in accelerator physics, see Sect.~\ref{sec:Bell-accelerator-physics}. In 1951, the accelerator group at Malvern moved to the Atomic Energy Research Establishment (AERE) at Harwell, Oxfordshire.\\

In 1953, one of the few persons who had got that privilege, John was selected to spend some time at a university to perform a PhD Thesis while keeping his usual AERE salary. He had eventually chosen the University of Birmingham, where Rudolf Peierls was the head of Theoretical Physics. There he could study quantum field theory having much contact with Paul T. Matthews. John succeeded already as a student to write a fundamental work, namely his PhD Thesis. It consisted of two parts: \emph{``Time reversal in field theory''} and \emph{``Some functional methods in field theory''}. In the first part, published in 1955 in the Proceedings of the Royal Society~\cite{Bell-CPTtheorem}, he established the celebrated $CPT$ Theorem ($C$, the charge conjugation operator, which replaces particles by anti-particles; $P$, the parity operator, which performs a reflection; and $T$, which carries out a time reversal). The theorem states that any Lorentz invariant, local quantum field theory with a Hermitian Hamiltonian must be $CPT$ symmetric. It is commonly believed that Nature is $CPT$ invariant, no $CPT$ violating phenomena have been found so far. For a long time Gerhart L\"uders~\cite{Lueders-CPT} and Wolfgang Pauli~\cite{Pauli-CPT}, who proved the theorem a little ahead of Bell, received all the credit but nowadays Bell's `elementary' derivation, being more accessible than the formal mathematical arguments of L\"uders and Pauli, is also rightly recognized.\\

After his PhD, about 1955, John turned to nuclear and particle physics, a growing field at that time. At AERE he closely worked with Tony Skyrme, the head of the Nuclear Physics Group, whose soliton model for nucleons, called skyrmions, became well-known later on. Together they investigated the \emph{``Nuclear spin-orbit coupling''}~\cite{Bell-Skyrme-Phil-Mag}, the \emph{``Magnetic moments of nuclei and the nuclear many body problem''}~\cite{Bell-Eden-Skyrme-Nucl-Phys, Bell-Nucl-Phys-1957} and the \emph{``Anomalous magnetic moments of the nucleons''}~\cite{Bell-Skyrme-Proc-Roy-Soc}. His mastery in field theory he also showed in a paper about a \emph{``Variational method in field theory''}~\cite{Bell-var-method}, where he found a new form of Skyrme's variational principle for the one-nucleon propagator.

John also contributed various papers to the many-body theory of nucleons~\cite{Bell-many-body-1957, Bell-many-body-ProcPhysSoc-1959, Bell-particle-hole, Bell-BlinStoyle} and together with Euan Squires he derived an effective one-body potential, the \emph{``Formal optical model''} for the scattering of a particle incident on a complex target~\cite{Bell-Squires, Bell-lecture-many-body}.

An other one of John's colleagues at Harwell was Franz Mandl. They harmoniously worked together in particle physics, for instance, on the \emph{``Polarization-asymmetry equality''}~\cite{Bell-Mandl} for elastic scattering of spin $\frac{1}{2}$ particles off unpolarized targets, but concerning the interpretation of quantum mechanics Mandl was John's strong opponent. At that time David Bohm's papers on the \emph{``Interpretation of quantum theory in terms of hidden variables''}~\cite{Bohm1952} had appeared, which were a \emph{``revelation''} for John. After John had reflected upon the papers he also gave a talk about them in the Theory Division and, as Mary Bell remembered, there were many interruptions from Mandl \emph{``with whom he had many fierce arguments''}~\cite{MaryBell2002}.\\

In 1960, John together with Mary joined CERN (Conseil Europ\'een pour la Recherche Nucl\'eaire), the European Organization for Nuclear Research which attracted the preeminent scientists, and he worked there until the rest of his life. His interest was both in phenomenological aspects of particle physics, where he frequently interacted with experimentalists, and in more formal, mathematical features of the theory, which often had no relation to experiment. As I could experience, John always was open to discuss and study any topic in physics, no matter how speculative it was. He liked to test his thoughts by basic examples. \emph{``Always test your general reasoning against simple models!''} was his maxim.\\

In the 1960s, the theory of weak interactions became a hot topic. John also entered in this field and collaborated with several physicists. With his experimental colleague Jack Steinberger, who received the Nobel Prize in 1988 \emph{``for the discovery of the muon neutrino''}, John has written the influential review on \emph{``Weak interaction of kaons''}~\cite{Bell-Steinberger}. The so-called \emph{Bell-Steinberger unitarity relations} for the kaon decay amplitudes belong still to the standard achievements in this area.\\

Most important, I think, was John's scientific exchange with his colleague and friend Martinus Veltman, just called with his nickname `Tini', who was a Fellow at CERN. They had a fruitful and each other appreciative collaboration in particle physics, however, concerning John's quantum work Tini was, as most people at that time, quite reluctant to recognize its value (see Veltman's book~\cite{Veltman-book} or Mary Bell's essay~\cite{MaryBell-talk2014}).

John and Tini started to investigate the carriers of the weak interactions, the intermediate bosons or W-bosons, which were just hypothetical particles at that time. But they had faith in their existence, encouraged and assisted the experimentalists at CERN to find these particles, and published important papers on the W-boson production by neutrinos on nuclei~\cite{Bell-Lovseth-Veltman, Bell-Veltman-1, Bell-Veltman-2}, which served as a basis for the neutrino experiments. As we know, it took two decades to discover the W-bosons, due to their big mass.

They also had numerous discussions about the characteristics of a quantum field theory, the issue of symmetries turned up and the feature of currents in a quantized theory. In these discussions already the seed was planted that led John finally to his most important discovery in particle physics. In case of a modern quantum field theory the \emph{gauge symmetry} formed the basis for the so-called \emph{gauge theory}, i.e., the Lagrangian and the basic equations of the theory were invariant with respect to possible gauge transformations. To each generator of the transformation corresponded a gauge field whose quanta were called gauge bosons, the W-bosons in case of weak interactions. Veltman successfully pursued these ideas further with his former student Gerald 't Hooft and both were awarded with the Nobel Prize in 1999 for \emph{``elucidating the quantum structure of electroweak interactions''}.\\

However, such a quantized theory for the weak interaction, what we call nowadays \emph{standard model}, the unification of the electromagnetic and weak interactions, was not yet developed in the 1960s and physicists probed different ideas. Gell-Mann's \emph{current algebra}, where instead of the conventional fields the currents were considered, was quite popular. Using plausible ideas from group theory Murray Gell-Mann postulated a canonical structure for the commutators of the current components which were involved in the physical process~\cite{Gell-Mann-current-algebra}. Conserved vector current (CVC) and partially conserved axial current (PCAC) were assumed. Physical processes could be studied by calculating the corresponding matrix elements of the currents. Amusingly, Gell-Mann published his current algebra, the \emph{``Symmetry group of vector and axial-vector currents''}~\cite{Gell-Mann-current-algebra}, in the now-defunct journal \emph{Physics}, in which Bell just some pages behind published his seminal \emph{``On the Einstein Podolsky Rosen paradox''} paper~\cite{Bell-Physics1964}. Due to the success of current algebra many particle physicists began to research its foundations and applications. So did John, and it is within this area that he made his outstanding contribution to particle physics.\\

In quantum field theory infinities occur and quantities like currents must be renormalized. It was not clear at that time whether the postulated relations of the canonical commutators for the current components survived a proper renormalization procedure. John illuminated this problem by studying an \emph{``Equal-time commutator in a solvable model''}~\cite{Bell-NC47}, the unrealistic but completely solvable Lee model. He demonstrated that indeed the canonical commutation relations must be taken with care since in a certain case, in the calculation of a related sum rule, the canonical values do not agree with the summation of the explicitly calculated amplitudes. On the other hand, relying on a work by Veltman about gauge invariance of sum rules~\cite{Veltman-PRL17}, John showed in \emph{``Current algebra and gauge invariance''}~\cite{Bell-NC50}  that the desired commutation relations of the currents are achieved if gauge invariance is imposed in the matrix elements of interest.

Next, trusting current algebra and PCAC an analysis revealed that the decay of the eta meson into three pions, $\eta \rightarrow 3\pi\,$, is forbidden~\cite{Sutherland-PL23, Bell-Sutherland-NP-B4}, even though the decay is experimentally seen. Moreover, with the same assumptions the calculation of the decay of the neutral pion into two photons, $\pi^{0} \rightarrow \gamma \gamma\,$, yielded a vanishing result~\cite{Sutherland-NP-B2, Veltman-ProcRoySci1967}, again in contradiction to experiment.\\

\begin{figure}
\begin{center}
\includegraphics[width=0.55\textwidth]{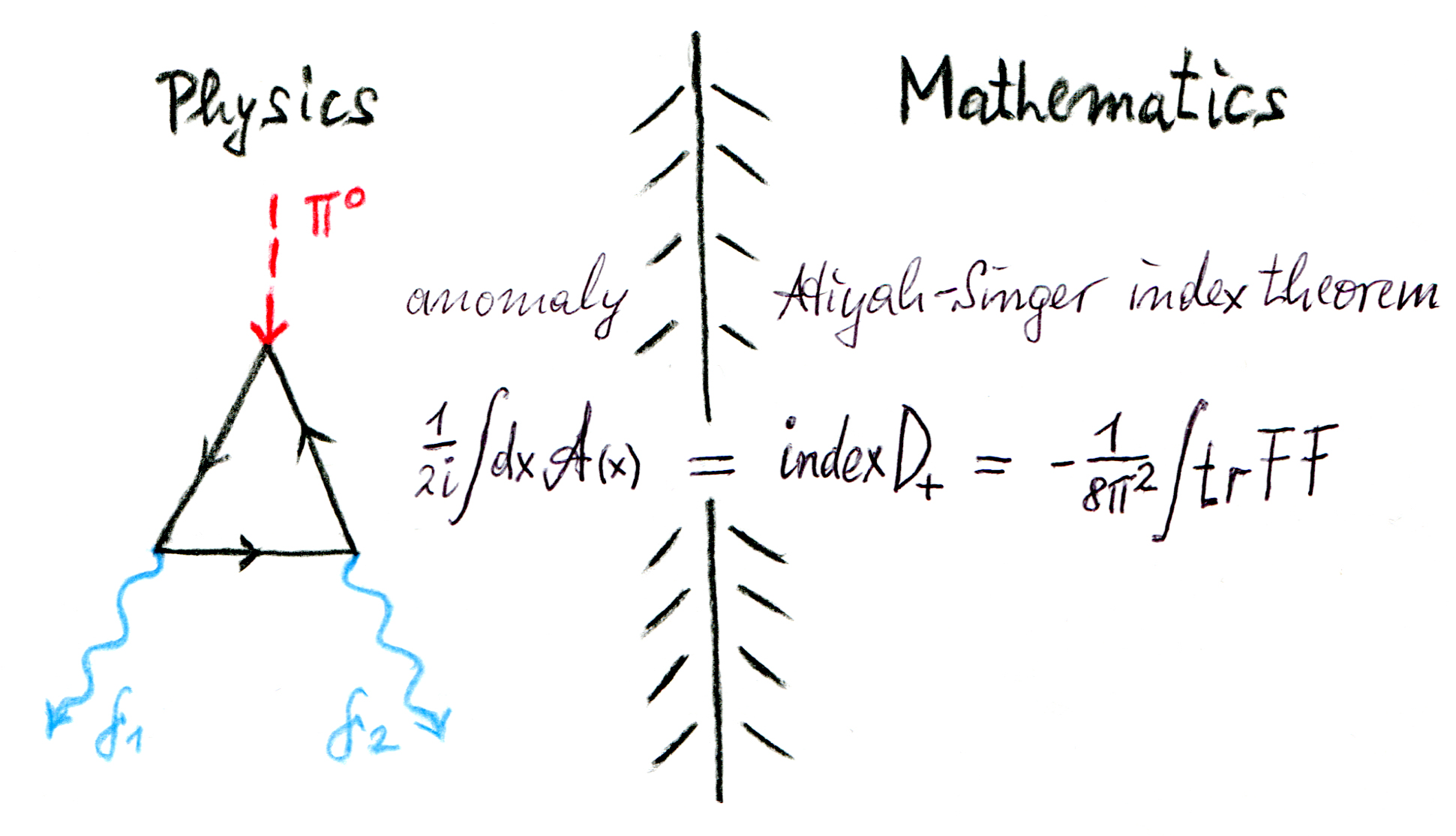}
\normalsize{
\caption{An interplay of physics and mathematics: The Abelian anomaly is responsible for the decay $\pi^0 \rightarrow \gamma\gamma\,$. It is represented by the triangle diagram with two vector current vertices that couple to the two photons and one axial vertex linking to the $\pi^0\,$. The anomaly is related to the Atiyah-Singer index theorem in topology.}
\label{fig:anomaly}}
\end{center}
\end{figure}

These features were generally considered as shortcomings of the otherwise successful theory of current algebra. But John, dissatisfied with incompleteness and deficiency of a theory, always kept these current algebra defects in his mind. When Roman Jackiw, a postdoc from MIT visiting CERN in 1967 - 1968, asked John for a research problem he suggested to analyze the failure of current algebra in the $\pi^{0} \rightarrow \gamma \gamma$ decay (see Jackiw's contribution to Ref.~\cite{Shimony-Jackiw2001}). The result then turned out, as we know now, to be John's most far-reaching contribution to particle physics and his most-quoted paper.

But how to tackle the problem? Interestingly, John's colleague Steinberger with whom he had frequently contact, in a discussion during a coffee break in the CERN cafeteria pointed the right way. In 1949, Steinberger~\cite{Steinberger1949} had already calculated in his PhD a Feynman diagram, a triangle diagram with two vector current vertices and one axial vertex (see Fig.~\ref{fig:anomaly}), in an at that time fashionable pion - nucleon model in order to describe the decay $\pi^{0} \rightarrow \gamma \gamma\,$. He obtained a nonvanishing result that, moreover, agreed well with experiment. John and Jackiw recognized immediately that Steinberger's procedure could be pursued in the sigma model ($\sigma$-model)~\cite{Gell-Mann--Levy1960}, a field theory based on current algebra and PCAC. So they calculated the pion decay amplitude directly and found that the $\sigma$-model did not satisfy the requirements of PCAC, the effective coupling constant of the decay did not vanish. Their paper was entitled \emph{``A PCAC puzzle: $\pi^0 \rightarrow \gamma\gamma$ in the $\sigma$-model''}~\cite{Bell-Jackiw}.

Independently, in the same year, Stephan L. Adler from the Institute of Advanced Study in Princeton investigated the axial-vector vertex in spinor electrodynamics~\cite{Adler1969}. He found that the axial Ward identity failed in case of the triangle diagram, which led him to modify PCAC by an extra term, the \emph{anomaly} as he phrased it. When applying this modified PCAC equation to calculate the pion decay rate Adler obtained an excellent agreement with experiment, when the fermions propagating in the triangle diagram are interpreted as quarks with their fractional charges, which occur in three species, \emph{colours} as we know now. Thus the \emph{modified PCAC equation}
\beq\label{modifiedPCAC}
\partial^{\mu} j^{5(3)}_{\mu}(x) \;=\; f_{\pi} m^2_{\pi} \phi^{(3)}_{\pi}(x) \,+\, \A \;,
\eeq
states that axial current $j^{5\,a}_{\mu} = \bar{\psi}\gamma_{\mu}\gamma_5 \frac{\sigma^a}{2}\psi$ is not conserved in massless limit of the theory but represents the celebrated \emph{ABJ anomaly} (in honour of Adler, Bell and Jackiw)
\beq\label{ABJ-anomaly}
\A \;=\; \frac{e^2}{16\pi^2} \,\varepsilon^{\mu\nu\alpha\beta}F_{\mu\nu}F_{\alpha\beta} \;,
\eeq
where $F_{\mu\nu}$ is the electromagnetic field strength tensor.

Now the ice was broken, it turned out that the anomaly was not just a pathology of the quantization procedure but opened the door to a deeper understanding of quantum field theory. A new era of field theory research began (for an overview see Ref.~\cite{Bertlmann-anomaly-book}).\\

Quantum field theories with non-Abelian fields had been studied subsequently and anomalies were found there. In terms of differential geometry the anomalies could be formulated very concisely:\\

\emph{Singlet anomaly} (corresponding to the Abelian- or ABJ anomaly with $e = 1$)
\beq\label{singlet-anomaly}
\A \;=\; d \ast j^5 \;=\; \frac{1}{4\pi^2}\, \tr \,FF \;=\; \frac{1}{4\pi^2}\,d\,\tr (AdA \,+\, \frac{2}{3}A^3) \;,
\eeq

\emph{non-Abelian anomaly}
\beq\label{nonAbelian-anomaly}
G^a[A] \;=\; -(D\ast j)^a \;=\; \pm \frac{1}{24\pi^2}\,\tr\, T^a d\,(AdA \,+\, \frac{1}{2}A^3) \;,
\eeq
with $A = A^a_{\mu} T^a \,dx^{\mu}$ the non-Abelian 1-form (or connection, geometrically), $F = \frac{1}{2} F^a_{\mu\nu}T^a \,dx^{\mu}\wedge dx^{\nu}$ the field strength 2-form (or curvature), $D$ the covariant derivative and $T^a$ the generators of the gauge group. The signs $\pm$ corresponded to positive or negative chiral fields. Expression (\ref{nonAbelian-anomaly}) was determined by a simple equation, the Wess-Zumino consistency condition
\beq
s \,G(v,A) \;=\; s \int v^a G^a[A] \;=\; 0 \;,
\eeq
where $v = v^a T^a$ denoted the Faddeev-Popov ghost and $s$ the BRST operator with $s^2 = 0\,$, a gauge variation with respect to the gauge fields \emph{and} the Faddeev-Popov ghosts (for literature, see the book~\cite{Bertlmann-anomaly-book}).

Particularly interesting was the connection of the anomaly to topology in mathematics. Several authors, among them Roman Jackiw, had discovered that the singlet anomaly was determined by the distinguished Atiyah-Singer index theorem (see Fig.~\ref{fig:anomaly}). The reason was that the anomaly could be expressed by a sum of eigenfunctions of the Dirac operator, where only the zero-modes of a given chirality ($n_{+} , n_{-}$) survived
\beq
\frac{1}{2i}\int dx \,\A(x) \;=\; \int dx \sum_{n} \varphi^{\dag}_{n}(x) \gamma_5 \,\varphi_{n}(x) \;=\; n_{+} \,-\, n_{-} \;=\; {\rm{index}} \,D_{+} \;.
\eeq
The difference of the chirality zero modes represented the index of the Weyl operator $D_{+}$ for positive chirality, which was expressed via the Atiyah-Singer index theorem by a Chern character
\beq
{\rm{index}} \,D_{+} \;=\; - \frac{1}{8\pi^2} \int \tr \,FF \;.
\eeq

Furthermore, when gravitation was considered as a gauge theory, where the gauges were the general coordinate transformations or the rotations in the tangent frame, then the classical conservation law of the energy-momentum tensor could be broken in the quantum case, an Einstein- or Lorentz anomaly occurred (see Refs.~\cite{Bertlmann-anomaly-book, Bertlmann-Kohlprath-AnnPhys}).\\

Thus quantum anomalies play a vital role in all quantum theories and it is their double feature which makes them so important for physics. On one hand, anomalies are needed to describe certain experimental facts but, on the other hand, they must be avoided since they violate a classical conservation law and signal the breakdown of gauge invariance, which ruins the consistency of the theory. This avoidance of the anomaly, which may be possible, leads to severe constraints on the physical content of the theory. For example, the \emph{standard model} of electro-weak interaction is constructed such that no anomalies occur (actually, they canceled each other), which has led to the prediction of the top quark that has been discovered much later on.\\

It is interesting that John did not participate actively in these further developments of the anomaly, whereas Jackiw, Adler and many other physicists did. Roman Jackiw once asked me in a letter (from May 17th, 1996~\cite{Jackiw-letter1996}) why this might be the case:

\emph{``I was very interested in your `Preface'} [of the book~\cite{Bertlmann-anomaly-book}] \emph{where you reminisce about conversations with Bell on anomalies a decade ago. I was very happy to read about this, because I always felt a certain surprise (disappointment) that he did not take an active role in subsequent developments, after our paper was published. Did you ever learn why this was so?''}

My assessment of John's missing engagement in the further developments I expressed in a return letter to Jackiw (from July 21st, 1996~\cite{Bertlmann-letter1996}) as follows:

\emph{``At first side it seems indeed surprising that John Bell did not take an active role in the further developments of anomalies after your common paper which had such a great influence on QFT. He never mentioned explicitly a reason for that. But my impression is that one aspect of his character or of his attitude towards physics was to find out the fundamental weaknesses of a theory. It was his criticism which has led him to important discoveries. This was so in his famous works on quantum mechanics and also I experienced this attitude in our collaboration. After having found the crucial point (error, ...) he was not so interested any more in working out the further details.}

\emph{In the case of the anomalies John got interested again in the middle of the 80ties since he was puzzled by the fact that several types of anomalies are linked in different dimensions (descent equations). In this connection John always spoke with high respect of your works.''}\\

Of course, after the anomaly paper John did further important works. Let me just mention a few.

Together with Eduardo de Rafael~\cite{Bell-deRafael} he calculated an upper bound on the hadronic contribution to the anomalous magnetic moment of the muon, which is satisfied by today's accepted value. Nevertheless, the topic is still of high interest due to the present discrepancy between experimental- and theoretical value.

More phenomenological work, also in view of the experiments at CERN, John carried out in collaboration with Christopher LLewellyn Smith who later on became Director General of CERN (1994 - 1998) and was knighted in 2001. They studied several effects in neutrino - nucleus interactions~\cite{Bell-LlewellynSmith1970, Bell-LlewellynSmith1971}.

More formal, mathematical topics on \emph{hadronic symmetry classification schemes, Melosh transformations, and all that ... }, John presented in his Lectures Notes of the Schladming Winter School 1974~\cite{Bell-SchladmingLectures}, Austria, and in further papers in collaboration with Anthony Hey~\cite{Bell-Hey} and Henri Ruegg~\cite{Bell-Ruegg-NuclPhysB93, Bell-Ruegg-NuclPhysB104}.

About John's last interests, QCD, gluon condensate and potential models, I have reported already in the previous Section; that was the topic I had the great joy to collaborate with him.\\

John received several prestigious awards, among them the Dirac Medal of the Institute of Physics (1988), the Dannie Heinemann Prize of the American Physical Society (1989), and the Hughes Medal of the Royal Society (1989):

\emph{``For his outstanding contributions to our understanding of the structure and interpretation of quantum theory, in particular demonstrating the unique nature of its predictions.''}

Moreover, in 1988 he received honorary degrees from both Queen's University Belfast and Trinity College Dublin. John obtained these honours, as Mary Bell emphasized, mainly for his contributions in particle physics. I personally think that the discovery of the anomaly had such an enormous impact on several branches in physics that sooner or later John, if he had lived longer, would have been awarded the Nobel Prize together with Jackiw and Adler.

\section{John Bell -- the Accelerator Physicist}\label{sec:Bell-accelerator-physics}

As already mentioned John started his scientific career at Malvern in 1949. There he joined the group of William Walkinshaw and was mainly concerned with the theory of particle accelerators. Walkinshaw highly appreciated Bell's abilities (recorded in Ref.~\cite{Burke-Percival}):

\emph{``Here was a young man of high caliber who soon showed his independence on choice of project, with a special liking for particle dynamics. His mathematical talent was superb and elegant.''}\\

John began with the study of a dielectric-loaded LINAC (Linear Accelerator) for electrons~\cite{Bell-accelerator-1, Bell-accelerator-2}, which at the beginning were used for medical purposes and later on for basic science. In 1951 Walkinshaw's group moved to Harwell and was one of those groups that established the Theory Division of AERE. In the course of the setting-up of CERN they also began to investigate proton accelerators. John himself contributed to \emph{``Scattering-''} and \emph{``Phase debunching by focussing foils in a proton linear accelerator''}~\cite{Bell-accelerator-8, Bell-accelerator-9}. Although at that time John's works appeared as internal reports they were highly appreciated and read by the accelerator community, in particular, his report on \emph{``Basic algebra of the strong focussing system''}~\cite{Bell-accelerator-14} received much attention. Just 2 works out of 21 papers (see Bell's collected works in Ref.~\cite{Bell-collectedPapers}), the topic was on strong focussing, John's special interest, have been published in journals~\cite{Bell-accelerator-12, Bell-accelerator-18}. In fact, an other work about \emph{``Linear accelerator phase oscillations''}~\cite{Bell-accelerator-21}, John was pleased with, he submitted to a journal and got back a `typical' referee report, as Mary Bell remembered~\cite{MaryBell2002}: \emph{``One referee said it was too short, the other that it was too long.''} So John gave up, he was leaving the accelerator field anyhow and turned to particle physics.\\

In the beginning of the 1980s John became interested again in accelerator physics. It was the time of the construction of the SPS (Super Proton Synchrotron) and LEAR (Low Energy Antiproton Ring) at CERN. There he collaborated with Mary on \emph{electron cooling, quantum beamstrahlung and bremsstrahlung}, which will be reported in Sect.~\ref{sec:JointWorks-Bell}.\\

A particularly attractive work, in my opinion, was Bell's combination of the Unruh effect of quantum field theory with accelerator physics. According to Unruh, a uniformly accelerated observer through the electromagnetic vacuum will experience a black body radiation at a temperature proportional to the acceleration~\cite{Unruh}. There is a close connection between the Unruh effect and the Hawking radiation of a black hole, which can be seen already in the temperature formula. From the Unruh temperature $T_U$ for a given acceleration $a$
\beq
k\, T_U \;=\; \frac{\hbar \,a}{2\pi c} \quad \longrightarrow \quad \frac{\hbar \,\kappa}{2\pi c} \;=\; \frac{\hbar \,c^3}{8\pi GM} \;=\; k\,T_H
\eeq
follows directly the Hawking $T_H$ temperature~\cite{Hawking} for the \emph{surface gravity} $\kappa = \frac{c^2}{2R_S}$ of a black hole with Schwarzschild radius $R_S = \frac{2GM}{c^2}\,$.

John discussed this subject with Jon Leinaas, a Norwegian CERN Fellow in the early 1980s, who got interested in the Unruh effect. Starting from a paper of John David Jackson~\cite{Jackson1976} about the spin polarization effect of electrons circulating in a storage ring, they tried to find a connection to the Unruh effect. Their idea was to consider accelerated electrons as detectors, with the spin degree of freedom used to measure the temperature of the radiation. Indeed, what John and Leinaas found was that the spin depolarization effect of the electrons in a storage ring is closely related to the thermal effect of linearly accelerated electrons~\cite{Bell-Leinaas1983}. The effect is small but in agreement with the measured values at a storage ring. However, there are complications due to the circular motion (Thomas precession), a resonance occurs, etc. ...  For this reason \emph{``the measurements cannot be considered as a direct demonstration of the (circular) Unruh effect. Therefore a measurement of the vertical fluctuations would be of interest, as a more direct demonstration of this effect. However, these fluctuations are small, and it is not clear whether it would be possible to separate this effect from other perturbations in the orbit.''}~\cite{Bell-Leinaas1987}. These subtle points show their great effort and achievement to test a quite sophisticated theory in an advanced experiment of accelerator physics. A discussion of the Unruh effect for the case of extended thermometers is given in Ref.~\cite{Bell-Hughes-Leinaas1985}.

\section{John and Mary Bell}

\begin{figure}
\begin{center}
\includegraphics[width=0.66\textwidth]{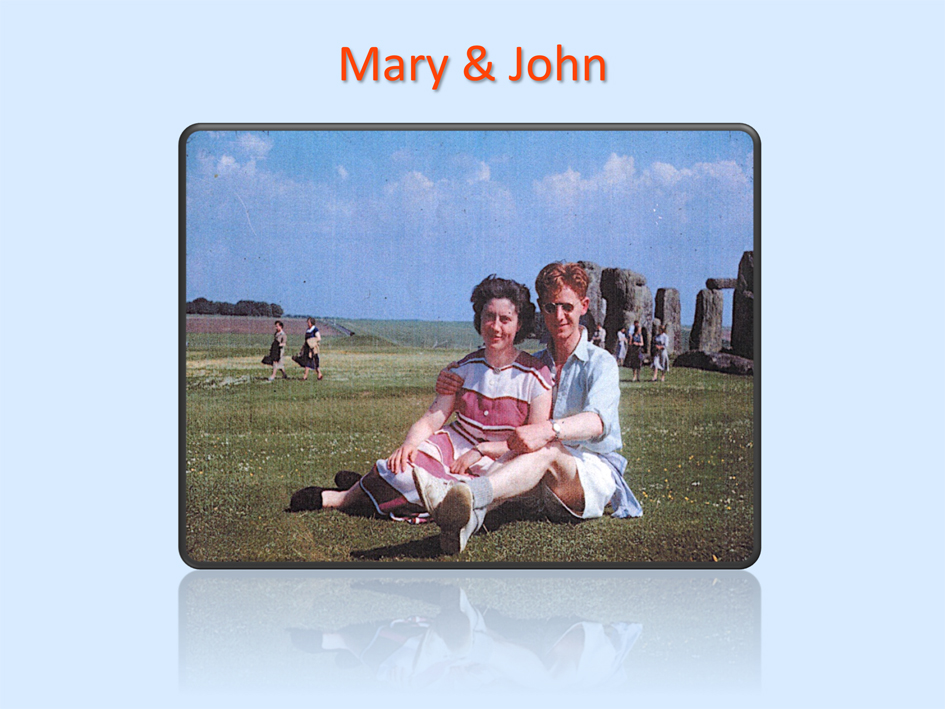}
\normalsize{
\caption{Mary and John Bell at Stonehenge in the 1950s. Foto: \copyright Mary Bell.}
\label{fig:Mary-John-Stonehenge}}
\end{center}
\end{figure}

John and Mary met for the first time in 1949 when John came to Harwell, where Mary had worked on reactors since 1947. At 1949/1950, both moved for a year to Malvern, first John then Mary followed, and they returned again to Harwell. At that time John was still without a beard as one could see from an excursion of the couple to Stonehenge in the 1950s, see Fig.~\ref{fig:Mary-John-Stonehenge}. They obviously enjoyed both their scientific and private life. In 1954, they were married and pursued their careers together. In course of their life they collaborated several times on issues of accelerator physics.

In 1960, both moved to CERN and lived in Geneva. End of 1963, they took a sabbatical year staying in the USA. There John had time to work on his \emph{`hobby'}, the foundations of quantum mechanics, and it was at SLAC where he had written his celebrated \emph{`inequality paper'}.\\

\begin{figure}
\begin{center}
\includegraphics[width=0.6\textwidth]{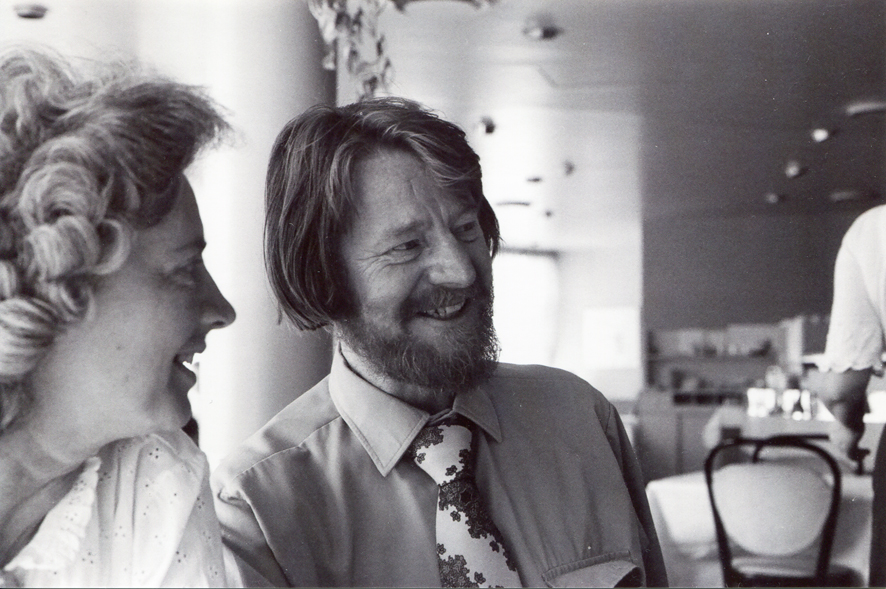}
\normalsize{
\caption{Mary and John Bell in the former \emph{`Haas Haus Restaurant'} in Vienna in 1982. Foto: \copyright Renate Bertlmann.}
\label{fig:Mary-John1982}}
\end{center}
\end{figure}

\begin{figure}
\begin{center}
\includegraphics[width=0.73\textwidth]{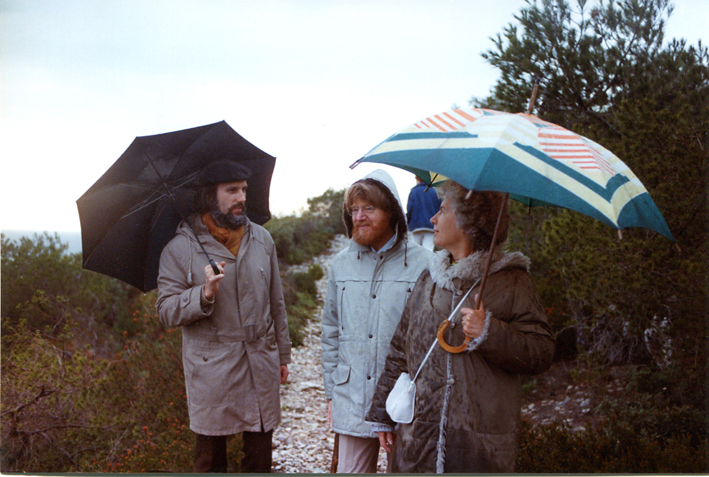}
\normalsize{
\caption{Reinhold, John \& Mary on a walk in the Calanque de Port d'Alon, South of France, 1983. Foto: \copyright Renate Bertlmann.}
\label{fig:Bell-walk1983}}
\end{center}
\end{figure}

When I think of John I always remember both, John and Mary, the couple. In lasting memory are the many pleasant events, the lunches, walks, ... , Renate and myself spent together with the Bells: For instance, the lunch in the former \emph{`Haas Haus Restaurant'} in Vienna in 1982, see Fig.~\ref{fig:Mary-John1982},
when John was the distinguished \emph{`Schr\"odinger Guest Professor'} at our Institute of Theoretical Physics; Or the beautiful walks in the Calanque de Port d'Alon in 1983, where we even could arrange a typically British weather for John and Mary, which was not so easy to get in South of France, see Fig.~\ref{fig:Bell-walk1983}.
I was at that time a Visiting Professor at the University of Marseille, Luminy, and the Bells visited us in Bandol; Or, finally, our last reunion in Bures sur Yvette in April 1990, when I stayed at the CNRS (Centre Nationale de la Recherche Scientific).\\

Mary always was present, what John nicely had expressed in the preface of his book \emph{Speakable and Unspeakable in Quantum Mechanics} \cite{Bell-book1987}, when he wrote the moving tribute to Mary:

\emph{``... I here renew very especially my warm thanks to Mary Bell. When I look through these pages again I see her everywhere.''}\\

\subsection{Mary Bell}

Mary, with maiden name Ross, originates from a Scottish family resident in Glasgow. The parents were Alexander Munro Ross, a commercial manager in a ship-building firm, and Catherine Brown Wotherspoon, an elementary school teacher. Her father was quite enthusiastic to send Mary and her two older sisters to university.

As a teenager Mary attended the \emph{Hyndland Secondary School} which was situated at a two minutes walk from where she lived. There she spent all her school life except for the year 1939--1940, when her father had sent her to the \emph{Kingussie High School} in Iverness-shire to stay with his two sisters. This was to elude the bombing in Glasgow at the beginning of World War II. There Mary's talents as a storywriter had been already discovered, for instance, her amusing agent short-story \emph{``The Secret Service Agent''} had been printed in the \emph{Kingussie Secondary School Magazine}~\cite{Mary-KingussieMagazine}.

Mary returned to Hyndland for the year 1940--1941. The Hyndland Secondary School catered for \emph{all} types of pupils, academic and non-academic, but in different classes. Mary was quite fortunate that this school had a strong science department. Two of the headmasters were science teachers: There was an older building, but also a fine airy new building with a number of laboratories, art rooms, and an apartment for domestic science. Hyndland School was without fees, in contrast to the \emph{`Girls' High School'} (in the centre of the city) which had fees. Hyndland had pupils, boys \emph{and} girls, from a wide area. There was an examination which all pupils (at all schools) had to take on a given day for getting money for the Girls' High School, which was considered as a good school. Mary told me in a letter~\cite{Mary-letter-Reinhold-16-2-2015}:

\emph{``Unfortunately, I won this, but my father insisted that I stay at the local school, instead of travelling into the town each day. He was right. The Girls' High School had only botany among the science subjects (no physics).''}

At Hyndland, science was a strong subject. Mary even obtained a prize for \emph{``General Excellence in Dynamics''} in the season 1938--1939, a phrase John found quite amusing, and she still remembers learning Newton's Laws of Motion.\\

Mary also won an open \emph{`Bursary Competition for the University'}, in particular for a female studying science, and got enough money to pay the university fees. \emph{``But money for fees was not really a problem''} as she said.

The university course was four years in \emph{Mathematics and Natural Philosophy} (Natural Philosophy was the former term for Physics). Due to World War II many students were called up after two years and could qualify for a wartime degree. Mary was allowed to stay three years because of the need for demonstrators. Eventually in 1944, she was called up and sent to TRE in Malvern, which was the big government establishment for radio navigation, radar, infra-red detection and related subjects. After a year the war ended and she returned to the University. When she graduated in 1947, it was quite natural for her to apply for AERE at Harwell.\\

At that time, there were only a few female scientists in the Theory Division, about 3 of 20 persons as Mary remembers. Science was not considered as an appropriate profession for a woman. Thus it was quite exceptional, in my opinion, that Mary had chosen already as a young girl a scientific subject for her career and not just a \emph{`girl subject'}. When I once asked her why, she replied in her beautiful handwriting~\cite{Mary-letter-Reinhold2013}:

\emph{``I always liked to solve little problems, even when I knew only arithmetic.''}

\begin{figure}
\begin{center}
\includegraphics[width=0.25\textwidth]{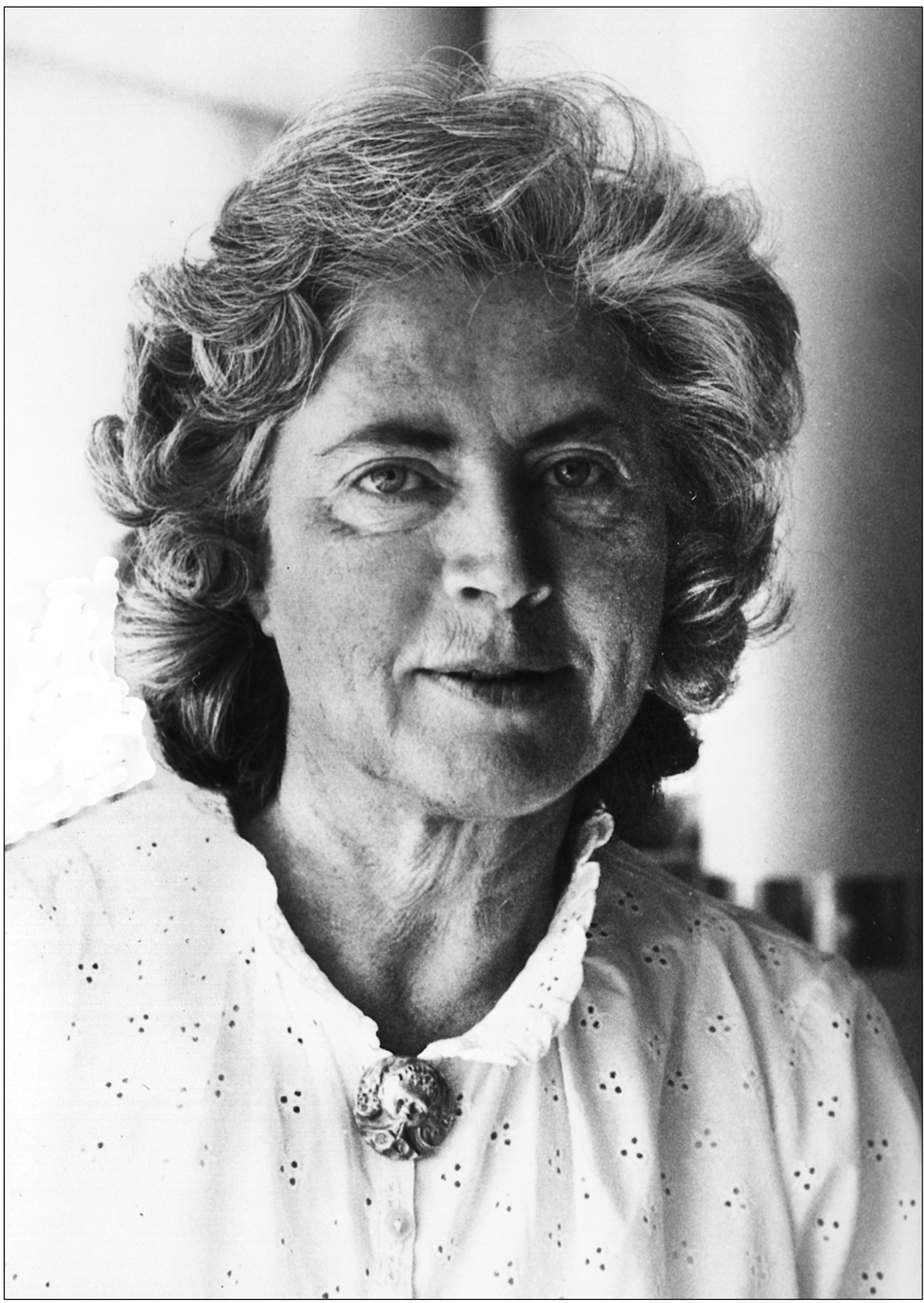}
\normalsize{
\caption{Mary Bell in 1982. Foto: \copyright Renate Bertlmann.}
\label{fig:MaryBell-1982}}
\end{center}
\end{figure}

So already at an early stage her vision for a career was science and not so much being just a `house wife', may be also her good scientific education in the Hyndland School was quite encouraging for her.\\

Mary's appointment at Harwell started in February 1947. The research was on nuclear reactors but soon it broadened including nuclear physics, accelerators and related topics. The head of the theory division was the well-known Klaus Fuchs, a left-wing refugee from Germany with British citizenship, who had worked before on the \emph{`Manhatten Project'} in Los Alamos and on the \emph{`Tube Alloys'} programme, the British atomic bomb project. Fuchs had close contact with his compatriot Rudolf Peierls (later on Sir Rudolf) who was a consultant to AERE and had much influence there.

\emph{``At the beginning the work at Harwell was not very interesting''}, Mary remembered~\cite{Mary-letter-Reinhold2013}. But then Fuchs told her \emph{``to move to more interesting problems.''} She was instructed to work on treating the control rods in a fast fission reactor. Some calculations she published in the paper: \emph{``Slow neutron absorption cross-section of the elements''}~\cite{MaryBell-Pub-1949}. In fact, she had to carry out a special perturbation calculation devised by Klaus Fuchs. Soon after this work was completed Fuchs confessed, in 1950, that he was a spy (out of conviction) supplying information from the American, British, Canadian atomic bomb project to the Soviet Union. He was sentenced to 14 years imprisonment, got arrested but released in 1959 and emigrated to the former German Democratic Republic.\\

As already mentioned, in 1950, Mary and John were sent to TRE in Malvern to join the accelerator group. After a year working there, for a second time,  Mary and all the other physicists moved back to Harwell to the Theory Division which was then headed by Brian Flowers (later on Lord Flowers). In the next years Mary worked there on several subjects of accelerator physics. A fairly complete list of papers, mainly AERE reports, are given by the Refs.~\cite{MaryBell-Pub-1949, MaryBell-AERE-T/M-55, MaryBell-AERE-T/M-58, MaryBell-AERE-T/M-66, MaryBell-AERE-T/M-70, MaryBell-AERE-T/M-83, MaryBell-AERE-T/M-99, MaryBell-AERE-T/M-101, MaryBell-AERE-T/M-112, MaryBell-AERE-T/M-125, MaryBell-AERE-T/M-128, MaryBell-AERE-T/M-139, MaryBell-AERE-T/R-2077, MaryBell-AERE-T/R-2209, MaryBell-AERE-T/M-155, MaryBell-Pub-1957, MaryBell-AERE-R-3326}. Some works, whose topics I personally find very interesting, are:

\begin{itemize}
\item{\emph{``Series impedance of double-ridged wave guides.''}~\cite{MaryBell-AERE-T/M-58}}
\item{\emph{``Focussing system for the 600 MeV proton linear accelerator.''}~\cite{MaryBell-AERE-T/M-112}}
\item{\emph{``Nonlinear equations of motion in the synchrotron.''}~\cite{MaryBell-AERE-T/M-125}}
\item{\emph{``Injection into the 7 GeV synchrotron.''}~\cite{MaryBell-AERE-T/M-155}}
\end{itemize}

After working for a decade at Harwell John and Mary felt somehow ready for a change, mainly since John had been attracted by particle physics. In 1960, they moved to CERN, the big European center for particle physics, John to the Theory Division and Mary to the Accelerator Research Group.\\

At CERN Mary wrote a lot of computer programmes representing the orbits of the circulating particles in different accelerator machines. These orbits, of course, had to be extremely accurate. In the 1970s, when particle accelerators for higher energy  (some GeV region) were planed and constructed the question of cooling particles in a storage ring became a hot issue. Two methods were debated at CERN for further use in the accelerators in order to reduce the large phase space spread of an ion, proton or antiproton beam in the ring: \emph{`Electron cooling'} proposed by Gersh Itskovich Budker~\cite{Budker1967} and successfully applied at the Novosibirsk storage ring~\cite{Budker1976} and \emph{`stochastic cooling'} invented by Simon van der Meer at CERN.

Mary was studying electron cooling (a picture of her at about that time can be seen in Fig.\ref{fig:MaryBell-1982}) which is based on repeated interactions of protons circulating in the storage ring with a dense and cold electron beam. She published a paper on \emph{``Electron cooling with magnetic field''}~\cite{MaryBell-PartAccelerators1980}, where she discussed the basic equations for the slowing of a proton in an electron gas with magnetic field. She was also part of a working group headed by Frank Kienen, which studied the cooling of protons by means of electrons in the ICE (Initial Cooling Experiment) storage ring in order to test the feasibility of a high luminosity proton - antiproton collider. Kienen often had discussions with John. It raised John's interest again in accelerator physics and he began to collaborate with Mary on these topics, which will be reported in Sect.~\ref{sec:JointWorks-Bell}. The results of the ICE measurements of the momentum spread, beam profile, beam lifetime, etc. ... , were quite in agreement with theory and were published in a common paper~\cite{MaryBell-etal-1979}, where also Carlo Rubbia (appointed to Director-General of CERN in 1989) and Simon van der Meer contributed. Further details of the cooling method in the ICE were published in Ref.\cite{MaryBell-etal-1981} and Mary had to prepare the paper for publication. As she remembers~\cite{MaryBell2002}:

\emph{``This involved a lot of condensing to make the paper a reasonable size, and John kindly helped me. He particularly liked to be acknowledged at the end for `helping with the typing'.''}\\

Also stochastic cooling was explored in the ICE storage ring and in these tests it turned out that van der Meer's stochastic cooling method was sufficient to be implemented in the newly constructed SPS (Super Proton Synchrotron) to collide protons and antiprotons in the same ring. Rubbia and van der Meer were the leading figures in constructing the SPS and detected in 1983 the W and Z bosons. In 1984, they received the Nobel Prize \emph{``for their decisive contributions to the large project, which led to the discovery of the field particles W and Z, communicators of weak interactions.''} Nevertheless, electron cooling was used further on in the redesigned ring LEAR (Low Energy Antiproton Ring) to decelerate and store antiprotons.

\subsection{Joint Works of the Bells}\label{sec:JointWorks-Bell}

When John stayed at Malvern also Mary joined the accelerator group of Walkinshaw. Having previously worked on nuclear reactors she now switched to accelerator physics. Clearly, Mary and John had many discussions on accelerator issues, and back again at Harwell they wrote a common paper in 1952 (where Mary still signed with her maiden name Mary Ross) about: \emph{``Heating of focussing foils by a proton beam''}~\cite{MaryBell-AERE-T/M-70}. Since John was leaving the accelerator field and turned to particle physics there was a long pause of collaboration of about 30 years. John's interest in accelerator physics got revived about 1980 at CERN when Mary was working on electron cooling.

John and Mary's first common paper at CERN was devoted to \emph{``Electron cooling in storage rings''}~\cite{John-MaryBell-PartAccelerators1981}, where they calculated the effect of \emph{`flattening'} of the electron velocity distribution, which meant that the longitudinal velocity spread was suppressed, to increase the rate of cooling of small betatron oscillations. This paper they devoted to the nuclear and accelerator physicist Yuri Orlov who was imprisoned at that time in the Soviet Union for his human rights activism, freed later on and deported to the USA. Such an act of solidarity was quite typical for the Bells.

A paper on a similar issue: \emph{``Capture of cooling electrons by cool protons''}~\cite{MaryBell-John-PartAccelerators1982} followed. There Mary and John presented formulae for the capture of low-energy electrons by stationary protons using Maxwellian and flattened electron velocity distributions. The latter was more appropriate for the electron beams used in the accelerator proton beam cooling experiments. They found out that the flattening increased the capture rate by a factor of about two. Also similar formulae for the capture of antiprotons by protons were mentioned.\\

In a further paper \emph{``Radiation damping and Lagrange invariants''}~\cite{MaryBell-John-PartAccelerators1983} Mary and John proposed a general formula for the damping of small oscillations about closed orbits. It was applied to derive the results for the effects of classical radiation damping on storage ring orbits.\\

Next, Mary and John turned to \emph{`beamstrahlung'} which is the radiation of the whole beam of charged particles in a storage ring or LINAC, when the beam interacts with the electromagnetic field of the other beam. In their paper \emph{``Quantum beamstrahlung''}~\cite{MaryBell-John-PartAccelerators1988-Vol22} they showed that the well-known Blankenbecler-Drell formula~\cite{Blankenbecler-Drell1987} for the bremsstrahlung energy loss of a relativistic electron passing through the field of a cylindric charge agrees with the popular \emph{`Russian formula'}~\cite{Sokolov-Ternov1971, Nikishov-Ritus1967, Matveev1956-1957, Baier-Katkov1968-1969} if the spin-flip contributions were added to Blankenbecler-Drell approach.

In \emph{``End effects in quantum beamstrahlung''}~\cite{MaryBell-John-PartAccelerators1988-Vol24} Mary and John also included so-called \emph{`end effects'} in their beamstrahlung calculations. These effects occurred at the ends of a sharply bounded cylindrical charge bunch. Their conclusion was that end effects were indeed negligible when the bunch length is large compared to its typical quantum length.

In their last joint work \emph{``Quantum bremsstrahlung in almost uniform fields''}~\cite{MaryBell-John-NuclInstr1989} Mary and John studied inhomogeneity effects in quantum bremsstrahlung in the extreme quantum limit, in the case of a weakly nonuniform deflecting field.

\section{Out of the Blue}

At CERN Bell was a kind of \emph{`Oracle'} for particle physics, consulted by many colleagues who wanted to get his approval for their ideas. Of course, I had heard that he was also a leading figure in quantum mechanics, specifically, in quantum foundations. But nobody could actually explain to me his work in this quantum area, neither at CERN nor anywhere else. The standard answer was: \emph{``He discovered some `relation' whose consequence was that quantum mechanics turned out alright. But we knew that anyway, so don't worry.''} And I didn't. John, on the other hand, never mentioned his quantum works to me in the first years of our collaboration. Why? This I understood later on, John was reluctant to push somebody into a field that was quite unpopular at that time.\\

\begin{figure}
\begin{center}
\includegraphics[width=0.65\textwidth]{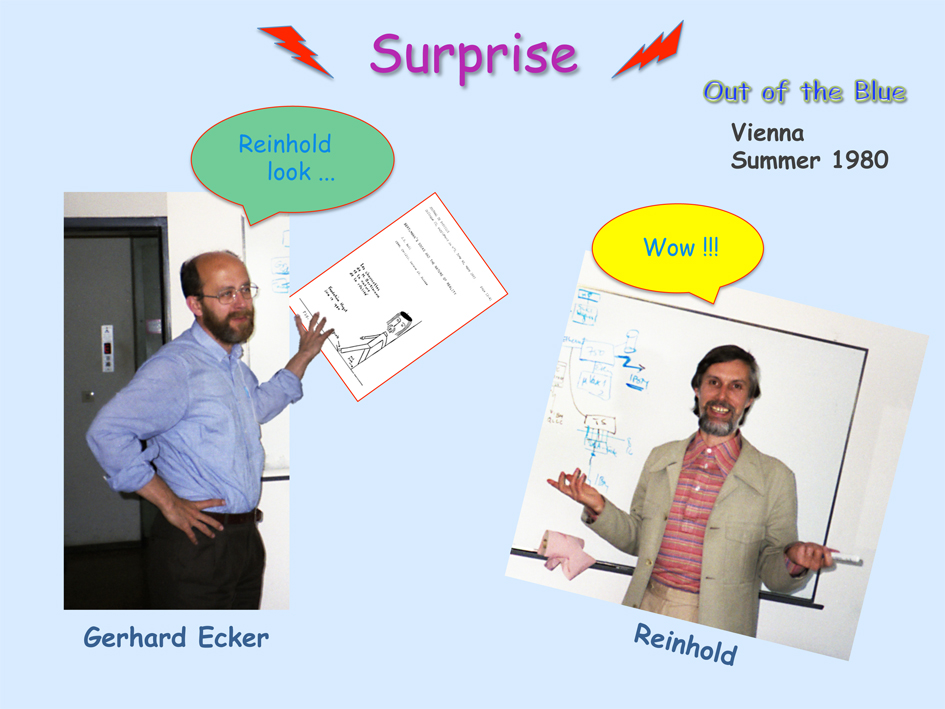}
\normalsize{
\caption{Gerhard Ecker standing in front of Reinhold Bertlmann holds a preprint in his hands with the title \emph{``Bertlmann's socks and the nature of reality''}~\protect\cite{Bell-Bsocks-CERNpreprint}. Fotos: \copyright Renate Bertlmann.}
\label{fig:Out-of-the-blue}}
\end{center}
\end{figure}

At the end of summer 1980, I returned for some time to my home institute in Vienna to continue our collaboration on the gluon condensate potential from there. At that time, there was no internet, and it was a common practice to send preprints (typed manuscripts) of the work, prior to publication, to all main physics institutions in the world. Also we in Vienna had such a preprint shelf where each week the new incoming preprints were exhibited.

One day, on the 15th of September, I was sitting in the Institute's computer room, handling my computer cards, when my colleague Gerhard Ecker, who was in charge of receiving the preprints, rushed in waving a preprint in his hands (see Fig.~\ref{fig:Out-of-the-blue}). He shouted, \emph{``Reinhold look -- now you're famous\,!''} I hardly could believe my eyes as I read and reread the title of a paper by John S. Bell~\cite{Bell-Bsocks-CERNpreprint}: \emph{``Bertlmann's socks and the nature of reality''}.\\

I was totally excited. Reading the first page my heart stood still:\\

\begin{figure}
\begin{center}
\setlength{\fboxsep}{2pt}\setlength{\fboxrule}{0.8pt}\fbox{
\includegraphics[angle = 0, width = 75mm]{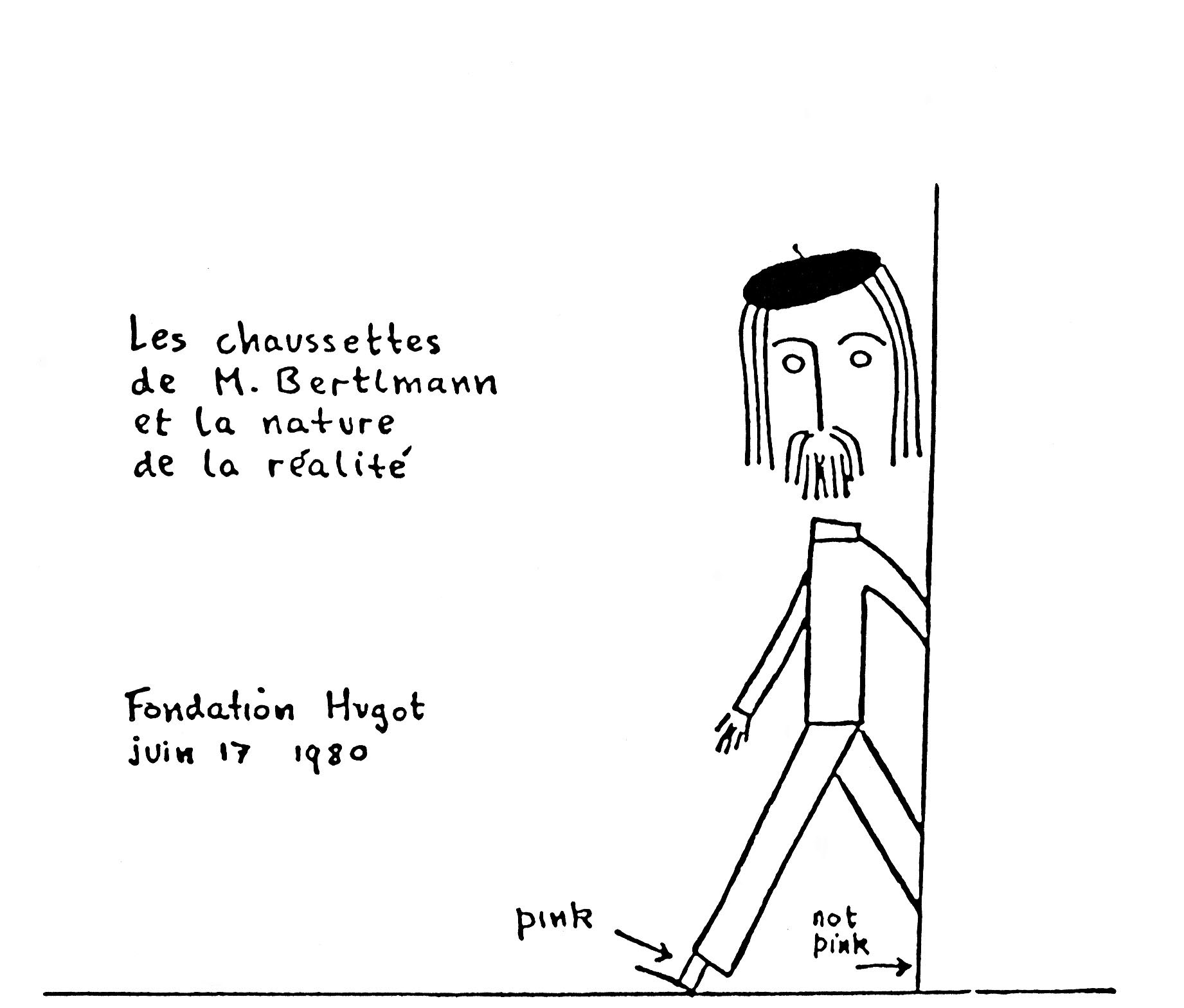}}
\normalsize{
\caption{Cartoon in the CERN preprint Ref.TH.2926-CERN \emph{``Bertlmann's socks and the nature of reality''} of John Bell from 18th July 1980~\protect\cite{Bell-Bsocks-CERNpreprint}. The article is based on an invited lecture John Bell has given at le Colloque sur les \emph{``Implications conceptuelles de la physique quantique''}, organis\'{e} par la Foundation Hugot du Coll\`{e}ge de France, le 17 juin 1980, published in Journal de Physique~\protect\cite{Bell-Bsocks-Journal-de-Physique}.}
\label{fig:Bell-Bsocks-CERNpaper-cartoon}}
\end{center}
\end{figure}

\emph{``The philosopher in the street, who has not suffered a course in quantum mechanics, is quite unimpressed by Einstein-Podolsky-Rosen correlations~\cite{EPR}. He can point to many examples of similar correlations in every day life. The case of Bertlmann's socks is often cited. Dr. Bertlmann likes to wear two socks of different colours. Which colour he will have on a given foot on a given day is quite unpredictable. But when you see (Fig.~\ref{fig:Bell-Bsocks-CERNpaper-cartoon}) that the first sock is pink you can be already sure that the second sock will not be pink. Observation of the first, and experience of Bertlmann, gives immediate information about the second. There is no accounting for tastes, but apart from that there is no mystery here. And is not the EPR business just the same ?''}\\

Seeing the cartoon John has sketched by himself (see Fig.~\ref{fig:Bell-Bsocks-CERNpaper-cartoon}), showing me with my odd socks, nearly knocked me down. All this came so unexpectedly -- I had not the slightest idea that John had noticed my habits of wearing socks of different colours, a habit I cultivated since my early student days, my special \emph{`generation-68'} protest. This article pushed me \emph{instantaneously} into the quantum debate, which changed my life. Since then \emph{`Bertlmann's socks'} had developed a life of its own. You can find Bertlmann's socks everywhere on the internet, in popular science debates, and even in the fields of literature and art.\\

Now the time had come for diving into the quantum world of John, to understand why the \emph{``EPR business''} was \emph{not} just the same as \emph{``Bertlmann's socks''}, and to appreciate his profound insight. It was John who pushed the rather philosophical Einstein-Bohr discussions of the 1930s about realism and incompleteness of quantum mechanics onto physical grounds. His axiom of \emph{locality} or \emph{separability} was the essential ingredient of a \emph{hidden variable theory} and illuminated the physical difference between all such hidden variable theories and the predictions of quantum mechanics. Due to \emph{`Bell's Theorem'} we can distinguish \emph{experimentally} between quantum mechanics and all local realistic theories with hidden variables. I was impressed by the clarity and depth of John's thoughts. From this time on we had fruitful discussions about the foundations of quantum mechanics and this was a great fortune and honour for me. It was just about the time when Alain Aspect \cite{Aspect-Dalibard-Roger1982} finished his time-flip experiments on Bell inequalities and the whole field began to attract the increasing interest of physicists. For me a new world opened up -- the universe of John Bell -- and caught my interest and fascination for the rest of my life.

\section{Bell and the Quantum}

John never was satisfied with the interpretations of quantum mechanics. Already as a student at Queen's University Belfast he disliked the so-called \emph{Copenhagen Interpretation} with its distinction between the quantum world and the classical world. He wondered, \emph{``Where does the quantum world stop and the classical world begin?''} He wanted to get rid of that division.

For him it was clear that hidden variable theories, where quantum particles do \emph{have} definite properties governed by hidden variables, would be appropriate to reformulate quantum theory. \emph{``Everything has definite properties\,!''} I remember John saying.

\subsection{Contextuality}

Hidden variable theories (HVT) as well as quantum mechanics describe an ensemble of individual systems. Whereas in QM the orthodox (Copenhagen) doctrine tells us that measured properties, e.g. the spin of a particle, have no definite values before measurement, the HVT in contrast postulate that the properties of individual systems do \emph{have} pre-existing values revealed by the act of measurement.

Given a set of observables $\{A, B, C, ...\}$ then a hidden variable theory assigns to each individual system a set of values (eigenvalues) $\{v(A), v(B), v(C), ...\}$ to the corresponding observables~\cite{Mermin-RevModPhys}. The hidden variable theory provides a rule how to distribute the values over all individual systems of an ensemble. Such states, specified by the quantum mechanical state vector \emph{and} by the additional hidden variable, are called \emph{dispersion-free}.

If a functional relation, $f(A, B, C, \ldots ) \,=\, 0$, is satisfied by a set of mutually commuting observables $A, B, C, \ldots$, then the same relation must hold for the values in the individual systems, $f\big(v(A), v(B), v(C), \ldots \big) \,=\, 0$. Amazingly, just by relying on above two conditions a contradiction can be constructed, a so-called \emph{No-Go Theorem}.\\

Bell started his investigation \emph{``On the problem of hidden variables in quantum mechanics''}~\cite{Bell-RevModPhys1966} in 1964~\footnote{Due to a delay in the Editorial Office of Review of Modern Physics the paper was published not until 1966.} by criticizing John von Neumann who had given already in 1932 a proof~\cite{vonNeumann1932} that dispersion-free states, and thus hidden variables, are incompatible with quantum mechanics. What was the criticism? Consider three operators $A, B, C$ with condition $C \,=\, A + B \,$, then the correspondingly attached values also have to satisfy $v(C) \,=\, v(A) + v(B)\,$, since the operators $A, B$ are supposed to commute.

Von Neumann, however, also imposed the additivity property for \emph{noncommuting} operators. \emph{``This is wrong!''} John grumbled and illustrated his dictum, before giving a general proof, with the example of spin measurement. Measuring the spin operator $\sigma_x$ for a magnetic particle requires a suitably oriented Stern-Gerlach apparatus. The measurements of $\sigma_y$ and $(\sigma_x + \sigma_y)$ demand different orientations. Since the operators cannot be measured simultaneously, there is no reason for imposing additivity. Of course, for the quantum mechanical expectation values we have additivity in the mean $\langle \psi | A + B | \psi \rangle \,=\, \langle \psi | A | \psi \rangle + \langle \psi | B | \psi \rangle \,$, irrespective whether $A, B$ commute or not.

Interestingly, already in 1935 the mathematician and philosopher Grete Hermann~\cite{GreteHermann1935} raised her objection to von Neumann's assumption but she was totally ignored. Also Simon Kochen and Enst Specker~\cite{Kochen2002}, when reading von Neumann's proof in 1961, had their doubts about the additivity for noncommuting operators.

John being aware of Gleason's Theorem~\cite{Gleason1957}, which was not explicitly addressed to HVT but aimed instead to reduce the axioms for QM, established the following corollary~\cite{Bell-RevModPhys1966}, more directed to HVT:

\begin{corollary}[Bell's Corollary]\ \
Consider a state space $\V$. If $dim \,\V > 2$ then the additivity requirement for expectation values of commuting operators cannot be met for dispersion-free states.
\label{corollary:Bell}
\end{corollary}

Corollary~\ref{corollary:Bell} states that for $dim \V > 2$ it is in general impossible to assign a definite value for each observable in each individual quantum system. Thus Bell pointed to another class of hidden variable models, where the results may depend on different settings of the apparatus. Such models are called \emph{contextual} and may agree with quantum mechanics. Corollary~\ref{corollary:Bell}, on the other hand, states that all \emph{noncontextual} HVT are in conflict with QM (for $dim > 2$). Hence the essential feature for the difference between HVT and QM is \emph{contextuality}.\\

In 1967, Simon Kochen and Ernst Specker published their famous paper on \emph{``The problem of hidden variables in quantum mechanics''} \cite{Kochen-Specker}, where they established their No-Go Theorem that noncontextual hidden variable theories are incompatible with quantum mechanics.

\begin{theorem}[Kochen-Specker Theorem]\ \
In a Hilbert space $\Ha$ of $dim \,\Ha > 2$ it is impossible to assign values to all physical observables while simultaneously preserving the functional relations between them.
\label{theorem:Kochen-Specker}
\end{theorem}

Since then, contextuality has become an important issue in the research of quantum systems (see, e.g., Refs.~\cite{Cabello-Guehne-etal_PRL2013, Canas-Cabello-etal2013, Cabello-etal_PRL-112-040401-2014, Cabello-etal_PRL-111-180404-2013, Guehne-Cabello-etal2013, Acin-etal2012, Cabello-talk2014}, and references therein).

\subsection{Nonlocality}

The starting point of John's quantum studies was Bohm's~\cite{Bohm1952} reinterpretation of quantum theory as a deterministic, realistic theory with hidden variables. Although Bohm's work was not at all appreciated by the physics community, neither by Einstein nor by Pauli, John was very much impressed by Bohm's work and often remarked, \emph{``I saw the impossible thing done''.} To me John continued, \emph{``In every quantum mechanics course you should learn Bohm's model!''}

John examined Bohm's model quite carefully and analyzed a system of two particles with spin $\frac{1}{2}$~\cite{Bell-RevModPhys1966}, interacting with the external magnetic fields $\vec{B}$ of two magnets that analyze the spins. The argumentation goes as follows. The hidden variables in this two-particle system are two vectors $\vec{X}_1$ and $\vec{X}_2$ which yield the results for position measurements. The variables are supposed to be distributed in configuration space with probability density
\begin{equation}\label{probability-density}
\rho (\vec{X}_1, \vec{X}_2) \;=\; \sum_{i,j} |\psi_{ij}(\vec{X}_1, \vec{X}_2)|^2 \;,
\end{equation}
which describes the quantum mechanical state; $\psi_{ij}$ is the solution of the Schr\"odinger equation. The position operators, the hidden variables, for the two-particle system then vary in time according to ($\hbar = 1, 2m = 1$)
\beq\label{position-operators-time-evolution}
\frac{d\vec{X}_1}{dt} &\;=\;& \frac{1}{\rho (\vec{X}_1,\vec{X}_2,t)} \,{\rm{Im}}\, \sum_{i,j} \psi_{ij}^\ast (\vec{X}_1,\vec{X}_2,t)\frac{\partial}{\partial\vec{X}_1}\psi_{ij} (\vec{X}_1,\vec{X}_2,t) \nonumber\\
\frac{d\vec{X}_2}{dt} &\;=\;& \frac{1}{\rho (\vec{X}_1,\vec{X}_2,t)} \,{\rm{Im}}\, \sum_{i,j} \psi_{ij}^\ast (\vec{X}_1,\vec{X}_2,t)\frac{\partial}{\partial\vec{X}_2}\psi_{ij} (\vec{X}_1,\vec{X}_2,t) \;.
\eeq
The strange feature now is that the trajectory equations (\ref{position-operators-time-evolution}) for the operators, the hidden variables, have a highly nonlocal character. Only in case of a factorizable wave function for the quantum system
\begin{equation}\label{wave-function-factorizable}
\psi_{ij}(\vec{X}_1, \vec{X}_2, t) \;=\; \eta_i (\vec{X}_1,t) \cdot \chi_j (\vec{X}_2,t)  \;,
\end{equation}
the trajectories decouple
\beq\label{position-operators-time-evolution-decoupled}
\frac{d\vec{X}_1}{dt} &\;=\;& \frac{1}{\sum_{i} |\eta_i (\vec{X}_1,t)|^2} \,{\rm{Im}}\, \sum_{i} \eta_{i}^\ast (\vec{X}_1,t)\frac{\partial}{\partial\vec{X}_1}\eta_i (\vec{X}_1,t) \nonumber\\
\frac{d\vec{X}_2}{dt} &\;=\;& \frac{1}{\sum_{j} |\chi_j (\vec{X}_2,t)|^2} \,{\rm{Im}}\, \sum_{j} \chi_{j}^\ast (\vec{X}_2,t)\frac{\partial}{\partial\vec{X}_2}\chi_j (\vec{X}_2,t) \;,
\eeq
and the trajectories of $\vec{X}_1$ and $\vec{X}_2$ are determined separately by involving the magnetic fields $\vec{B}(\vec{X}_1)$ and $\vec{B}(\vec{X}_2)$ respectively. However, in general, this is not the case. The trajectory of particle $1$ depends in a complicated way on the trajectory and wave function of particle $2\,$, and thus on the analyzing magnetic field acting on particle $2\,$, no matter how remote the particles are. Therefore, as John remarked~\cite{Bell-RevModPhys1966}: \emph{``In this theory an explicit causal mechanism exists whereby the disposition of one piece of apparatus affects the results obtained with a distant piece.''}

John, realizing the importance of this \emph{nonlocal feature}, wondered if it was just a defect of this particular hidden variable model, or is it somehow intrinsic in a hidden variable theory reproducing quantum mechanics. After playing around a bit to find a local account for the quantum results he could construct an impossibility proof, a \emph{Bell inequality}.

\subsection{Bell Inequalities}

In his paper \emph{``On the Einstein-Podolsky-Rosen paradox''}~\cite{Bell-Physics1964} John reconsidered the at that time totally disregarded paper of Albert Einstein, Boris Podolsky and Nathan Rosen (EPR)~\cite{EPR}. Therein the authors argued that quantum mechanics is an \emph{incomplete} theory and that it should be supplemented by additional parameters, the hidden variables. These additional variables would restore causality and locality in the theory. What was John's essence when considering Bohm's spin version~\cite{Bohm-Aharanov-EPRspin1957} of EPR?

\begin{figure}
\centering
\includegraphics[width=0.65\textwidth]{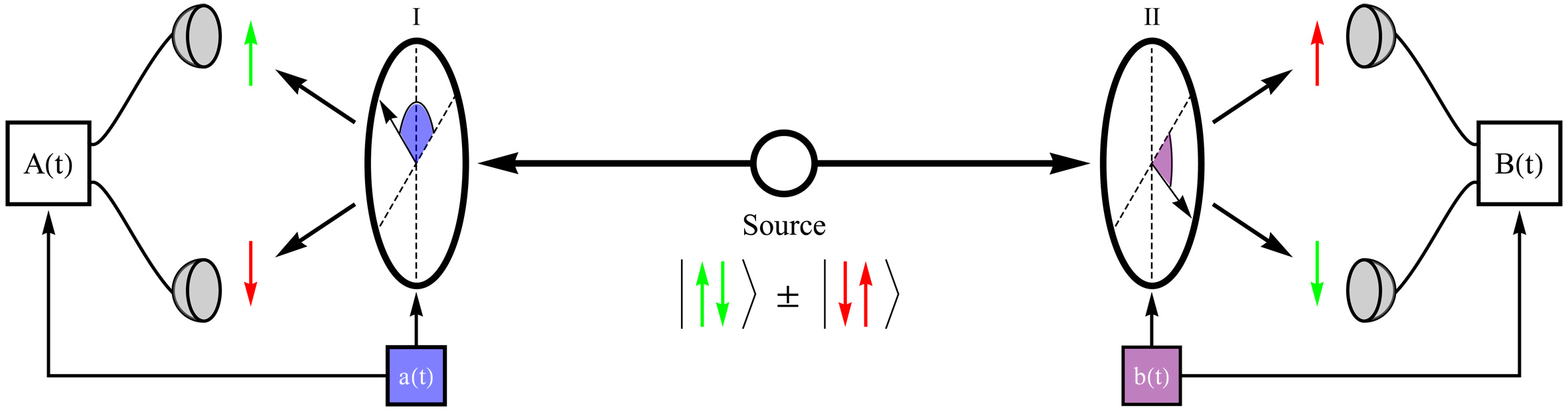}
\caption{In a Bohm-EPR setup a pair of spin $\frac{1}{2}$ particles, prepared in a spin singlet state, propagates freely in opposite directions to the measuring stations called Alice and Bob. Alice measures the spin in direction $\vec{a}$, whereas Bob measures simultaneously in direction $\vec{b}$.}
	\label{fig:Bell-setup-experiment}
\end{figure}

Let us analyze such a Bohm-EPR setup, where a pair of spin $\frac{1}{2}$ particles is produced in a spin singlet state and propagates freely into opposite directions (see Fig.~\ref{fig:Bell-setup-experiment}).
The spin measurement on one side, called Alice, performed by a Stern-Gerlach magnet along some direction $\vec{a}$ is described by the operator $\vec{\sigma}_A \cdot \vec{a}\,$ and yields the values $\pm 1\,$. Since we can predict in advance the result of $\vec{\sigma}_B \cdot \vec{b}\,$ on the other side, Bob's side, the result must be predetermined. This predetermination we specify by the additional variable $\lambda\,$. In such an extended theory we denote the measurement result of Alice and Bob by
\begin{equation}
A(\vec{a},\lambda) \;=\; \pm 1,0 \qquad \mbox{and} \qquad B(\vec{b},\lambda) \;=\; \pm 1,0 \;.
\end{equation}
We also include $0$ for imperfect measurements to be more general, i.e., what we actually require is
\begin{equation}
|A| \;\leq\; 1 \qquad \mbox{and} \qquad |B| \;\leq\; 1 \;.
\end{equation}
Then the expectation value of the joint spin measurement of Alice and Bob is
\begin{equation}\label{expectation-value-Bell-locality}
E(\vec{a},\vec{b}) \;=\; \int\!d\lambda\,\rho(\lambda)\,A(\vec{a},\lambda)\cdot B(\vec{b},\lambda) \;.
\end{equation}
The choice of the product in expectation value (\ref{expectation-value-Bell-locality}), where $A$ does not depend on Bob's settings, and $B$ does not depend on Alice's setting, is called \emph{`Bell's Locality Hypothesis'}. It is the obvious definition of a physicist as an engineer, and \emph{must not} be confused with other locality definitions, like local interactions or locality in quantum field theory.

The function $\rho(\lambda)$ represents some probability distribution for the variable $\lambda$, and does not depend on the measurement settings $\vec{a}$ and $\vec{b}\,$ which can be chosen truly free or random. This is essential! The distribution is normalized $\int\!d\lambda\,\rho(\lambda) \,=\, 1 \,$.\\

Now it is quite easy to derive \emph{Bell's original inequality}~\cite{Bell-Physics1964} by assuming perfect (anti-)correlation $E(\vec{a},\vec{a}) \,=\, -1$
\beq\label{Bells-inequality}
S_{\rm{Bell}} \;:=\; |E(\vec{a},\vec{b}) \,-\, E(\vec{a},\vec{b}^{\,'})| \,-\, E(\vec{b}^{\,'},\vec{b}) \;\le\; 1 \;.
\eeq

The quantum mechanical expectation value for the joint measurement when the system is in the spin singlet state $\ket{\psi^{\,-}} \,=\, \frac{1}{\sqrt{2}}(\ket{\Uparrow} \otimes \ket{\Downarrow} \,-\, \ket{\Downarrow} \otimes \ket{\Uparrow})\,$, also called Bell state, is given by
\beq\label{expectation-value-QM-joint-measurement}
E(\vec{a},\vec{b}) &\;=\;& \langle \psi^{\,-} | \,\vec{a} \cdot \vec{\sigma}_A \otimes \vec{b} \cdot \vec{\sigma}_B \,| \psi^{\,-} \rangle \nonumber\\
&\;=\;& -\, \vec{a}\cdot \vec{b} \;=\; -\, \cos(\alpha - \beta)\;,
\eeq
where $\alpha,\beta$ are the angles of the orientations in Alice's and Bob's parallel planes. Inserting expectation value (\ref{expectation-value-QM-joint-measurement}) then inequality (\ref{Bells-inequality}) is violated maximally for the choice of $(\alpha , \beta , \beta^{\,'}) \,=\, (0, 2\frac{\pi}{3}, \frac{\pi}{3})\,$, the \emph{`Bell angles'}
\beq\label{Bells-inequality-max-violation}
S_{\rm{Bell}}^{\rm{QM}} \;=\; \frac{3}{2} \;=\; 1.5 \;>\; 1 \,.
\eeq

Well adapted to experiment is an other inequality, the familiar \emph{CHSH inequality}, named after Clauser, Horne, Shimony, and Hold who published it in 1969 \cite{CHSH}
\beq\label{CHSH-inequality}
S_{\rm{CHSH}} \;:=\; |E(\vec{a},\vec{b}) \,-\, E(\vec{a},\vec{b}^{\,'})| \,+\, |E(\vec{a}^{\,'},\vec{b}) \,+\, E(\vec{a}^{\,'},\vec{b}^{\,'})| \;\le\; 2 \;.
\eeq
As we know, in case of quantum mechanics (\ref{expectation-value-QM-joint-measurement}) the CHSH inequality (\ref{CHSH-inequality}) is violated maximally
\beq\label{CHSH-max-violation}
S_{\rm{CHSH}}^{\rm{QM}} \;=\; 2 \sqrt{2} \;=\; 2.828 \;>\; 2 \,,
\eeq
for the choice of the Bell angles $(\alpha , \beta , \alpha^{\,'} , \beta^{\,'}) \;=\; (0, \frac{\pi}{4}, 2 \frac{\pi}{4}, 3 \frac{\pi}{4})\,$. Inequality (\ref{CHSH-inequality}) has been tested experimentally, e.g., by Zeilinger's group~\cite{Weihs-Zeilinger-Bell-experiment1998, Weihs-thesis-Bell-experiment1998} by using entangled photons in the Bell state $\ket{\psi^{\,-}}\,$. In the photon case, however, the expectation value (\ref{expectation-value-QM-joint-measurement}) changes to $E(\vec{a},\vec{b}) \,=\, -\, \cos2(\alpha - \beta)\,$, i.e., the Bell angles become a factor of $2$ smaller as compared to the spin case.\\

Other types of Bell inequalities, often used in experiments, are:\\

Firstly, \emph{Wigner's inequality} derived by Eugene P. Wigner in 1970~\cite{Wigner1970}. He focused on probabilities which are proportional to the number of clicks in a detector. In terms of probabilities $P$ for the joint measurements the expectation value can be expressed by
\beq\label{expectation-value-intermsof-probabilities}
E(\vec{a},\vec{b}) \;=\; P(\vec{a}\Uparrow,\vec{b}\Uparrow) \,+\, P(\vec{a}\Downarrow,\vec{b}\Downarrow) \,-\, P(\vec{a}\Uparrow,\vec{b}\Downarrow) \,-\, P(\vec{a}\Downarrow,\vec{b}\Uparrow)\,,
\eeq
and assuming that $P(\vec{a}\Uparrow,\vec{b}\Uparrow) \equiv P(\vec{a}\Downarrow,\vec{b}\Downarrow)$ and $P(\vec{a}\Uparrow,\vec{b}\Downarrow) \equiv P(\vec{a}\Downarrow,\vec{b}\Uparrow)$ together with $\sum P = 1$ the expectation value becomes
\beq\label{expectation-value-probability}
E(\vec{a},\vec{b}) \;=\; 4 \,P(\vec{a}\Uparrow,\vec{b}\Uparrow) \,-\, 1 \;.
\eeq
Inserting expression (\ref{expectation-value-probability}) into Bell's inequality (\ref{Bells-inequality}) yields \emph{Wigner's inequality} for the joint probabilities, where Alice measures spin $\Uparrow$ in direction $\vec{a}$ and Bob also $\Uparrow$ in direction $\vec{b}$ (we drop from now on the spin notation $\Uparrow$ in the formulae)
\beq\label{Wigners-inequality}
P(\vec{a},\vec{b}) \;\leq\; P(\vec{a},\vec{b}^{\,'}) \,+\, P(\vec{b}^{\,'},\vec{b}) \;,
\eeq
or rewritten
\beq\label{Wigners-inequality-for-S}
S_{\rm{Wigner}} \;:=\; P(\vec{a},\vec{b}) \;-\; P(\vec{a},\vec{b}^{\,'}) \,-\, P(\vec{b}^{\,'},\vec{b}) \;\leq\; 0 \;.
\eeq
For the Bell state $\ket{\psi^{\,-}}$ the quantum mechanical probability gives
\beq\label{joint-probability-for-Bell-state-psi-minus}
P(\vec{a},\vec{b}) \;=\; | \big(\bra{\vec{a}\Uparrow} \otimes \bra{\vec{b}\Uparrow} \big) \ket{\psi^{\,-}} |^2 \;=\; \frac{1}{2}\,\sin^2 \frac{1}{2}(\alpha - \beta) \;,
\eeq
and leads to a maximal violation of inequality (\ref{Wigners-inequality-for-S})
\beq\label{Wigner-inequality-max-violation}
S_{\rm{Wigner}}^{\rm{QM}} \;=\; \frac{1}{8} \;=\; 0.125 \;>\; 0 \,,
\eeq
for $(\alpha , \beta , \beta^{\,'}) \,=\, (0, 2\frac{\pi}{3}, \frac{\pi}{3})\,$, the same choice as in Bell's original inequality.\\

Secondly, the \emph{Clauser-Horne inequality} of 1974 \cite{Clauser-Horne1974}. It relies on weaker assumptions and is very well suited for photon experiments with absorptive analyzers. Clauser and Horne work with relative counting rates, i.e., number of registered particles in the detectors. More precisely, the quantity $ N(\vec{a},\vec{b})$ is the rate of simultaneous events, coincidence rate, in the photon detectors of Alice and Bob after the photons passed the corresponding polarizers in direction $\vec{a}$ or $\vec{b}$ respectively. The relative rate $N(\vec{a},\vec{b})/N \,=\, P(\vec{a},\vec{b})\,$, where $N$ represents all events when the polarizers are removed, corresponds in the limit of infinitely many events, which is practically the case, to the joint probability $P(\vec{a},\vec{b})\,$. If one polarizer is removed, say on Bob's side, then expression $N_{A}(\vec{a})/N \,=\, P_{A}(\vec{a})$ stands for the single probability at Alice's (or the correspondingly at Bob's) side.

Starting from a pure algebraic inequality $-XY \;\leq\; x_1y_1 - x_1y_2 + x_2y_1 +x_2y_2 - Yx_2 - Xy_1 \,=\, S \;\leq\; 0$ for numbers $0 \,\leq\, x_1,x_2 \,\leq\, X$ and $0 \,\leq\, y_1,y_2 \,\leq\, Y\,$, it is now easy to derive the corresponding inequality for probabilities, which is the \emph{Clauser-Horne inequality}
\beq\label{Clauser-Horne-inequality}
-1 \;\leq\; P(\vec{a},\vec{b}) \,-\, P(\vec{a},\vec{b}^{\,'}) \,+\, P(\vec{a}^{\,'},\vec{b}) \,+\, P(\vec{a}^{\,'},\vec{b}^{\,'})
\,-\, P_{A}(\vec{a}^{\,'}) \,-\, P_{B}(\vec{b}) \;:=\; S_{\rm{CH}} \;\leq\; 0 \;.
\eeq

Inequality (\ref{Clauser-Horne-inequality}) has been used by Aspect in his time-flip experiment \cite{Aspect-Dalibard-Roger1982}. The two-photon state produced was the symmetrical Bell state $\ket{\phi^{\,+}} \,=\, \frac{1}{\sqrt{2}}(\ket{R}\otimes\ket{L} + \ket{L}\otimes\ket{R}) \,=\, \frac{1}{\sqrt{2}}(\ket{H}\otimes\ket{H} + \ket{V}\otimes\ket{V})\,$, where $\ket{R}, \ket{L}$ denote the right and left handed circularly polarized photons and $\ket{H}, \ket{V}$ the horizontally and vertically polarized ones.

In case of $\ket{\phi^{\,+}}$ entangled photons the quantum mechanical probability to detect a linear polarized photon with an angle $\alpha$ on Alice's side, and simultaneously an other linear polarized one with angle $\beta$ on Bob's side, is given by
\beq\label{joint-probability-for-Bell-state-phi-plus}
P(\vec{a},\vec{b}) \;=\; \big| \big[ \big( \bra{H}\cos\alpha \,+\, \bra{V}\sin\alpha \big) \otimes \big(\bra{H}\cos\beta \,+\, \bra{V}\sin\beta \big) \big] \ket{\phi^{\,+}} \big|^2 \;=\; \frac{1}{2}\,\cos^2 (\alpha - \beta) \;.
\eeq
Choosing now for the Bell angles $(\alpha , \beta , \alpha^{\,'} , \beta^{\,'}) \;=\; (0, \frac{\pi}{8}, 2 \frac{\pi}{8}, 3 \frac{\pi}{8})$ the quantum mechanical probabilities (\ref{joint-probability-for-Bell-state-phi-plus}) violate the Clauser-Horne inequality (\ref{Clauser-Horne-inequality}) maximally
\beq\label{Clauser-Horne-inequality-max-violation}
S_{\rm{CH}}^{\rm{QM}} \;=\; \frac{\sqrt{2} - 1}{2} \;=\; 0.207 \;>\; 0 \,.
\eeq

The violation of the above discussed Bell inequalities is expressed by the following theorem:

\begin{theorem}[Bell's Theorem 1964]\ \
In a certain experimental situation \emph{all} local realistic theories are incompatible with quantum mechanics\,!
\label{theorem:Bell}
\end{theorem}

For a thorough discussion we refer to Ref.~\cite{Wiseman-Cavalcanti2015} and further literature can be found in the review article \cite{Brunner-etal-Bell-nonlocality2013, Guehne-Toth2009}.\\

\noindent \textbf{Conclusions:}\\

I remember very well, when I had derived Bell's inequality (\ref{Bells-inequality}) for the first time, I was totally astonished and fascinated that quantum mechanics contradicted an inequality that relied on such general and quite `natural' assumptions. It was impressive to see how John could turn the pure philosophical debate of Einstein and Bohr into exact mathematical terms. And, this formulation could be tested experimentally!

What are the conclusions? In all Bell inequalities the essential ingredient is Bell's locality hypothesis, Eq.~(\ref{expectation-value-Bell-locality}), i.e., Einstein's vision of reality and Bell's concept of locality, therefore we have to conclude as expressed in Theorem~\ref{theorem:Bell}:

\begin{center}
\emph{Local realistic theories are incompatible with quantum mechanics\,!}
\end{center}

Bell in his seminal work~\cite{Bell-Physics1964}  realized the far reaching consequences of a realistic theory as an extension to quantum mechanics and expressed it in the following way:

\emph{``In a theory in which parameters are added to quantum mechanics to determine the results of individual measurements, without changing the statistical predictions, there must be a mechanism whereby the setting of one measuring device can influence the reading of another instrument, however remote. Moreover, the signal involved must propagate instantaneously, so that such a theory could not be Lorentz invariant.''}\\

He continued and stressed the crucial point in such EPR-type experiments:
\emph{``Experiments ... , in which the settings are changed during the flight of the particles, are crucial.''}

Thus it is of utmost importance \emph{not} to allow some mutual report by the exchange of signals with velocity less than or equal to that of light.

\section{Historical Experiments}

\noindent \textbf{First Generation Experiments of the Seventies:}\\

After John had published his paper about the inequality there was practically no interest in this field. It was the \emph{`dark era'} of the foundations of quantum mechanics. Pauli's opinion was often cited~\cite{Pauli-letter-to-Born1954}:

\emph{``One should no more rack one's brain about the problem of whether something one cannot know anything about exists at all, than about the ancient question of how many angels are able to sit on the point of a needle.''}\\

The first who got interested in the subject was John Clauser, a young graduate student from Columbia University, in the late sixties. When he studied Bell's inequality paper~\cite{Bell-Physics1964} that contained a bound for all hidden variable theories, he was astounded by its result. As a true experimentalist he wanted to see the experimental evidence for it. So he planed to do the experiment. However, experiments of this type were not appreciated at that time. When Clauser had an appointment with Richard Feynman at Caltech to discuss an experimental EPR configuration for testing the predictions of QM, he immediately threw him out of his office saying~\cite{Clauser2002}:

\emph{``Well, when you have found an error in quantum-theory's experimental predictions, come back then, and we can discuss your problem with it.''}

But, fortunately, Clauser remained stubborn, he belonged to the revolting generation, and prepared the experiment. He sent an Abstract to the Spring Meeting of the American Physical Society proposing the experiment~\cite{Clauser1969}. Soon afterwards, Abner Shimony called and told him that he and his student Michael Horne had the same ideas. So they joined and wrote together with Richard Holt, a PhD student of Francis Pipkin from Harvard, the famous CHSH paper~\cite{CHSH}, where they proposed an inequality that was well adapted to experiments.

Clauser carried out the experiment in 1972 together with Stuart Freedman~\cite{Clauser-Freedman1972}, a graduate student at Berkeley, who received his PhD with this experiment. As pointed out in the CHSH paper~\cite{CHSH}, pairs of photons emitted in an atomic radiative cascade would be suitable for a Bell inequality test. Clauser and Freedman chose Calcium atoms pumped by lasers, where the excited atoms emitted the desired photon pairs. The signals were very weak at that time, a measurement lasted for about 200 hours. For comparison with theory a very practical inequality was used, which was derived by Freedman~\cite{Freedman1972}. The outcome of the experiment is well known, they obtained a clear violation of the Bell inequality very much in accordance with QM.\\

In 1976, at Houston Edward Fry and his student Randall Thompson set up an experiment by using mercury atoms. As in Clauser's experiment the correlated photons were produced in a radiative cascade from by lasers excited atomic levels. Due to the much better signals with improved lasers they could collect enough data already in 80 minutes. The result was in excellent agreement with QM, the Bell inequality was violated by 4 standard deviations~\cite{Fry-Thompson1976}.\\

Thus at that time, it was already convincing that hidden variable theories did not work but quantum mechanics was correct. However, the experiments were not perfect yet, the analyzers were static, only a small amount of photon pairs were registered, etc. There still existed several loopholes, the detection efficiency or fair sampling loophole, and the communication absence or freedom of choice loophole, just to mention some important ones. To close these loopholes was the challenge of the future experiments.\\

\noindent \textbf{Second Generation Experiments of the Eighties:}\\

In the late 1970s and beginning of the 1980s, the general atmosphere in the physics community was still such:

\emph{``Quantum mechanics works very well, so don't worry!''}\\

I remember, in 1980 I stayed for some time at the Rockefeller University. There I met Abraham Pais, an outstanding particle physicist, with whom I had several stimulating discussions. He had published a bestseller \emph{``Subtle is the Lord: The Science and the Life of Albert Einstein''}~\cite{Pais-SubtleIsTheLord1982}, where he described very thoroughly all the works of Einstein. However, the EPR paper was, in my opinion, treated a bit poor and not with his usual enthusiasm for Einstein. Pais summarized (on p.456 of his book~\cite{Pais-SubtleIsTheLord1982}):

\emph{``The content of this paper has been referred to on occasion as the Einstein-Podolsky-Rosen paradox. It should be stressed that this paper contains neither a paradox nor any flaw of logic. It simply concludes that objective reality is incompatible with the assumptions that quantum mechanics is complete. This conclusion has not affected subsequent developments in physics, and it is doubtful that it ever will.''}\\

Having read this I felt somehow unease about his EPR assessment. So I asked him frankly: \emph{``You don't appreciate the EPR paper?''} And with an impish smile Pais responded: \emph{``The EPR paper was the only slip Einstein made.''} How wrong can be sometimes the judgement and prophecy of a physical work!\\

Alain Aspect, on the other hand, when reading Bell's inequality paper~\cite{Bell-Physics1964}, was so strongly impressed that he immediately decided to do his \emph{``th\`ese d'\' etat''} on this fascinating topic. He visited John Bell at CERN to discuss his proposal. John's first question was, as Alain told me, \emph{``Do you have a permanent position?''} Only after Aspect's positive answer the discussion could begin. Aspect's goal was to include variable analyzers.

Aspect and his collaborators performed a whole series of experiments~\cite{Aspect1976, Aspect-etal1981, Aspect-etal1982, Aspect-Dalibard-Roger1982, Aspect-etal1985} with an improved design and approached step by step the `ideal' setup configuration. As Clauser, they chose a radiative cascade in calcium that emitted photon pairs in the Bell state $\ket{\phi^{\,+}}\,$. For comparison with theory the Clauser-Horne inequality (\ref{Clauser-Horne-inequality}) was used, which was significantly violated in each experiment.\\

In the final time-flip experiment~\cite{Aspect-Dalibard-Roger1982} together with Jean Dalibard and G\' erard Roger a clever acoustic-optical switch mechanism was incorporated. It worked such that the switching time between the polarizers, as well as the lifetime of the photon cascade, was much smaller than the time of flight of the photon pair from the source to the analyzers. That implied a space-like separation of the event intervals. However, the time flipping mechanism was still not ideal, i.e., truly random, but \emph{``quasi-periodic''}, as they called it. The mean for two runs which lasted about 2 hours yielded the result $S_{\rm CH}^{\rm exp} \,=\, 0.101 \pm 0.020$ in very good agreement with the quantum mechanical prediction $S_{\rm CH}^{\rm QM} \,=\, 0.113 \pm 0.005$ that had been adapted for the experiment (recall the ideal value is $S_{\rm CH}^{\rm QM} \,=\, 0.207$ (\ref{Clauser-Horne-inequality-max-violation})).\\

This time-flip experiment of Aspect received much attention in the physics community and also in popular science, and Alain was the best apologist. In my opinion, it caused a turning point, the physics community began to realize that there was something essential in it. The research started and flourished into a new direction, into what is called nowadays quantum information and quantum communication~\cite{nielsen-chuang2000, bertlmann-zeilinger02}.\\

\noindent \textbf{Third Generation Experiments of the Nineties and Beyond:}\\

\begin{figure}
\centering
\includegraphics[width=0.8\textwidth]{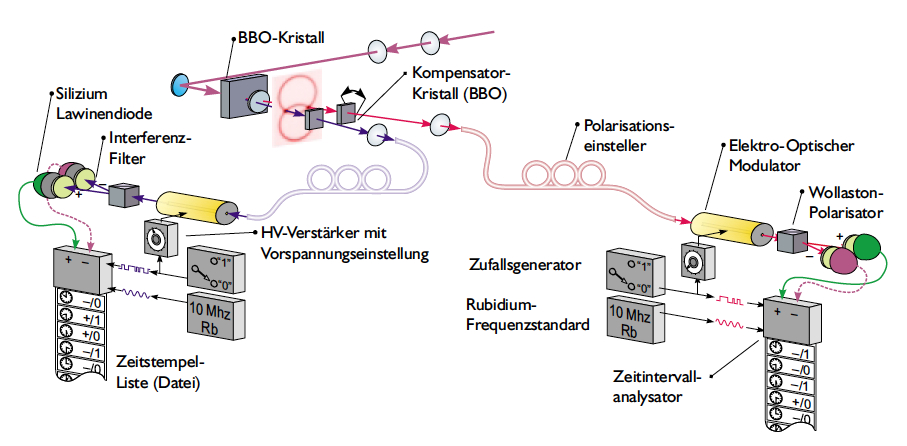}
\caption{The timing experiment of Weihs et al.~\protect\cite{Weihs-Zeilinger-Bell-experiment1998}. The EPR source is a so-called BBO crystal pumped by a laser, the outgoing photons are vertically and horizontally polarized on two different cones and in the overlap region they are entangled. This entangled photons are led separately via optical fibres to the measurement stations Alice and Bob. During the photon propagation the orientations of polarizations are changed by an electro-optic modulator which is driven by a truly random number generator, on each side. In this way the strict Einstein locality condition---no mutual influence between the two observers Alice and Bob---is achieved in the experiment. The figure is taken from Ref.~\protect\cite{Weihs-thesis-Bell-experiment1998}, \copyright Gregor Weihs.}
	\label{fig:Weihs-Bell-experiment}
\end{figure}

In the 1990s, the spirit towards foundations in quantum mechanics totally changed since quantum information gained increasing interest, Bell inequalities and quantum entanglement were the basis.

Meanwhile, the technical facilities improved considerably too, the electronics and the lasers. Most important was the invention of a new source for creating two entangled photons. That was spontaneous parametric down conversion, where a nonlinear crystal was pumped with a laser and the pump photon was converted into two photons that propagated vertically and horizontally polarized on two different cones. In the overlap region they were entangled.

Such an EPR source was used by Anton Zeilinger and his group, when they performed their Bell experiment in 1998~\cite{Weihs-Zeilinger-Bell-experiment1998}. Zeilinger's student Gregor Weihs obtained his PhD with this experiment~\cite{Weihs-thesis-Bell-experiment1998} (see Fig.~\ref{fig:Weihs-Bell-experiment}). Their challenging goal was to construct an ultra-fast and truly random setting of the analyzers at each side of Alice and Bob, such that strict Einstein locality -- no mutual influence between the two observers Alice and Bob -- was achieved in the experiment. The data were compared with the CHSH inequality (\ref{CHSH-inequality}) and the experimental result was: $S_{\rm{CHSH}}^{\rm{exp}} \,=\, 2.73 \pm 0.02\,$, which corresponded to a violation of the inequality of $30$ standard deviations. Due to this high efficiency photon source the measurement could be performed already in $3-4$ minutes. It was \emph{the} experiment that truly included the vital time factor, John Bell insisted upon so strongly.\\

About the same time other groups investigated energy correlated photon pairs to test Bell inequalities~\cite{Brendel-etal1992, Tapster-etal1994}. A record was set by the group of Nicolas Gisin~\cite{Tittel-Gisin-etal1998}, by using energy-time entangled photon pairs in optical fibres. They managed to separate their observers Alice and Bob by more than 10 km and could show that this distance had practically no effect on the entanglement of the photons. The investigated Bell inequalities had been violated by 16 standard deviations.

Fascinating experiments on quantum teleportation~\cite{Bouwmeester-Zeilinger-etal1997, Ma-Zeilinger-etal-teleportation-144km-Nature2012} and quantum cryptography~\cite{Jennewein-Zeilinger-etal-QuCrypto2000, Gisin-etal-QuCrypto2002} followed.

Then a race started in achieving records of entanglement based long distance quantum communication. The vision was to install a global network, in particular via satellites or the International Space Station, that provided an access to secure communication via quantum cryptography at any location.

It was again Zeilinger's group that pushed the distance limits further and further. Firstly, in an open air experiment in the City of Vienna over a distance of $7.8$ km the group~\cite{Resch-Blauensteiner-Zeilinger-etal-freespace-City2005} could violate a CHSH inequality (\ref{CHSH-inequality}) by more than $13$ standard deviations. Secondly, this is presently the world record, the group~\cite{Ursin-Blauensteiner-Zeilinger-etal-Tenerife2007} successfully carried out an open air Bell experiment over $144$ km between the two Canary Islands La Palma and Tenerife.\\

In search of closing loopholes a recent Bell experiment of the group~\cite{Giustina-Zeilinger-fairsampling-Nature2013} closed the fair-sampling loophole, i.e., their results of violating an inequality \` a la Eberhard~\cite{Eberhard-inequ-1993} were valid without assuming that the sample of measured photons accurately represented the entire ensemble.

Another loophole, the detection efficiency loophole, could be closed with ion traps. Working with ions the group Rowe et al.~\cite{Rowe-Wineland-Bell-ions-Nature2001} tested a Bell inequality with perfect detection efficiency.\\

Finally, I also want to refer to Bell inequality tests of the group of Helmut Rauch~\cite{Hasegawa-etal-KochenSpecker2003, Hasegawa-etal-KochenSpecker2010, Hasegawa-etal-contextuality2009}. These neutron interferometer experiments were of particular interest since in this case the quantum correlations were explored in the degrees of freedom of a single particle, the neutron. Physically, it meant that rather contextuality was tested than nonlocality in space.\\

It is quite interesting and amusing to see the development of Bell experiments in the history of time. Beginning in the 1970s, where one had to overcome huge technical difficulties and the enormous resistance of the physics community, the development ended in the 2010s in such a way that a Bell experiment belonged already to the standard educational program \emph{``Laboratory Quantum Optics''} for the students at the Faculty of Physics of the University of Vienna. It would have been nice if John Bell could have seen that!\\

For a detailed description of all Bell-type experiments the reader may consult \emph{Quantum [Un]speakables}~\cite{bertlmann-zeilinger02}.\\

\section{Some Memories and John's Quantum Legacy}

When I think back again and recall the sometimes lively discussions I had with John about quantum mechanics and its interpretations, in particular, about the meaning of contextuality and nonlocality, then interestingly, John was never so much concerned about contextuality and its implications. Whereas, I always thought that contextuality was \emph{the} important quantum feature and had a profound rooting in Nature, and I am still convinced that some day it will have technical applications.\\

John, on the other hand, was deeply disturbed by the nonlocality feature of quantum mechanics since for him it was equivalent to a \emph{``breaking of Lorentz invariance''} in an extended theory for quantum mechanics, what he hardly could accept. He often remarked:
\emph{``It's a great puzzle to me ... behind the scenes something is going faster than the speed of light."}

At the end of his Bertlmann's socks paper John expressed again his concern~\cite{Bell-Bsocks-Journal-de-Physique}:

\emph{``It may be that we have to admit that causal influences \textsl{do} go faster than light. The role of Lorentz invariance of a completed theory would then be very problematic. An `ether' would be the cheapest solution. But the unobservability of this ether would be disturbing. So would be the impossibility of `messages' faster than light, which follows from ordinary relativistic quantum mechanics in so far as it is unambiguous  and adequate  for procedures we can actually perform. The exact elucidation of concepts like `message' and `we', would be a formidable challenge.''}\\

When John gave a talk on the foundational issues, there often arose a great tension between him and the audience, especially, about the item of nonlocality. People didn't want to listen, didn't want to accept what John was saying, John was like \emph{`a lone voice in the wilderness'}. Once, in the late 1980s, he gave a talk at the ETH Z\"urich and I asked him afterwards, \emph{``How was it?''} John replied clenching his fists, \emph{``I could beat them!''}\\

On a summer afternoon in 1987, John and I were sitting outside in the garden of the CERN cafeteria, drinking our late \emph{4~o'clock tea}\,, and talked as so often about the implications of nonlocality. In this tea-atmosphere I spontaneously said: \emph{``John, you deserve the Nobel prize for your theorem.''} John, for a moment puzzled, replied quite strictly: \emph{``No, I don't. ... it's like a null experiment, and you don't get the Nobel prize for a null experiment. ... for me, there are Nobel rules as well, it's hard to make the case that my inequality benefits mankind.''} I countered: \emph{``I disagree with you! It's not a null result. You have proved something new, nonlocality! And for that I think you deserve the Nobel Prize.''} John, although feeling somehow pleased, raised slowly his arms, shrugged his shoulders and mumbled sadly: \emph{``Who cares about this nonlocality?''}\\

\begin{figure}
\centering
\includegraphics[width=0.4\textwidth]{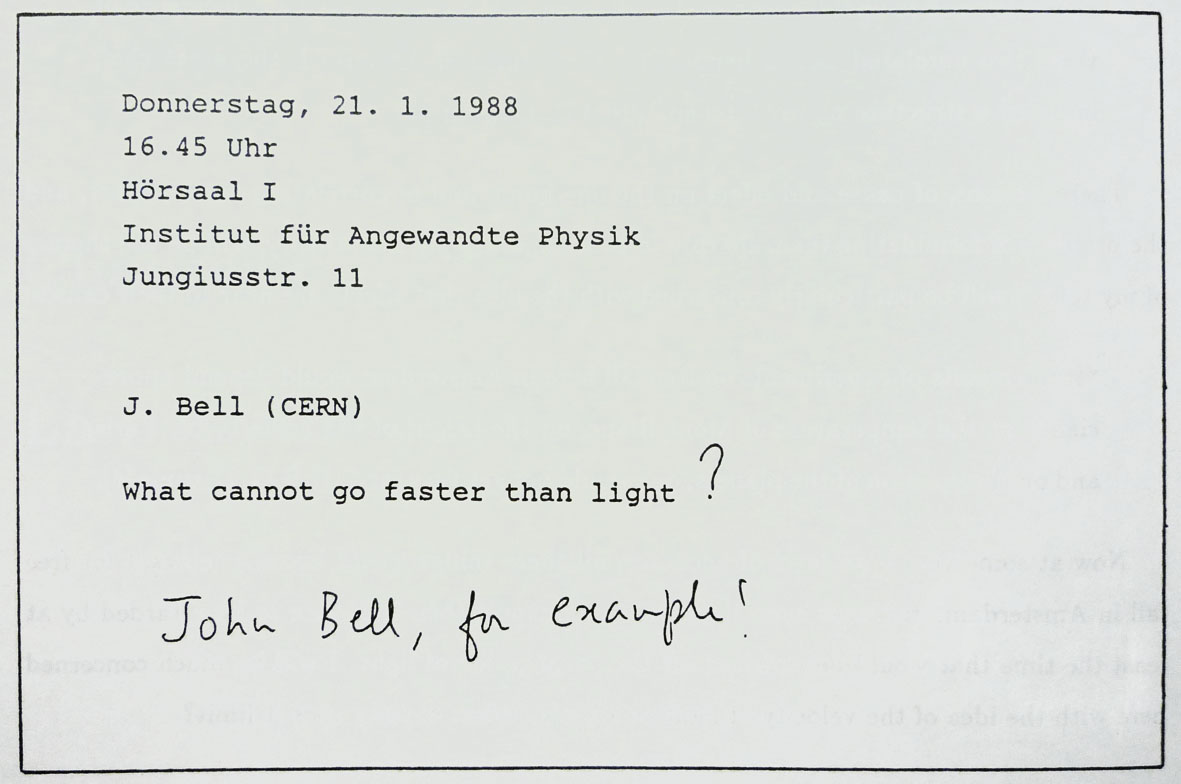}
\caption{Announcement of John Bell's talk: \emph{``What cannot go faster than light?''} at the University of Hamburg in 1988. Somebody with Hanseatic humor added to the announcement by hand: \emph{``John Bell, for example!''}}
	\label{fig:Bell-Vortragsankuendigung}
\end{figure}

In John's last paper \emph{``La nouvelle cuisine''}, published in 1990 \cite{Bell-nouvelle-cuisine} (and see his collected quantum works \cite{Bell-book}), he still remained profoundly concerned with this nonlocal structure of Nature. The paper was based on a talk he gave at the University of Hamburg, in 1988, about the topic: \emph{``What cannot go faster than light?''}. Somebody with Hanseatic humor added to the announcement by hand: \emph{``John Bell, for example!''} (see Fig.~\ref{fig:Bell-Vortragsankuendigung}). This remark made John thinking, what exactly that meant, his whole body or just his legs, his cells or molecules, atoms, electrons ... Was it meant that none of his electrons go faster than light?\\

In our modern view of Nature the concepts of a classical theory changed, the sharp location of objects had been dissolved by the fuzzyness of the wave function or by the fluctuations in quantum field theory. As John remarked:
\emph{``The concept `velocity of an electron' is now unproblematic only when not thought about it.''}\\

Finally, John discussed \emph{``Cause and effect''} in this paper. As Einstein \cite{Einstein1907} already pointed out, if an effect follows its cause faster than the propagation of light, then there exists an inertial frame where the effect happens before the cause. Such a thing was unacceptable for both, Einstein and Bell. Therefore, sticking to \emph{``no signals faster than light''} John defined locally causal theories and demonstrated, via an EPR-Bell type experiment, that \emph{``ordinary quantum mechanics is not locally causal''}, or more precisely, \emph{``quantum mechanics cannot be embedded into a locally causal theory''}. It was essential in his argumentation that the measurement settings $\vec{a}$ at Alice's side and $\vec{b}$ at Bob's side could be chosen totally free, i.e., at random. \emph{``But still, we cannot signal faster than light''} John noted at the end.\\

Let me finally cite John's point of view concerning the existence, the realism of Nature, John in his own words, taken from an interview he gave in the late 1980s~\footnote{The whole interview with John Bell can be seen on a DVD available at the Austrian Central Library for Physics and Chemistry, Vienna.}.\\

\noindent John's confession:

\emph{``Oh, I'm a realist and I think that idealism is a kind of ... it's a kind of ... I think it's an artificial position which scientists fall into when they discuss the meaning of their subject and they find that they don't know what it means. I think that in actual daily practice all scientists are realists, they believe that the world is really there, that it is not a creation of their mind. They feel that there are things there to be discovered, not a world to be invented but a world to be discovered. So I think that realism is a natural position for a scientist and in this debate about the meaning of quantum mechanics I do not know any good arguments against realism.''}\\

John was totally convinced that \emph{realism} is the right position of a scientist. He believed that the experimental results are predetermined and not just induced by the measurement process. Even more, in John's EPR analysis reality is not assumed but inferred! Otherwise (without realism), he said, \emph{``It's a mystery if looking at one sock makes the sock pink and the other one not-pink at the same time.''}

So he did hold on the hidden variable program continuously, and was not discouraged by the outcome of the EPR-Bell experiments but rather puzzled. For him \emph{``The situation was very intriguing that at the foundation of all that impressive success} [of quantum mechanics] \emph{there are these great doubts''}, as he once remarked.\\

\begin{figure}
\begin{center}
\includegraphics[width=0.4\textwidth]{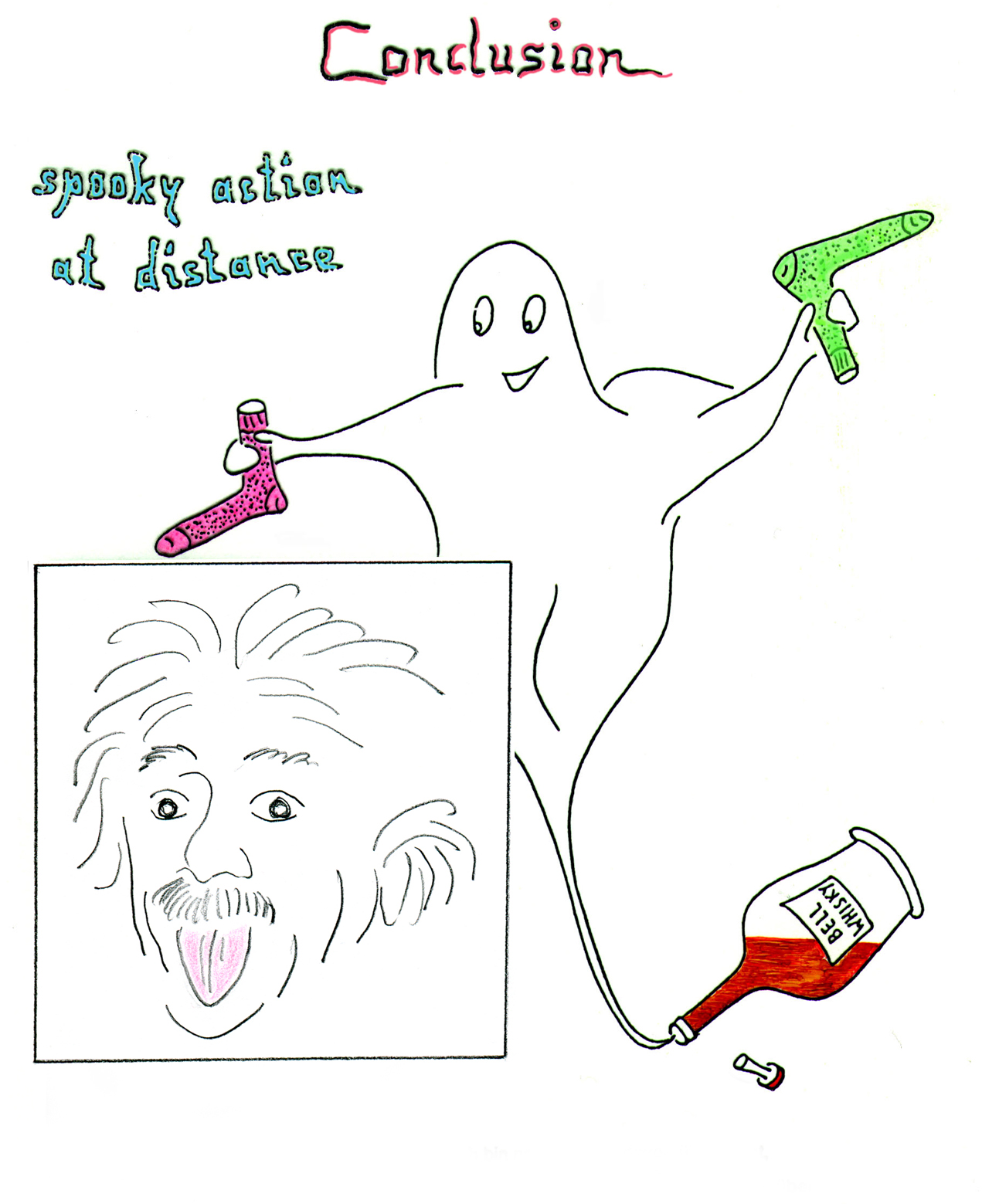}
\normalsize{
\caption{Sketch of my conclusions in the paper \emph{``Bell's theorem and the nature of reality''}, which I dedicated John Bell in 1988 on occasion of his 60th birthday~\protect\cite{Bertlmann-BellsTheorem-FoundationOfPhysics}. Cartoon: \copyright Reinhold A. Bertlmann.}
\label{fig:spooky-ghost}}
\end{center}
\end{figure}

I got back at John for \emph{``Bertlmann's socks''} in paper \emph{``Bell's theorem and the nature of reality''}~\cite{Bertlmann-BellsTheorem-FoundationOfPhysics} that I dedicated to him in 1988 on occasion of his 60th birthday. I sketched my conclusions in a cartoon, shown here as Fig.~\ref{fig:spooky-ghost}. John, as a strict teetotaler, was very much amused by my illustration, since the \emph{spooky, nonlocal ghost} emerged from a bottle of \emph{Bell's Whisky}, a brand that \emph{really} did exist.\\

In the 1990s, Bell's huge impact on the developments of quantum information became widely appreciated. It is known that in 1990 (the year of John's unexpected death) John was appointed for the Nobel Prize. Nowadays physicists agree that John would have definitely received the prize for his outstanding contributions to the foundations of quantum mechanics if he had lived longer. For instance, Daniel Greenberger expressed it explicitly in an interview given at the Conference \emph{Quantum [Un]Speakables II} in Vienna~\cite{Greenberger-interview2014}:

\emph{``Of course, people more and more appreciate John Bell's beautiful work. He was essentially starting the field, his work was totally seminal, and if he were alive he certainly would have won the Nobel Prize\,!''}

\section{Turn to Quantum Mechanics}

A year after John's death Franco Selleri, Professor at the Universit\'a degli Studi di Bari Aldo Moro, and his crew organized the Conference \emph{``Bell's Theorem and the Foundations of Modern Physics''} (7 -- 10 October, 1991, Cesena). It was an international conference in memory of John Bell. Many leading scientists of the field participated and gave talks to honour posthumously John Bell. Among them where: Jeffrey Bub, James Cushing, Bernard d'Espagnat, Giancarlo Ghirardi, Daniel Greenberger, Max Jammer, Leonard Mandel, Sir Roger Penrose, Franco Selleri, Euan Squires, Anton Zeilinger, myself and many others.

\begin{figure}
\begin{center}
\includegraphics[width=0.65\textwidth]{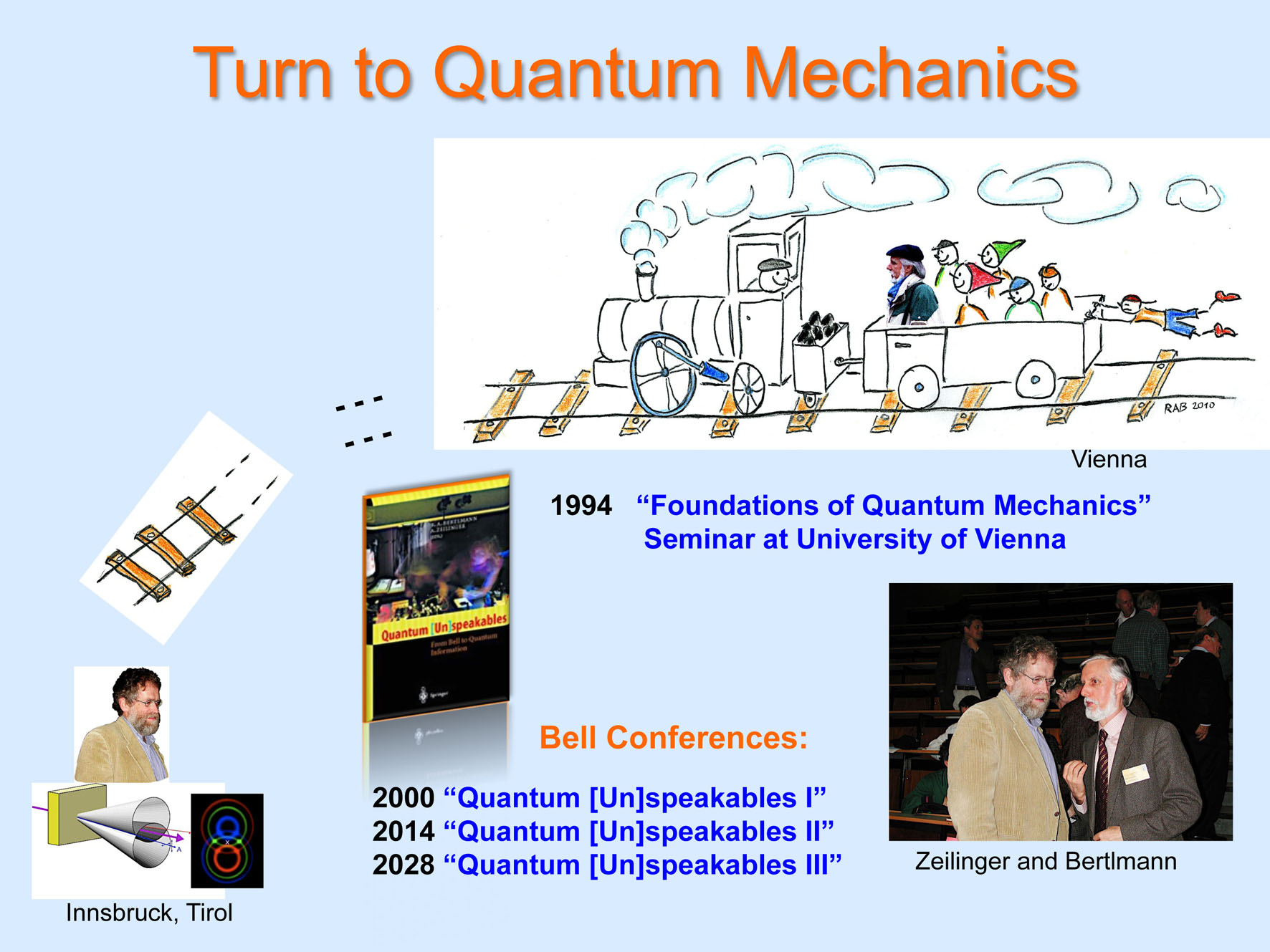}
\normalsize{
\caption{Activities of Anton Zeilinger and Reinhold Bertlmann: Establishing in 1994 \emph{``Foundations of Quantum Mechanics''}, a Joint Seminar between the Universities of Vienna and Innsbruck. Organizing the Conferences \emph{``Quantum [Un]Speakables I''} 2000, \emph{``Quantum [Un]Speakables II''} 2014, and \emph{``Quantum [Un]Speakables III''} 2028. Cartoon: \copyright Reinhold A. Bertlmann, Foto: \copyright Renate Bertlmann.}
\label{fig:Turn-to-QM}}
\end{center}
\end{figure}

There I met Anton Zeilinger, he was one of the two people who responded to my \emph{`Bell-revenge'} paper~\cite{Bertlmann-BellsTheorem-FoundationOfPhysics}. Anton just became a Professor at the University of Innsbruck and was establishing a quantum group. At that time I still worked in particle physics, but we found an overlapping interest in the foundations of quantum mechanics. Since we both were fascinated by this topic we thought that it would be a good idea to educate the young Austrian generation in this field. So we intensified the contact and exchange of our universities, which, in 1994, resulted in founding officially the Joint Seminar \emph{``Foundations of Quantum Mechanics''} between the universities of Vienna and Innsbruck. For the meetings Anton and his group came to Vienna, and alternatingly, the Vienna students were travelling by train to Innsbruck, see Fig.~\ref{fig:Turn-to-QM}.

This Seminar immediately became very popular among the Viennese students since Anton's group reported on fascinating experiments carried out in Innsbruck, and the experiments were performed by young scientists, in fact by students, what impressed our students very much. Also the quite informal, familiar character of the Seminar -- we always served coffee \& cake -- added to the success of the Seminar. One event is unforgettable, when Alois Mair, a student of Anton's group reported on the first experiment of \emph{``Entangled states of orbital angular momentum of photons''}(published in Ref.~\cite{Mair-etal-2001}). In such states the phase surface of the wave resembled a screw in direction of the wave propagation. At the filters it impressively looked like a doughnut. After the talk we had a big doughnut party, where 70 \emph{real} doughnuts, delivered by the nearby bakery \emph{``Ritz''}, had been served. Meanwhile the Seminar belongs to the regular student educational programme of our Faculty.

Of course, after Anton's move from Innsbruck to Vienna in 1999, we further intensified our collaboration in several areas, for instance, in organizing conferences like \emph{``Quantum [Un]Speakables''} in 2000 in commemoration of John Bell, \emph{``Quantum [Un]Speakables II''} in 2014 to celebrate 50 years of Bell's Theorem, see Fig.~\ref{fig:Turn-to-QM}. In 2014 there were celebrations of Bell's quantum achievements all over the world, also Queen's University Belfast, John's home university, organized an exhibition \emph{``Action at a Distance''} in The Naughton Gallery at Queen's~\cite{BelfastExhibition} and the Belfast City Council named a street \emph{``Bell's Theorem Crescent''} in Belfast's Titanic Quater to honour John Bell as \emph{``One of the Northern Ireland's most eminent scientists''}. Finally, for 2028 a third conference \emph{``Quantum [Un]Speakables III''} had been announced to commemorate Bell's 100th birthday~\cite{announcement-QuantumUnspeakablesIII}, see Fig.~\ref{fig:Turn-to-QM}.\\

This fascination for the foundations of quantum mechanics also stimulated my research interest in this field and I began to study this topic in the area I was familiar with, that was particle- and mathematical physics. About this research I want to report next.

\section{Entanglement in Particle Physics}

\subsection{Decoherence of Entangled Particle--Antiparticle Systems}

The second person who responded to my \emph{`Bell-revenge'} paper~\cite{Bertlmann-BellsTheorem-FoundationOfPhysics} was Walter Grimus, a distinguished particle physicist of our Institute in Vienna. He was an expert in \emph{strangeness systems}, the $K$-mesons, and in \emph{beauty systems}, the $B$-mesons. So it was quite natural that we discussed the phenomena of quantum information, the peculiar quantum correlations of bipartite and multipartite systems, within these systems in particle physics.

The difference to the photon systems, discussed so far, was that particle systems had entirely different and additional properties, which the photons did not have. First of all, the investigated particles were very massive, they decayed into other particles, they oscillated between their flavour content, i.e., between their particle and antiparticle nature, and they could regenerate, e.g., once the short-lived kaon state had decayed it could be regenerated from the surviving long-lived component. In addition, they possessed internal symmetries, like the $CP$ symmetry (charge conjugation and parity), which turned out to be essential. For these reasons, I think, that it was, and still is, of great importance to investigate such systems, particularly, with regard to the EPR-Bell quantum correlations.

\begin{figure}
\begin{center}
\includegraphics[width=0.65\textwidth]{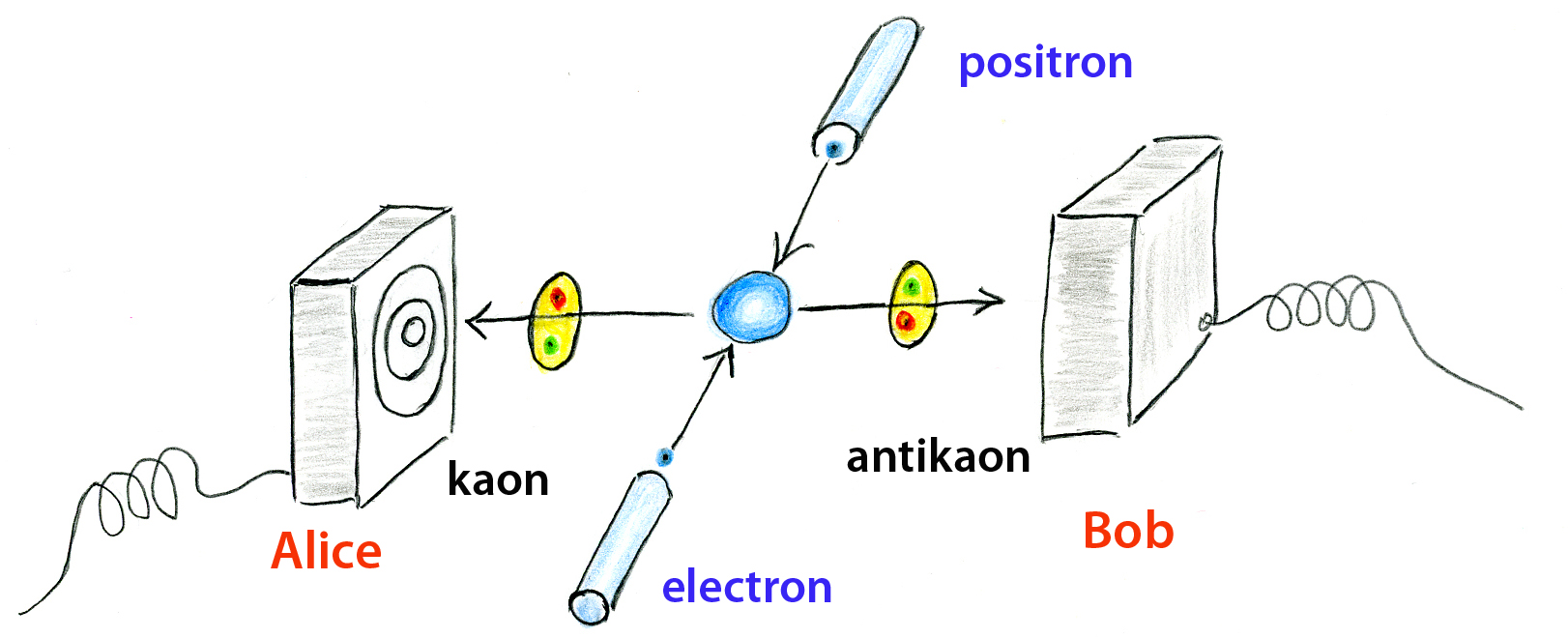}
\normalsize{
\caption{Entanglement of matter and antimatter created in an accelerator of particle physics. Cartoon: \copyright Reinhold A. Bertlmann.}
\label{fig:K-antiK-creation}}
\end{center}
\end{figure}

Experimentally, the particle--antiparticle systems that were generated in the huge particle accelerators were already entangled due to conservations laws. For example, the $K^0 \bar{K}^0$ system was produced at the $\Phi$ resonance in the $e^{+} e^{-}$ machine DA$\Phi$NE at Frascati, for a sketch see Fig.~\ref{fig:K-antiK-creation}, and the $B^0 \bar{B}^0$ system at the $\Upsilon$(4S) resonance in machines like DORIS II (Doppel Ring Speicher) at DESY (Deutsches Elektronen Synchrotron), in CESR (Cornell Electron Storage Ring) at Cornell, or nowadays in the KEK B-factory in Japan.\\

The first system Walter and I investigated was the $B^0 \bar{B}^0$ system since there existed already usable data from DORIS II and CESR. We found that the $B^0 \bar{B}^0$ state generated in the decay of the resonance $\Upsilon$(4S) at $10.6$ GeV is very well suited to perform tests of the EPR correlations over macroscopic distances. Using measurements of the ratio $R =$ (\emph{No. like-sign dilepton events})/(\emph{No. opposite-sign dilepton events}) we could show that already presently existing data strongly favoured the contribution of the interference term to $R\,$, as it was required by the rules of quantum mechanics~\cite{Bertlmann-Grimus-PL1997, Bertlmann-Grimus-PRD58-1998}.

The next system we explored was the $K^0 \bar{K}^0$ system, where data was available from the CPLEAR experiment at CERN~\cite{CPLEAR1998}. It was precisely the time, when a young enthusiastic graduate student, with name Beatrix Hiesmayr, approached me searching for a Diploma Thesis. I thought, that's a good topic for her, so she joined and it was the starting point for fruitful and still ongoing collaboration.

Our aim was to show, if a pair of particle--antiparticle had been created by any kind of interaction in an entangled state, the two-particle wave function retained its non-separable character even if the particles were space-like separated over large macroscopic distances (about 9 cm). To describe quantitatively spontaneous factorization and decoherence of the $K^0 \bar{K}^0$ system we modified the quantum-mechanical interference term of the
entangled 2-kaon state by a multiplication with the term $(1-\zeta)$. Thus we changed the quantum mechanical expression in such a way that the effective decoherence parameter $\zeta$ quantified the deviation from quantum mechanics (corresponding to $\zeta = 0$) and provided a measure for the distance of the total system from its total decoherence or spontaneous factorization ($\zeta = 1$).

The relevant quantities we had to calculate were the probabilities to measure \emph{like-strangeness} and \emph{unlike-strangeness} events at a time $t_l$ on the left side and $t_r$ on the right side. Starting from a produced asymmetric Bell state  $\ket{\psi^{\,-}} \,=\, \frac{1}{\sqrt{2}}(\ket{K_S} \otimes \ket{K_L} \,-\, \ket{K_L} \otimes \ket{K_S})\,$, where the short-lived kaon $K_S$ and the long-lived kaon $K_L$ played the role of spin up $\Uparrow$ and spin down $\Downarrow\,$, we found for the \emph{like-strangeness probability}\cite{Bertlmann-Grimus-Hiesmayr1999} (see also reviews~\cite{Bertlmann-LectureNotes2006, CarlaSchuler2014})
\begin{eqnarray}\label{plikezeta}
\lefteqn{P(K^0,t_l;K^0,t_r)\;=\;||\langle K^0|_l \otimes \langle K^0|_r \;
|\psi^-(t_l,t_r)\rangle||^2}\nonumber\\
&\longrightarrow& \quad P_{\zeta}(K^0, t_l; K^0, t_r) \;=\; \frac{1}{2}\;\biggl\lbrace
e^{-\Gamma_S t_l-\Gamma_L t_r} |\langle K^0|K_S\rangle_l|^2\;
|\langle K^0|K_L\rangle_r|^2\nonumber\\
& &+\; e^{-\Gamma_L t_l-\Gamma_S t_r}|\langle K^0|K_L\rangle_l|^2\; |\langle
K^0|K_S\rangle_r|^2  \;-\; 2\; \underbrace{(1-\zeta)}\; e^{-\Gamma(t_l+t_r)}\nonumber\\
& &\quad\hspace{6.2cm}\textrm{modification} \nonumber\\
& &\times\;\mathrm{Re}\big\{\langle K^0|K_S\rangle_l^* \langle K^0|K_L\rangle_r^* \langle
K^0|K_L\rangle_l \langle K^0|K_S\rangle_r\;
e^{-i \Delta m \Delta t}\big\} \biggr\rbrace \nonumber\\
&=&\frac{1}{8}\,\biggl\lbrace e^{-\Gamma_S t_l-\Gamma_L t_r} + e^{-\Gamma_L t_l-\Gamma_S
t_r} \;-\; 2\, \underbrace{(1-\zeta)}\, e^{-\Gamma(t_l+t_r)}\, \cos(\Delta m \Delta
t)\biggr\rbrace\,,
\nonumber\\
& &\hphantom{\frac{1}{8}\biggl\lbrace e^{-\Gamma_S t_l-\Gamma_L t_r} + e^{-\Gamma_L
t_l-\Gamma_S t_r}} \;\;\quad\textrm{modification}
\end{eqnarray}
and the \emph{unlike-strangeness probability} just changed the sign of the interference term.

The quantity directly sensitive to the interference term was the asymmetry
\begin{equation}\label{CPLEARasymetry}
A(t_r,t_l) \;=\; \frac{P_\mathrm{unlike}(t_r,t_l) - P_\mathrm{like}(t_r,t_l)}
                  {P_\mathrm{unlike}(t_r,t_l) + P_\mathrm{like}(t_r,t_l)}\;,
\end{equation}
that had been measured in the CPLEAR experiment. The comparison of the theoretical expression for the given basis $\{ K_S, K_L \}$~\cite{Bertlmann-Grimus-Hiesmayr1999}
\begin{equation}\label{zeta-asymetry}
A^{K_LK_S}_\zeta (t_r,t_l) \;=\; (1-\zeta) A^\mathrm{QM}(t_r,t_l) \qquad \mbox{with} \qquad A^\mathrm{QM}(t_r,t_l) \;=\;
\frac{\cos(\Delta m\, \Delta t)}{\cosh(\frac{1}{2} \Delta \Gamma \Delta t)} \,,
\end{equation}
where $\Delta m = m_L-m_S\,$, $\Delta \Gamma = \Gamma_L - \Gamma_S$ and $\Delta t = t_r - t_l\,$, with the CPLEAR data~\cite{CPLEAR1998} restricted the decoherence parameter $\zeta$ to the interval~\cite{Bertlmann-Grimus-Hiesmayr1999}
\begin{equation}\label{zeta-values}
\zeta \;=\; 0.13^{+0.16}_{-0.15}\;.
\end{equation}
This result confirmed nicely in a quantitative way the existence of entangled massive particles over macroscopic distances (9 cm).

When concentrating on a specific decay process of the kaon system $\phi \rightarrow K_S K_L \rightarrow \pi^{+} \pi^{-} \pi^{+} \pi^{-}$ the quantitative estimate of the $\zeta$ could be improved by orders of magnitude~\cite{KLOE-article-EPJ}
\begin{equation}\label{zeta-values-KLOE}
\zeta \;=\; 0.003 \pm 0.018_{\rm{stat}} \pm 0.006_{\rm{sys}}\;.
\end{equation}

We also found out that the decoherence parameter $\zeta\,$, which we had introduced by hand, also had a deeper physical basis. It was related to the decoherence strength $\lambda$ of a Lindblad~\cite{Lindblad1976} and Gorini-Kossakowski-Sudarshan~\cite{Gorini-Kossakowski-Sudarshan1976} master equation for the density matrix $\rho$ of the total quantum system
\begin{eqnarray}\label{Lindbladequation}
\frac{d\rho}{dt}\; &=&\; -\,i H\rho \,+\,i\rho H^\dagger\,-\,D[\rho] \;.
\end{eqnarray}
The \emph{dissipator} $D[\rho]$ was chosen as
\begin{eqnarray}\label{Dissipationterm}
D[\rho] \;=\; \lambda \, \big(P_1 \rho P_2 + P_2 \rho P_1\big) \;=\; \frac{\lambda}{2}
\sum_{j=1,2} \big[P_j,[P_j,\rho]\big] \;,
\end{eqnarray}
with the projectors $P_j\,=\,|e_j\rangle\langle e_j|$ ($j=1,2$) onto the states $|e_1\rangle\,=\,|K_S\rangle_l\otimes|K_L\rangle_r \;\textrm{and}\;
|e_2\rangle\,=\,|K_L\rangle_l\otimes|K_S\rangle_r \,$. Then the connection was~\cite{Bertlmann-Grimus-PRD64-2001, Bertlmann-LectureNotes2006}
\begin{equation}\label{zeta}
\zeta(t) \;=\; 1 \,-\; e^{-\lambda t} \;.
\end{equation}
The parameter $\lambda$ representing the strength of the interaction of the system with its environment had to be considered as the more fundamental one.\\

The increase of decoherence of the initially totally entangled $K^0 \bar  K^0$ system as time evolves means on the other hand a decrease of entanglement of the system, an entanglement loss. Interestingly, there is a direct relation between the decoherence parameter $\zeta\,$, or $\lambda\,$, which quantifies the spontaneous factorization of the wave function, and the entanglement loss of the system that is defined via the entropy. The amount of entanglement is defined by the familiar measures: \emph{Entanglement of formation E}~\cite{Bennett-etal-1996} or \emph{Concurrence C}~\cite{Hill-Wootters1997, Wootters1998, Wootters2001}.

Then we have the following proposition:

\begin{proposition}[Bertlmann-Durstberger-Hiesmayr~\cite{Bertlmann-Durstberger-Hiesmayr2003}]\ \ \\
The entanglement loss $(1 - C)$ or $(1- E)$ equals the amount of decoherence:
\begin{eqnarray}
1 - C\big(\rho(t)\big) \;&=&\; \zeta(t)\label{entanglementlossC} \;,\\
1 - E\big(\rho(t)\big) \;&\doteq&\; \frac{1}{\ln2}\;\zeta(t) \;\doteq\;
\frac{\lambda}{\ln2}\;t\label{entanglementlossE} \;.
\end{eqnarray}
\label{proposition:BDH}
\end{proposition}

\subsection{Bell Inequalities in Particle Physics}

Together with Beatrix I also turned to the topic of Bell inequalities in particle physics. The typical feature of these particle systems, e.g., of a kaon--antikaon system, is that the joint expectation value of a measurement at Alice's and Bob's detectors depends on both, on the flavour content, which corresponds to a quasi-spin property, and on the time of the measurement, once the system is created. More precisely, the expectation value for the combined measurement $E(k_a, t_a; k_b, t_b)$ is a function of the flavour $k_a$ measured on the left side at a time $t_a$ and on a (possibly different) $k_b$ on the right side at $t_b$. Relying on the usual argumentation for Bell inequalities we could derive the following \emph{Bell--CHSH inequality}~\cite{Bertlmann-Hiesmayr2001} (see Ref.~\cite{Bertlmann-LectureNotes2006}, for an overview in this field)
\begin{eqnarray}\label{chsh-inequ-kaon-gen}
|E(k_a, t_a; k_b, t_b) &-& E(k_a, t_a; k_{b'}, t_{b'})|\nonumber\\
&+& |E(k_{a'}, t_{a'}; k_b ,t_b) + E(k_{a'}, t_{a'}; k_{b'}, t_{b'})|\; \leq \; 2 \;,
\end{eqnarray}
which expressed both the freedom of choice in time \emph{and} in flavour. Identifying $E(k_a, t_a; k_b, t_b)\equiv E(\vec{a}, \vec{b})$ we are back at the inequality (\ref{CHSH-inequality}) for the spin--$\frac{1}{2}$ case.\\

Therefore we may choose in a Bell inequality:

I) \emph{varying the flavour (quasi-spin) or fixing the time},

I\!I) \emph{fixing the flavour (quasi-spin) or varying the time}.\\

However, the experimental test of Bell inequalities in particle physics is much more intricate than in photon physics. Active measurements have to be carried out, but they are difficult to achieve. Usually, the measurements are passive since they happen through the decays of the particles, for a detailed analysis see Ref.~\cite{Bertlmann-Bramon-Garbarino-Hiesmayr2004}.\\

Let me mention two important cases.\\

\textbf{Case I:}
By varying the flavour content in the particle--antiparticle system a Wigner-type inequality, like Eq.~(\ref{Wigners-inequality}), can be established for the kaon system~\cite{Uchiyama1997}
\begin{equation}\label{UchiyamaBI}
P(K_S,\bar K^0) \;\leq\; P(K_S,K_1^0) \,+\, P(K_1^0,\bar K^0) \; .
\end{equation}
Although inequality (\ref{UchiyamaBI}) cannot be tested directly, the $CP$-conserving kaon state $K_1^0$ does not exist in Nature, it can be converted into a Bell inequality for $CP$ violation when studying the \emph{leptonic charge asymmetry}
\begin{eqnarray}\label{asymlept}
\delta \;&=&\; \frac{\Gamma(K_L\rightarrow \pi^- l^+ \nu_l) - \Gamma(K_L\rightarrow \pi^+ l^-
\bar \nu_l)}{\Gamma(K_L\rightarrow \pi^- l^+ \nu_l) + \Gamma(K_L\rightarrow \pi^+ l^-
\bar \nu_l)} \;\quad\textrm{with}\quad\; l=\mu, e \;,
\end{eqnarray}
where $l$ represents either a muon or an electron.

Then Beatrix, Walter and myself~\cite{Bertlmann-Grimus-Hiesmayr2001} could convert the Wigner-type inequality (\ref{UchiyamaBI}) into the inequality
\begin{eqnarray}\label{inequaldelta}
\delta \;&\leq&\; 0 \;,
\end{eqnarray}
for the measurable leptonic charge asymmetry which is proportional to $CP$ violation. However, inequality (\ref{inequaldelta}) is in contradiction to the experimental value
\begin{equation}\label{deltaexp}
\delta_{\rm{exp}} \;=\; (3.27 \pm 0.12)\cdot 10^{-3} \;.
\end{equation}
In fact, considering further Bell inequalities~\cite{Bertlmann-Grimus-Hiesmayr2001} restricts the asymmetry to $\delta \,=\, 0\,$, which means strict $CP$ violation.\\

In conclusion, the premises of local realistic theories are \emph{only} compatible with strict $CP$ violation in $K^0 \bar  K^0$ mixing. Conversely, $CP$ violation in $K^0 \bar  K^0$ mixing always leads to a \textit{violation} of a Bell inequality. In this way, $\delta \neq 0$ is a manifestation of the entanglement of the considered state. I also want to remark that in case of Bell inequality (\ref{UchiyamaBI}), since it is considered at $t=0$, it is rather contextuality than nonlocality that is tested. This connection between the violation of a Bell inequality and the violation of an internal symmetry of a particle is quite remarkable and must have a deeper meaning, and probably will occur for other symmetries as well.\\

\textbf{Case I\!I:}
When fixing the flavour of the kaons and varying the time of the measurements, it turns out that due to the fast decay compared to the slow oscillation, which increases the mixedness of the total system, a Bell inequality is not violated anymore by quantum mechanics~\cite{Bertlmann-Hiesmayr2001}. However, Beatrix and a group of experimentalists and theorists~\cite{Hiesmayr-DiDomenico-Catalina-etal2012} succeeded to establish a generalized Bell inequality for the $K^0 \bar{K}^0$ system, which is violated by quantum mechanics in certain measurable time regions. In this case hidden variable theories are excluded. For such an experiment the preparations at DA$\Phi$NE for the KLOE-2 detector are in progress~\cite{DiDomenico}.\\

I also want to draw attention to possible experiments that test Bell inequalities by inserting a regenerator, that is a piece of matter,  into the kaon beam~\cite{Bramon-Escribano-Garbarino2006, Bramon-Escribano-Garbarino2007, Bramon-Garbarino-PRL88-2002, Bramon-Garbarino-PRL89-2002, Hatice2009}. These experiments are of particular interest since regeneration, a typical quantum feature of the $K$ meson, is directly related to a Bell inequality.

Furthermore, a Bell test for quite a different system, a hyperon system like the $\Lambda \bar{\Lambda}$ system, has been studied~\cite{Hiesmayr-hyperon2014, Hiesmayr-talk-Unspeakables} and is experimentally planned by the FLAIR collaboration, Darmstadt.

Last but not least, tests of local realism in the decay of a charmed particle into entangled vector mesons should be mentioned as well~\cite{Li-Qiao2010, Ding-Li-Qiao2007, Li-Qiao2009}.\\

These direct Bell-type tests of our basic concepts about matter are of utmost importance since there is always a slim chance of an unexpected result, despite the fact of the general success of quantum mechanics.\\

I think John Bell, who was both, a particle- and a quantum-physicist, had been pleased seeing the developments of this kind of experiments.\\

As a particle physicist I certainly was very much interested in studying the entanglement features in a relativistic setting. I was again lucky that an enthusiastic graduate student, Nicolai Friis whom I knew very well from my university lectures, approached me for a Diploma Thesis and showed his strong interest in the connection of quantum mechanics and relativity.\\

\begin{figure}
\begin{center}
\includegraphics[width=0.41\textwidth]{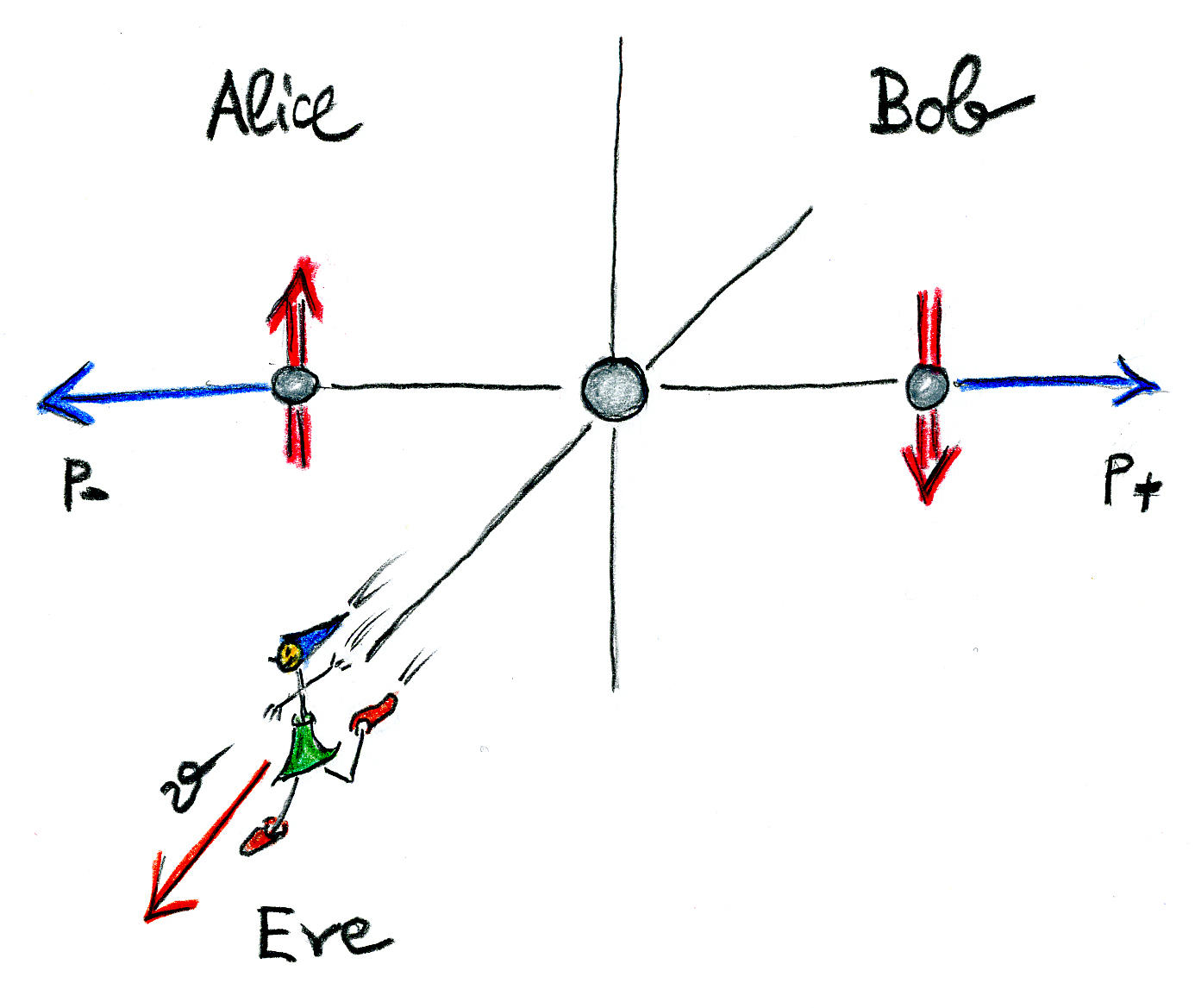}
\normalsize{
\caption{Eve, a relativistically moving observer, is studying the behaviour of the entanglement of a system of two massive spin--$\tfrac{1}{2}$ particles as a 4 qubit system, one qubit for each of the two momenta and each of the two spins. She is also testing the relativistic invariance of a Bell inequality in the Alice and Bob partition. Cartoon: \copyright Reinhold A. Bertlmann.}
\label{fig:Eve-running-Alice-Bob}}
\end{center}
\end{figure}

The first topic Nicolai and I investigated was \emph{``Relativistic entanglement of two massive particles''}~\cite{Bertlmann-Friis-relativ-entangle2010}, see Fig.\ref{fig:Eve-running-Alice-Bob}. We described the spin and momentum degrees of freedom of a system of two massive spin--$\tfrac{1}{2}$ particles as a 4 qubit system, one qubit for each of the two momenta and each of the two spins. Of course, relativistically spin and momentum of a particle were not independent of each other, what we had to take into account. Then we explicitly showed how the entanglement changed between different partitions of the qubits, when considered by different inertial observers~\cite{Bertlmann-Friis-relativ-entangle2010}. Although the two particle entanglement corresponding to a partition into Alice's and Bob's subsystems was, as often stated in the literature, invariant under Lorentz boosts, the entanglement with respect to other partitions of the Hilbert space on the other hand, was not. It certainly did depend on the chosen inertial frame and on the initial state considered. This surprising feature we could understand clearly. The change of entanglement arose, because a Lorentz boost on the momenta of the particles caused a Wigner rotation of the spin, which in certain cases entangled the spin- with the momentum states. We systematically investigated the situation for different classes of initial spin states and different partitions of the 4 qubit space.

Furthermore, we studied the behaviour of Bell inequalities for different observers and demonstrated how the maximally possible degree of violation, using the Pauli-Lubanski spin observable, could be recovered by any inertial observer, when Lorentz transforming both the states \emph{and} the observables such that each observer will measure the same expectation value if the correct measurement directions were chosen.\\

As a next step it was quite natural for Nicolai and me to consider non-inertial particle systems and to study the entanglement features there. For example, two observers shared a bipartite entangled state, where one observer was moving with uniform acceleration. There occurred an entanglement degradation in such a state by the accelerated motion, which was commonly attributed to the thermalization due to the Unruh effect. Whereas for bosonic modes the entanglement vanished in the infinite acceleration limit~\cite{fuentesschullermann05} there still remained a non-zero residual entanglement in case of (anti-) fermionic modes~\cite{bruschiloukomartinmartinezdraganfuentes10}. So we asked ourself if this residual entanglement could be used for quantum information tasks. The criterion for it was the inspection of a Bell inequality. The result published with our colleagues was \emph{``Residual entanglement of accelerated fermions is not nonlocal''}~\cite{Friis-Koehler-Martinez-Bertlmann2011}.

The statement, more precisely, was the following:
Two observers share a maximal entangled state of two fermions, where the entanglement decreases for increasing acceleration. The surviving entanglement, in the infinitely accelerated frame, however, cannot be used to violate the CHSH inequality, which is the optimal Bell inequality for this situation. Therefore no quantum information tasks using these correlations can be performed!

\begin{figure}
\begin{center}
\includegraphics[width=0.75\textwidth]{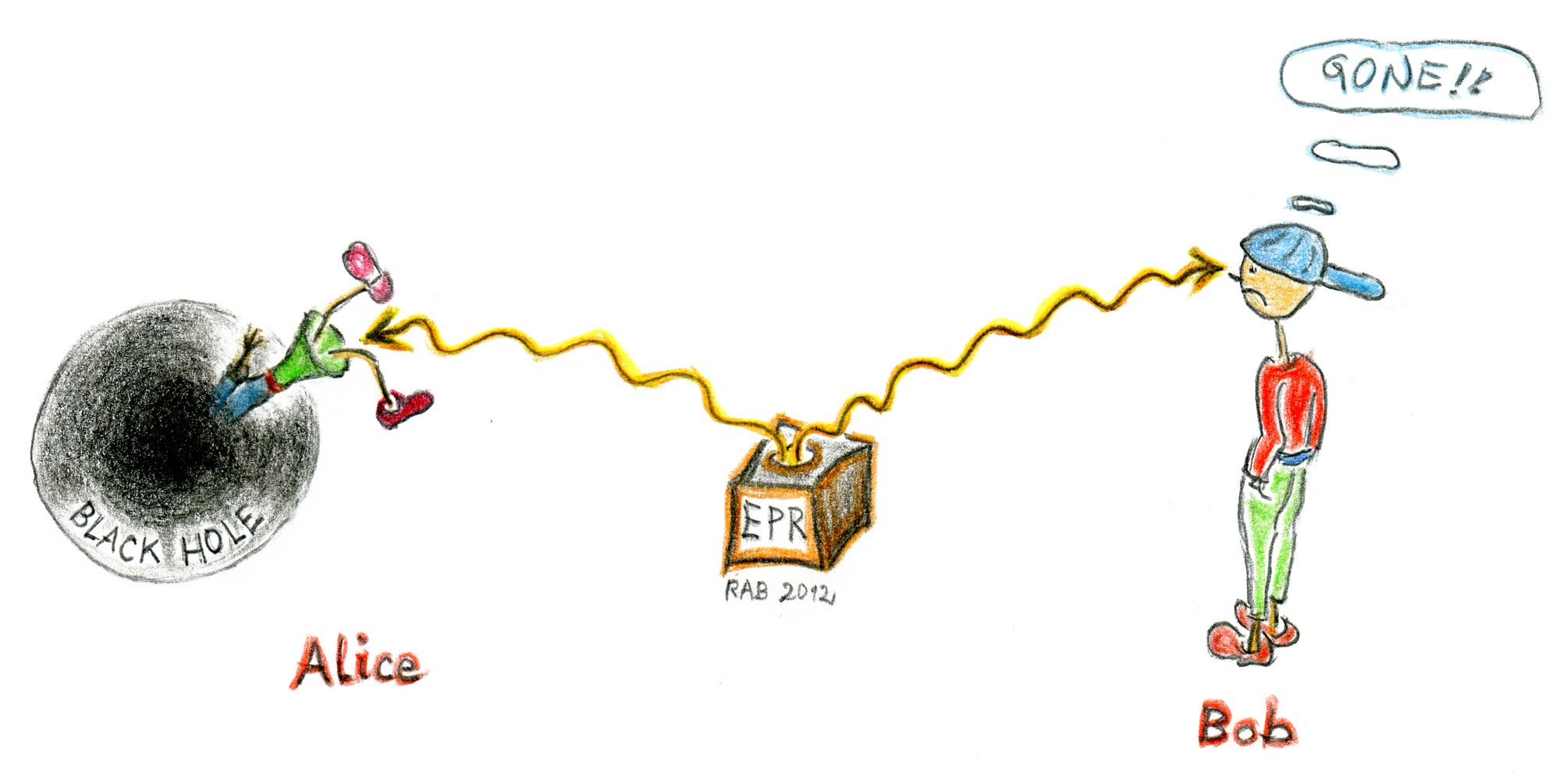}
\normalsize{
\caption{Relativistic entanglement: Alice, being an accelerated observer when falling into a black hole, cannot communicate on a quantum information level (via an EPR pair) with an observer, Bob, who is resting near the horizon. \emph{Thus quantum mechanics cannot overcome relativity\,!} Cartoon: \copyright Reinhold A. Bertlmann.}
\label{fig:Alice-falling-into-black-hole}}
\end{center}
\end{figure}

This is especially important not only for the results in the infinite acceleration limit but also if we identify this limit with a black hole situation, where an observer is freely falling and another observer is resting arbitrarily close to the event horizon. Alice, when falling into a black hole, cannot communicate on a quantum information level with an observer who is resting near the horizon, see Fig.\ref{fig:Alice-falling-into-black-hole}. Therefore we have to conclude:

\begin{center}
\emph{Quantum mechanics cannot overcome relativity\,!}
\end{center}

\section{Entanglement in Mathematical Physics}

\subsection{Entanglement and Bell Inequalities}

In mathematical physics the quantum states are described by density matrices. Then all quantum states can be classified into \emph{separable} or \emph{entangled} states. The \emph{set of separable states} is defined by the convex (and compact) hull of product states
\beq\label{set-separable-states}
S \;=\; \big\{\rho \,=\, \sum_{i} p_i \,\rho^A_i \otimes \rho^B_i \,|\; 0\leq p_i\leq 1\,, \sum_{i} p_i =1  \big\} \;.
\eeq
A state is called \emph{entangled} if it is not separable, i.e., $\rho_{\rm{ent}}\in S^c$ where $S^c$ denotes the complement of $S$, with $S \cup S^c = \widetilde{\cal H} \subset L(\Ha)\,$ and $\widetilde{\cal H} \,=\, {\widetilde{\cal H}}_A \otimes \widetilde{{\cal H}}_B$ represents the Hilbert-Schmidt space of linear operators $L(\cal H)$ on the finite dimensional bipartite Hilbert space $\cal H \,=\, {\cal H}_A \otimes {\cal H}_B$ of Alice and Bob, with dimension $D = d_A \times d_B$. For our discussion of qubits $d_A = d_B = 2\,$.\\

In terms of density matrices the CHSH inequality (\ref{CHSH-inequality}) can be rewritten in the following way
\beq\label{CHSH-inequality-density-matrix}
\langle \rho | \B_{\rm{CHSH}} \rangle \;=\; \rm{Tr} \,\rho \B_{\rm{CHSH}} \;\leq\; 2 \;,
\eeq
for all local states $\rho\,$, where the CHSH-Bell operator in case of qubits is expressed by
\beq\label{CHSH-Bell operator}
\B_{\rm{CHSH}} \;=\; \vec{a}\cdot\vec{\sigma}_A \otimes (\vec{b} - \vec{b}^{\,'})\cdot\vec{\sigma}_B \;+\; \vec{a}^{\,'}\cdot\vec{\sigma}_A \otimes (\vec{b} + \vec{b}^{\,'})\cdot\vec{\sigma}_B \;.
\eeq
Rewriting inequality (\ref{CHSH-inequality-density-matrix}) gives
\begin{equation}\label{CHSH-inequality-density-matrix-separable}
\langle \rho | \, 2\cdot\mathds{1} \,-\, \B_{\rm{CHSH}} \rangle \;\geq\; 0 \;.
\end{equation}
If we choose, however, the entangled Bell state $\rho^{-} = \ket{\psi^{\,-}}\bra{\psi^{\,-}}$ the inner product changes the sign
\begin{equation}\label{CHSH-inequality-density-matrix-Bellstate}
\langle \rho^{-} | \, 2\cdot\mathds{1} \,-\, \B_{\rm{CHSH}} \rangle \;<\; 0 \;.
\end{equation}

Now we can ask, is the inner product (\ref{CHSH-inequality-density-matrix-Bellstate}) negative for \emph{all} entangled states? The answer is \emph{yes} for all pure entangled states \cite{Gisin1991}, i.e., there exist measurement directions for which the CHSH inequality is violated. For mixed states, however, the situation is much more subtle (see, e.g., Section~5 of Ref.~\cite{Bertlmann-JPhysA2014}). Reinhard Werner \cite{werner89} discovered that a certain family of bipartite mixed states, which remained entangled, produced an outcome that admitted a local hidden variable model for projective measurements, that is, it satisfies all possible Bell inequalities.

This feature is nicely demonstrated by the so-called  \emph{Werner states}
\begin{equation}\label{Werner-state in Bloch notation}
\rho_{\,\rm{Werner}} \;=\; \alpha\, \rho^- \,+\, \frac{1-\alpha}{4}\, \mathds{1}_4 \;=\;
\frac{1}{4}\left(\mathds{1}\,\otimes\,\mathds{1} \,-\, \alpha\,\sigma_i\,\otimes\,\sigma_i\right)\;\equiv\; \rho_{\alpha} \;,
\end{equation}
written in terms of the Bloch decomposition with the parameter values $\alpha \in [0,1]\,$.\\

The region of separability is determined by the so-called \emph{PPT criterion} (positive partial transposition) of Peres~\cite{peres96} and the Horodecki family~\cite{horodecki96}. Given a general density matrix $\rho$ in Hilbert-Schmidt space $\widetilde{\cal H} \,=\, {\widetilde{\cal H}}_A \otimes \widetilde{{\cal H}}_B$ in its Bloch decomposition form
\begin{equation}\label{rho-general-decomposition-Bloch-form}
\rho \;=\; \frac{1}{4}\left(\,\mathds{1}\,\otimes\,\mathds{1} \,+\, r_i \,\sigma_i\,\otimes\,\mathds{1} \,+\,
u_i \,\mathds{1}\,\otimes\,\sigma_i \,+\, t_{ij} \;\sigma_i\,\otimes\,\sigma_j\right) \;,
\end{equation}
then a \emph{partial transposition} is defined by the operator $T$ acting in a subspace ${\widetilde{\cal H}}_A$ or ${\widetilde{\cal H}}_B$ and transposing there the off-diagonal elements of the Pauli matrices: $T\,(\sigma^i)_{kl} \,=\, (\sigma^i)_{lk}\,$. If and only if, in $2 \times 2$ and $2 \times 3$ dimensions, a state remains positive under partial transposition then the state is separable. In higher dimensions the PPT criterion is only necessary but not sufficient for separability.

In case of the Werner states (\ref{Werner-state in Bloch notation}) the partial transposition provides the following result: the states are separable for $\alpha \leq \frac{1}{3}$ and entangled for $\alpha > \frac{1}{3}\,$.\\

In order to find the states violating a Bell inequality an other theorem of the Horodecki family~\cite{horodecki95} is very powerful since we do not have to check all measurement directions $\vec{a}$ and $\vec{b}\,$. There one has to consider the square root of the two larger eigenvalues $t^2_1, t^2_2\,$ of the product of the $t$-matrices $(t_{ij})^T (t_{ij})\,$. If $B^{\rm{max}} \,=\, \frac{1}{2}\,\rm{max}_{\B}\,\rm{Tr}\,\rho\,\B_{\rm{CHSH}} \,=\, \sqrt{t^2_1 + t^2_2} \;>\; 1 \,$ then the CHSH inequality is maximal violated.

In case of the Werner states we can read off the coefficient matrix directly from the Bloch decomposition (\ref{Werner-state in Bloch notation}), which yields the maximal violation of the CHSH inequality by $B^{\rm{max}} \,=\, \sqrt{2 \alpha^2} \;>\; 1 \,$. Thus, for all $\alpha > 1/\sqrt{2}$ the CHSH-Bell inequality is violated.\\

\subsection{Entanglement Witness Inequality}

I remember, when Anton and I organized the conference \emph{``Quantum [Un]Speakables 2000''} Walter Thirring, our Doyen of Theoretical Physics, participated very actively. We had many discussions about Bell inequalities and their physical meaning. These resulted in an enjoyable collaboration and series of works together with Heide Narnhofer, a prominent mathematical physicist.

In \emph{``A geometric picture of entanglement and Bell inequalities''}~\cite{bertlmann-narnhofer-thirring02} we asked ourselves how to `detect' entanglement and to discriminate it from all separable quantum states. A Bell operator given by expression (\ref{CHSH-Bell operator}) was obviously not appropriate to find all entangled states. In order to locate entanglement accurately a different operator had to be constructed. This was a Hermitian operator, the so-called \emph{entanglement witness} $A$, that detected the entanglement of a state $\rho_{\rm ent}$ via an \emph{entanglement witness inequality}. So we arrived at the following theorem.

\begin{theorem}[Entanglement Witness Theorem~\cite{horodecki96, bertlmann-narnhofer-thirring02, terhal00}]\ \

A state $\rho_{\rm ent}$ is entangled if and only if there is a Hermitian operator $A$ -- the entanglement witness -- such that
\begin{eqnarray} \label{def-entwit}
    \left\langle \rho_{\rm ent}|A \right\rangle \;=\; \textnormal{Tr}\, \rho_{\rm ent} A
    & \;<\; & 0 \,,\nonumber\\
    \left\langle \rho|A \right\rangle = \textnormal{Tr}\, \rho A & \;\geq\; & 0 \qquad
    \forall \rho \in S \,,
\end{eqnarray}

where $S$ denotes the set of all separable states.
\label{theorem:entanglement-witness-theorem}
\end{theorem}

An entanglement witness is called \emph{optimal}, and denoted by $A_{\rm opt}\,$, if apart from Eq.~(\ref{def-entwit}) there exists a separable state $\rho_0 \in S$ such that
\begin{equation}
    \left\langle \rho_0 |A_{\rm opt} \right\rangle \;=\; 0 \,.
\end{equation}
The operator $A_{\rm opt}$ defines a tangent plane to the convex set of separable states $S$ (\ref{set-separable-states}), as illustrated in Fig.~\ref{fig:BNT-theorem-opt-entangle-witness}. Such an $A_{\rm opt}$ always exists due to the Hahn-Banach Theorem and the convexity of $S\,$.\\

On the other hand, with help of the Hilbert-Schmidt norm we can define the \emph{Hilbert-Schmidt distance} between two arbitrary states $\rho_1$ and $\rho_2$
\begin{equation}
    d_{\rm{HS}}(\rho_1,\rho_2) \;=\; \| \rho_1 - \rho_2 \| \;=\; \sqrt{<\rho_1 - \rho_2 | \rho_1 - \rho_2>} \;=\; \sqrt{\rm{Tr} \,(\rho_1 - \rho_2)^\dagger (\rho_1 - \rho_2)}  \;.
\end{equation}
We view the minimal distance of an entangled state $\rho_{\rm ent}$ to the set of separable states, the \emph{Hilbert-Schmidt measure}
\begin{equation} \label{def-HSmeasure}
    D(\rho_{\rm{ent}}) \;:=\; \min_{\rho \in S} \left\| \rho \,-\, \rho_{\rm{ent}} \right\|
    \;=\; \left\| \rho_0 \,-\, \rho_{\rm{ent}} \right\| \,,
\end{equation}
where $\rho_0$ denotes the nearest separable state, as a \emph{measure for entanglement}.\\

What we then discovered was an interesting connection between the Hilbert-Schmidt measure and the entanglement witness inequality. Let us rewrite entanglement witness inequality (\ref{def-entwit})
\beq\label{EWIrewritten}
\left\langle \rho|A \right\rangle \,-\, \left\langle \rho_{\rm ent}|A \right\rangle \;\geq\; 0 \qquad \forall \rho \in S \,,
\eeq
and define the maximal violation of inequality (\ref{EWIrewritten}) as follows ($\rho$ and $A$ are still free at our disposal):

\begin{definition}[Maximal violation of the entanglement witness inequality~\cite{bertlmann-narnhofer-thirring02}]\ \
\begin{eqnarray} \label{max-violation-EWI}
   B(\rho_{\rm ent}) \;=\; \max_{A} \,\big( \min_{\rho \in S} \left\langle \rho|A \right\rangle \,-\, \left\langle \rho_{\rm ent}|A \right\rangle  \big) \,.
\end{eqnarray}
\label{definiton:max-violation-of-EWI}
\end{definition}
The minimum is taken over all separable states and maximum over all possible entanglement witnesses $A\,$, suitably normalized. Then there holds the following theorem:

\begin{figure}
\begin{center}
\includegraphics[width=0.3\textwidth]{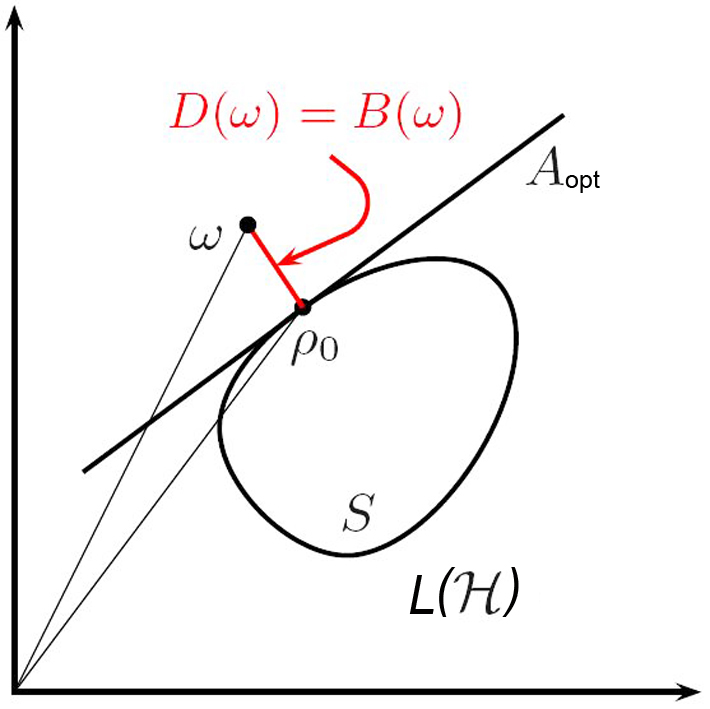}
\normalsize{
\caption{Illustration of the Bertlmann-Narnhofer-Thirring Theorem: $D(\omega) = B(\omega)$, the minimal distance $D$ of the entangled state $\omega$ to the set of separable states $S$ in the Hilbert-Schmidt space is equal to the maximal violation $B$ of the entanglement witness inequality. $A_{\rm{opt}}$ represents the optimal entanglement witness.}
\label{fig:BNT-theorem-opt-entangle-witness}}
\end{center}
\end{figure}

\begin{theorem}[Bertlmann-Narnhofer-Thirring Theorem~\cite{bertlmann-narnhofer-thirring02}]\ \
\begin{eqnarray} \label{B-equals-D}
    a)&& B(\rho_{\rm ent}) \;=\; D(\rho_{\rm{ent}}) \,,
\end{eqnarray}
    b) The maximal violation of the entanglement witness inequality is achieved when $\rho \rightarrow \rho_0$ and $A \rightarrow A_{\rm opt}\,$, then the optimal entanglement witness is given by \\
\begin{eqnarray} \label{Aopt-explicit-expression}
    A_{\rm opt} \;=\; \frac{\rho_0 \,-\, \rho_{\rm ent} \,-\, \left\langle \rho_0|\rho_0 \,-\, \rho_{\rm ent} \right\rangle \mathds{1}}{\left\| \rho_0 \,-\, \rho_{\rm ent} \right\|} \;.
\end{eqnarray}
\label{theorem:BNT-theorem}
\end{theorem}

In words:

\emph{The maximal violation of the entanglement witness inequality is equal to the Hilbert-Schmidt measure\,!}\\

As we regard the Hilbert-Schmidt measure (\ref{def-HSmeasure}) as a measure for entanglement, it means that the amount of entanglement is given by the amount of violation (\ref{max-violation-EWI}) of the witness inequality. This is a remarkable result that we have illustrated in Fig.~\ref{fig:BNT-theorem-opt-entangle-witness}. Furthermore, the optimal entanglement witness is given explicitly by expression (\ref{Aopt-explicit-expression}). However, we have to know the nearest separable state $\rho_0\,$, which is easy to find in low dimensions, but not in higher ones. Nevertheless, there exists an approximation procedure to approach $\rho_0$~\cite{bertlmann-krammer-AnnPhys09}. For a review, see Ref.~\cite{Guehne-Toth2009}.\\

For example, in case of Alice and Bob the Werner states are given by $\rho_{\alpha}$ (\ref{Werner-state in Bloch notation}), and the Bell state $\rho^{-} = \ket{\psi^{\,-}}\bra{\psi^{\,-}}$ by choosing the parameter value $\alpha = 1\,$, i.e., $\rho^{-} = \rho_{\alpha = 1}\,$.

The nearest separable state is easily found
\beq\label{nearest-sep-state-Alice-Bob}
\rho_0 \;=\; \frac{1}{4}\left(\mathds{1}\,\otimes\,\mathds{1} \,-\, \frac{1}{3}\,\sigma_i\,\otimes\,\sigma_i\right) \,,
\eeq
yielding the Hilbert-Schmidt measure
\beq\label{HSmeasure-for-Alice-Bob}
D(\rho_{\alpha}) \;=\; \left\| \rho_0 \,-\, \rho_{\alpha} \right\| \;=\; \frac{\sqrt{3}}{2}\,(\alpha \,-\, \frac{1}{3}) \,.
\eeq
The optimal entanglement witness we calculate from expression (\ref{Aopt-explicit-expression})
\begin{equation}\label{Aopt-for-Alice-Bob}
A_{\rm opt} \;=\; \frac{1}{2\sqrt{3}}\left(\mathds{1}\,\otimes\,\mathds{1} \,+\, \sigma_i\,\otimes\,\sigma_i\right) \,,
\end{equation}
and the maximal violation of the entanglement witness inequality from Eq.~(\ref{max-violation-EWI})
\beq\label{max-violat-of-EWI-for-Alice-Bob}
B(\rho_{\alpha}) \;=\; -\, \left\langle \rho_{\alpha}|A_{\rm opt} \right\rangle \;=\; \frac{\sqrt{3}}{2}\,(\alpha \,-\, \frac{1}{3}) \,.
\eeq
Clearly, both results (\ref{max-violat-of-EWI-for-Alice-Bob}) and (\ref{HSmeasure-for-Alice-Bob}) coincide as required by Theorem~\ref{theorem:BNT-theorem}.

\subsection{Geometry of Quantum States: Entanglement versus Separability}\label{subsec:geometry-of-quantum-states}

Let us next turn to a geometrical description of the quantum states, how they are distributed in the Hilbert-Schmidt space of the density matrices. The quantum states for a two-qubit system, the case of Alice and Bob, have very nice geometric features in the Hilbert-Schmidt space, more precisely, in the spin--spin space. Quite generally a quantum state can be decomposed as in Eq.~(\ref{rho-general-decomposition-Bloch-form}), where the last term, the spin--spin term, is the important one to characterize entanglement. If we parameterize the spin-spin space by
\beq\label{Bell-states-parametrisation-Alice-Bob}
\rho \;=\; \frac{1}{4}\left(\mathds{1}\,\otimes\,\mathds{1} \,+\, \sum_i\,c_i\,\sigma_i\,\otimes\,\sigma_i \right) \,,
\eeq
the Bell states have the coefficients $c_i = \pm 1$.

\begin{figure}
\begin{center}
(a)\;\setlength{\fboxsep}{2pt}\setlength{\fboxrule}{0.8pt}\fbox{
\includegraphics[angle = 0, width = 62mm]{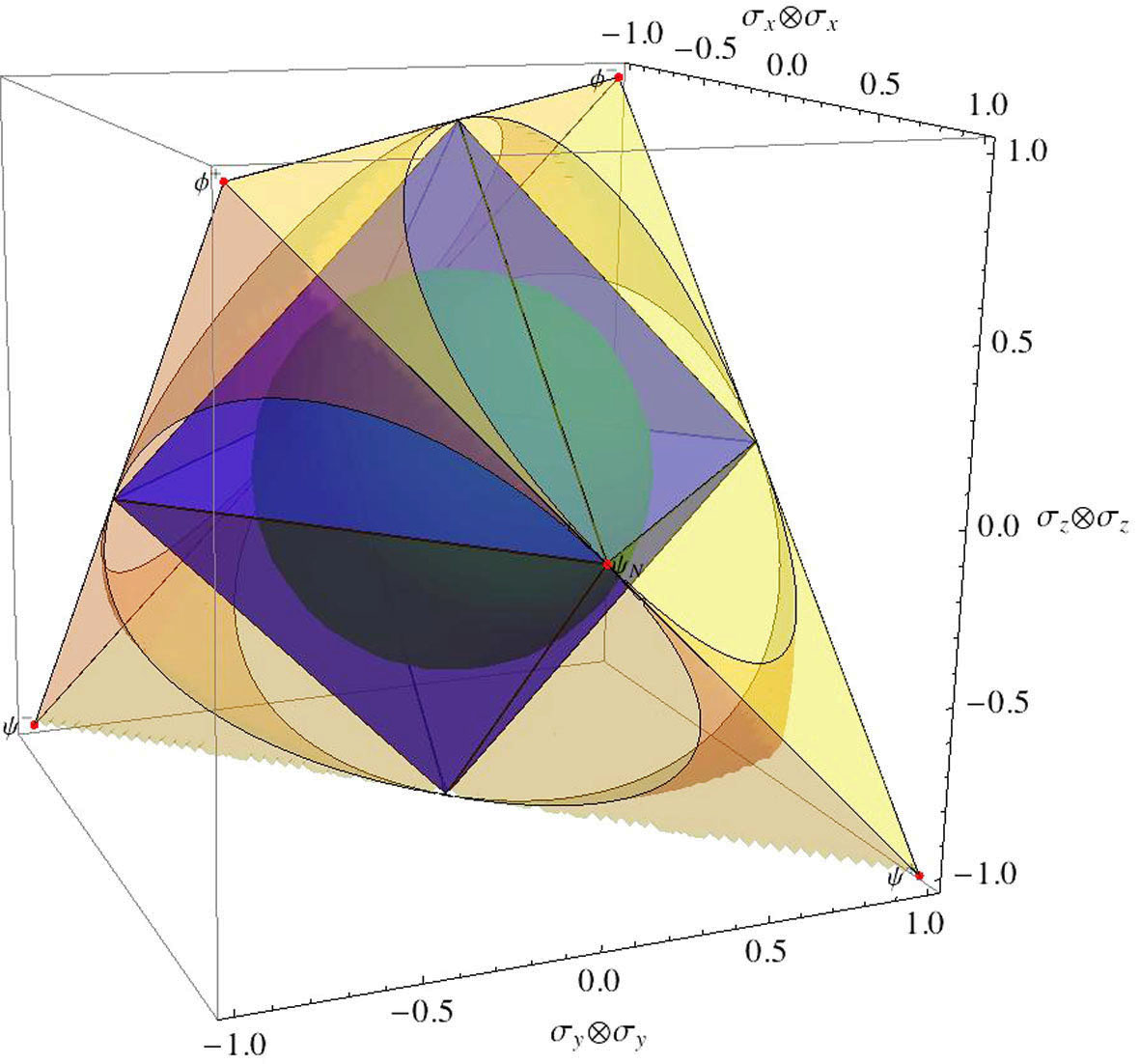}}
\hspace{5mm}
(b)\;\setlength{\fboxsep}{2pt}\setlength{\fboxrule}{0.8pt}\fbox{
\includegraphics[angle = 0, width = 59.9mm]{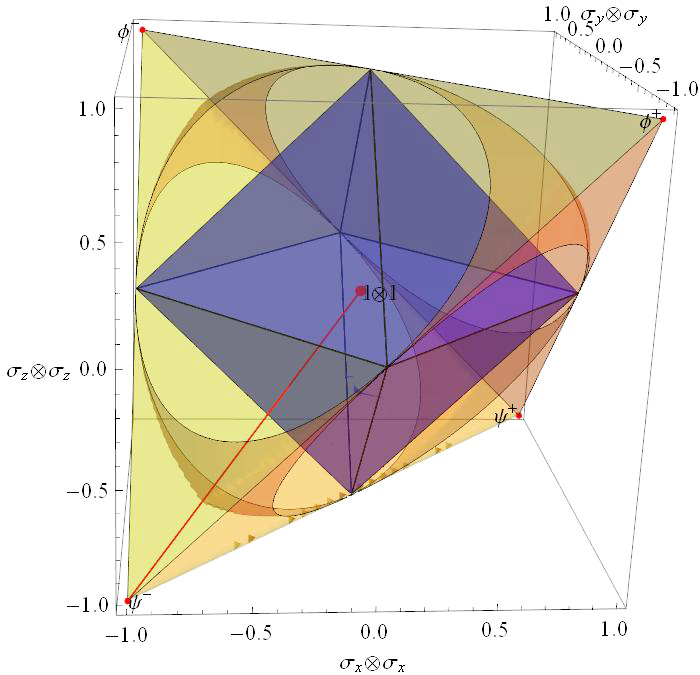}}
\normalsize{
\caption{(a) Tetrahedron of physical states in $2 \times 2$ dimensions spanned by the four Bell states $\psi^+ , \psi^- , \phi^+ , \phi^-\,$: The separable states form the blue double pyramid and the entangled states are located in the remaining tetrahedron cones. The unitary invariant Ku\'s-\.Zyczkowski ball (shaded in green), the maximal ball of absolutely separable states, is located within the double pyramid and the maximal mixture $\frac{1}{4}\mathds{1}_4$ is at the origin. Outside the ball at the corner of the double pyramid is the state $\psi_{\rm{N}}$, the separable (but not absolutely separable) state with maximal purity. The local states according to a Bell inequality lie within the dark-yellow surfaces containing all separable but also some entangled states. (b) Tetrahedron with the illustration of the Werner states (red line from the origin to the maximal entangled Bell state $\psi^-$) that pass through all regions of separability, locality and entanglement.}
\label{fig:Tetraeder-Simplex}}
\end{center}
\end{figure}

Due to the positivity of the density matrix the four Bell states $\psi^{-}, \psi^{+}, \phi^{-}, \phi^{+}$ set up a simplex, a tetrahedron, in this spin--spin space \cite{bertlmann-narnhofer-thirring02, vollbrecht-werner-PRA00, horodecki-R-M96}, as illustrated by the two figures in Fig.~\ref{fig:Tetraeder-Simplex}. The separable states, given by the PPT criterion, form an octahedron which lies inside, and the maximal mixed state $\frac{1}{4}\mathds{1}_4 \,=\, \frac{1}{4}(\mathds{1}\otimes\mathds{1})$ is placed at origin. The entangled states are located in the remaining cones. The local states, on the other hand, satisfying a CHSH-Bell inequality lie within the parachutes, the dark-yellow surfaces in the tetrahedrons of Fig.~\ref{fig:Tetraeder-Simplex}~\cite{TBKN}. They are determined by the Horodecki Theorem~\cite{horodecki95} and contain all separable but also a large amount of mixed entangled states. The Werner states (\ref{Werner-state in Bloch notation}), the red line in Fig.~\ref{fig:Tetraeder-Simplex}(b) from the origin to the maximal entangled Bell state $\psi^-\,$, show nicely how the states change from maximal mixed and separable to local, mixed entangled, and finally to nonlocal states, ending at $\psi^{-}$ which is pure and maximal entangled.\\

There is an important point I became aware of in my discussions with Walter and Heide. For a given quantum state we have the free choice of how to factorize the algebra of a density matrix, which implies either entanglement or separability of the quantum state. Only with respect to such a factorization it makes sense to talk about entanglement or separability. For instance, quantum teleportation precisely relies on this fact that we can think of different factorizations in which entanglement is localized with respect to the measurements that are carried out. Thus we may choose! Via global unitary transformations we can switch from one factorization to the other, where in one factorization the quantum state appears entangled, however, in the other not. These discussions we published in the paper \emph{``Entanglement or separability: The choice of how to factorize the algebra of a density matrix''}~\cite{TBKN}, where also Philipp K\" ohler, one of my last Diploma students, joined.

We realized that there is \emph{``total democracy between the different factorizations''}~\cite{TBKN}, no partition has ontologically a superior status over any other one (see also Ref.~\cite{zanardi2001}). For an experimentalist, however, a certain factorization is preferred and is clearly fixed by the set-up. Consequently, entanglement or separability of a quantum state depends on our choice of factorizing the algebra of the corresponding density matrix, where this choice is suggested either by the set-up of the experiment or by the convenience for the theoretical discussion. This was our basic message.\\

For pure states the status is quite clear. Any state can be factorized such that it appears separable up to being maximally entangled depending on the factorization. We can prove the following theorems~\cite{TBKN}:

\begin{theorem}[Factorization algebra]\ \
For any pure state $\rho$ one can find a factorization $M^D=\A_1 \otimes \A_2$ such that $\rho$ is separable with respect to this factorization and an other factorization $M^D=\B_1 \otimes \B_2$ where $\rho$ appears to be maximally entangled.
\label{theorem:factorization-algebra pure states}
\end{theorem}

The extension to mixed states requires some restrictions, as can be seen from the tracial state $\frac{1}{D}\mathds{1}_D$ which is separable for any factorization.

\begin{theorem}[Factorization in mixed states]\ \
For any mixed state $\rho$ one can find a factorization $M^D=\A_1 \otimes \A_2$ such that $\rho$ is separable with respect to this factorization. An other factorization $M^D=\B_1 \otimes \B_2$ where $\rho$ appears to be entangled exists only beyond a certain bound of mixedness.
\label{theorem:factorization-algebra mixed states}
\end{theorem}

It is interesting to search for this bound, for those states which are separable with respect to all possible factorizations of the composite system into subsystems $\A_1 \otimes \A_2\,$. This is the case if $\rho_{\rm{U}} \,=\, U\rho\,U^\dag$ remains separable for any unitary transformation $U$. Such states are called \emph{absolutely separable states} \cite{kus-zyczkowski, zyczkowski-bengtsson, bengtsson-zyczkowski-book}, the tracial state being the prototype. In this connection the \emph{maximal ball} of states around the tracial state $\frac{1}{d^2}\,\mathds{1}_{d^2}$ with a general radius $r = \frac{1}{d^2 - 1}$ of constant mixedness is considered, which can be inscribed into the separable states. This radius is given in terms of the Hilbert-Schmidt distance
\begin{equation}\label{radius by HS distance}
d\,(\rho , \mathds{1}_{d^2}) \;=\; \left\| \rho \,-\, \frac{1}{d^2}\,\mathds{1}_{d^2} \right\| \;=\; \sqrt{\rm{Tr}\,\big(\rho \,-\, \frac{1}{d^2}\,\mathds{1}_{d^2}\big)^2}\,.
\end{equation}

\begin{theorem}[Absolute separability of the Ku\'s-\.Zyczkowski ball \cite{kus-zyczkowski}]\ \
All states belonging to the maximal ball which can be inscribed into the set of mixed states for a bipartite system are not only separable but also absolutely separable.
\label{theorem:Kus-Zyczkowski ball}
\end{theorem}

The maximal ball of absolutely separable states we have illustrated by the green shaded ball in Fig.~\ref{fig:Tetraeder-Simplex}(a).\\

As illustration of Theorem~\ref{theorem:Kus-Zyczkowski ball} let us choose the following separable state
\beq\label{eq:narnhofer state in Bloch decomposition}
\rho_{\rm{N}} \;=\; \left|\,\psi_{\rm{N}}\,\right\rangle\left\langle\,\psi_{\rm{N}}\,\right| \;=\; \frac{1}{4}\left(\mathds{1} \otimes \mathds{1} \,+\, \sigma_x \otimes \sigma_x\right)\,.
\eeq
It is placed at the corner of the double pyramid of separable states (see Fig.~\ref{fig:Tetraeder-Simplex}(a)) and has the smallest possible mixedness or largest purity. The following unitary transformation~\cite{TBKN}
\begin{eqnarray}
U &\;=\;& \frac{1}{4} \big( (2+\sqrt{2})\,\mathds{1}\,\otimes\,\mathds{1} \,+\, i\sqrt{2}\,(
\sigma_x\,\otimes\,\sigma_y \,+\, \sigma_y\,\otimes\,\sigma_x\,)
\,-\, (2-\sqrt{2})\,\sigma_z\,\otimes\,\sigma_z \,\big)
\label{eq:unitary transform for narnhofer state}
\end{eqnarray}
transforms the state $\rho_{\rm{N}}$ (\ref{eq:narnhofer state in Bloch decomposition}) into
\begin{eqnarray}
\rho_{\rm{U}} &\;=\;& U\rho_{\rm{N}}\,U^\dag \;=\; \frac{1}{4} \big( \,\mathds{1}\,\otimes\,\mathds{1} \,+\, \frac{1}{2}
\,(\sigma_z\,\otimes\,\mathds{1} \,+\, \mathds{1}\,\otimes\,\sigma_z\,)
\,+\, \frac{1}{2}\,(\sigma_x\,\otimes\,\sigma_x \,+\, \sigma_y\,\otimes\,\sigma_y )\,\big) \,.
\label{eq:narnhofer state unitary transformed}
\end{eqnarray}
However, due to the occurrence of the term ($\sigma_z\,\otimes\,\mathds{1} \,+\, \mathds{1}\,\otimes\,\sigma_z$) the transformation $U$ (\ref{eq:unitary transform for narnhofer state}) leads to a quantum state that is located outside of the set of Weyl states which are pictured in Fig.~\ref{fig:Tetraeder-Simplex}. This new state $\rho_{\rm{U}}$ (\ref{eq:narnhofer state unitary transformed}) is not positive any more under partial transposition, $\rho_{\rm{U}}^{\rm{PT}} \not\geq 0\,$, where $\rho_{\rm{U}}^{\rm{PT}} = (\mathds{1}\,\otimes\,T_{\rm{B}} )\,\rho_{\rm{U}}$ and $T_{\rm{B}}$ means partial transposition on Bob's subspace. Therefore, due to the Peres-Horodecki criterion the state $\rho_{\rm{U}}$ (\ref{eq:narnhofer state unitary transformed}) is entangled and has a concurrence $C = \frac{1}{2}\,$. Transformation $U$ (\ref{eq:unitary transform for narnhofer state}) is already optimal, i.e., it entangles $\rho_{\rm{N}}$ maximally~\cite{TBKN}.\\

It is also quite instructive to illustrate Theorem~\ref{theorem:factorization-algebra pure states} by a specific example. General quantum states are expressed by Eq.~(\ref{rho-general-decomposition-Bloch-form}) and separable states can be decomposed into
\begin{equation} \label{rho-separable Bloch decomposition}
\rho_{\rm{sep}} \,=\, \frac{1}{4}\left(\,\mathds{1}\,\otimes\,\mathds{1} + r_i \,\sigma_i\,\otimes\,\mathds{1} +
u_i \,\mathds{1}\,\otimes\,\sigma_i + r_i u_j \;\sigma_i\,\otimes\,\sigma_j\right) \,,
\end{equation}
with ${\vec r}^{\,2} = {\vec u}^{\,2} = 1$. A specific separable state is
\beq\label{rho-updown Bloch decomposition}
\rho_{\Uparrow\Downarrow} \;=\; |\Uparrow \rangle \otimes |\Downarrow \rangle \langle \Uparrow | \otimes \langle \Downarrow |    \;&=&\;\frac{1}{4}\left(\mathds{1}\,\otimes\,\mathds{1} \,+\, \sigma_z\,\otimes\,\mathds{1} \,-\, \mathds{1}\,\otimes\,\sigma_z \,-\, \sigma_z\,\otimes\,\sigma_z\right) \,.
\eeq

Let us start with the maximal entangled Bell state
\begin{eqnarray} \label{rho-Bell-minus Bloch decomposition}
\rho^- \;&=&\; |\psi^- \rangle \langle \psi^- | \;=\;  \frac{1}{4}\left(\mathds{1}\,\otimes\,\mathds{1} \,-\, \vec\sigma\,\otimes\,\vec\sigma\right)\;.
\end{eqnarray}
Its optimal entanglement witness
\begin{equation} \label{entwit rho-Bell minus}
A_{\rm{opt}}^{\rho^-} \;=\; \frac{1}{2\sqrt{3}}\left(\mathds{1}\,\otimes\,\mathds{1} \,+\, \vec\sigma\,\otimes\,\vec\sigma\right)\,,
\end{equation}
provides the entanglement witness inequality
\begin{eqnarray} \label{EWI rho-Bell-minus}
    \left\langle \rho^-|A_{\rm{opt}}^{\rho^-} \right\rangle &\;=\;& \textnormal{Tr}\, \rho^- A_{\rm{opt}}^{\rho^-}
    \,=\, -\frac{1}{\sqrt{3}} \;<\; 0 \,,\nonumber\\
    \left\langle \rho_{\rm{sep}}|A_{\rm{opt}}^{\rho^-} \right\rangle &\;=\;& \textnormal{Tr}\, \rho_{\rm{sep}} A_{\rm{opt}}^{\rho^-} \,=\, \frac{1}{2\sqrt{3}}(1 + \cos \delta) \;\geq\; 0 \qquad
    \forall \rho \in S \,,
\end{eqnarray}
where $\delta$ represents the angle between the unit vectors $\vec r$ and $\vec u$.\\

Then there exists a global unitary matrix
\beq\label{eq:unitary-transform in Bloch notation}
U \;=\; \frac{1}{\sqrt{2}}\left(\mathds{1} \otimes \mathds{1} + i\,\sigma_x \otimes \sigma_y\right)\;,
\eeq
which transforms the Bell state $\rho^-$ into the separable state $\rho_{\Uparrow\Downarrow}$
\begin{eqnarray} \label{U-transform of rho-Bell minus}
    & & U\,\rho^-\,U^\dag \;=\; \frac{1}{4}\left(\mathds{1}\,\otimes\,\mathds{1} \,+\, \sigma_z\,\otimes\,\mathds{1} \,-\, \mathds{1}\,\otimes\,\sigma_z \,-\, \sigma_z\,\otimes\,\sigma_z\right) \;\equiv\; \rho_{\Uparrow\Downarrow}\,, \\
    & & \left\langle U\,\rho^-\,U^\dag|A_{\rm{opt}}^{\rho^-} \right\rangle \;=\; \textnormal{Tr}\; U\,\rho^-\,U^\dag \,A_{\rm{opt}}^{\rho^-} \,=\, 0 \;,
\end{eqnarray}
i.e., separability with respect to the algebra $\{ \sigma_i\,\otimes\,\sigma_j \}\,$. Thus the transformed state $U\,\rho^-\,U^\dag$ represents a separable pure state as claimed in Theorem~\ref{theorem:factorization-algebra pure states} and geometrically it has the Hilbert-Schmidt distance
\begin{equation} \label{HSdistance of rho-Bell minus to transformed state}
    d\,(\rho^-) \;=\; \left\| U\,\rho^-\,U^\dag \,-\, \rho^- \right\| \;=\; 1\,,
\end{equation}
to the state $\rho^-\,$. This distance represents the amount of entanglement of $\rho^-\,$.\\

Transforming on the other hand also the entanglement witness, i.e., choosing a different algebra,
\begin{equation} \label{U-transform of EW-Bell minus}
U\,A_{\rm{opt}}^{\rho^-}\,U^\dag \;=\; \frac{1}{4}\left(\mathds{1}\,\otimes\,\mathds{1} \,-\, \sigma_z\,\otimes\,\mathds{1} \,+\, \mathds{1}\,\otimes\,\sigma_z \,+\, \sigma_z\,\otimes\,\sigma_z\right) \,,
\end{equation}
we then get for the entanglement witness inequality
\begin{eqnarray} \label{U-transform of EWI}
    \left\langle U \rho^-\,U^\dag|U A_{\rm{opt}}^{\rho^-}\,U^\dag \right\rangle &\;=\;&
    \left\langle \rho^-|A_{\rm{opt}}^{\rho^-} \right\rangle \;=\; -\frac{1}{\sqrt{3}} \;<\; 0 \,,
\end{eqnarray}
and the transformed state is entangled again with respect to the other algebra factorization $\{ \sigma_i\,\otimes\,\mathds{1}, \mathds{1}\,\otimes\,\sigma_j , \sigma_i\,\otimes\,\sigma_j \}\,$. It demonstrates nicely the content of Theorem~\ref{theorem:factorization-algebra pure states}. It can be seen as an analogy to choosing either the Schr\"odinger picture or the Heisenberg picture in the characterization of the quantum states.\\

In an other collaboration Walter, Heide and myself studied \emph{``The time-ordering dependence of measurements in teleportation''}~\cite{Bertlmann-Narnhofer-Thirring2013}, where the phenomenon of \emph{``delayed-choice entanglement swapping''} could be traced back to the commutativity of the projection operators that were involved in the corresponding measurement process. We also proposed an experimental set-up which depended on the order of successive measurements corresponding to noncommutative projection operators.\\

Finally, I would like to mention a recent collaboration with Beatrix and Gabriele Uchida, an expert in scientific computing, where we investigated \emph{``Entangled Entanglement: The Geometry of Greenberger-Horne-Zeilinger States''}~\cite{Uchida-Bertlmann-Hiesmayr2014}. The familiar Greenberger-Horne-Zeilinger (GHZ) states could be rewritten by entangling the Bell states for two qubits with a state of the third qubit, which was named \emph{``entangled entanglement''}~\cite{Krenn-Zeilinger}. We showed that in a constructive way we could obtain all $8$ independent GHZ states that formed the simplex of entangled entanglement, the \emph{magic simplex}. The construction procedure allowed a generalization to higher dimensions both, in the degrees of freedom (considering qudits) as well as in the number of particles (considering $n$-partite states). Such bases of GHZ-type states exhibited a cyclic geometry, a \emph{Merry Go Round}, relevant for experimental and theoretical quantum information applications. We also discussed the inherent symmetries and the regions of (genuine) multi-partite entanglement within the simplex.

\section{As Time Goes By ...}

This Article is devoted to the memory of John Stewart Bell, the outstanding scientist and man with honest character and high moral. I had the great fortune to be close to him, to enjoy the fruitful collaboration and warm friendship. My aim was to show the large scope of Bell's Universe, Bell's deep insight into Nature, by describing his superb contributions in particle physics, accelerator physics and quantum physics. John himself had no preference for his works in the different fields, he was just as pleased with his particle physics papers as with his accelerator papers or with his quantum mechanics papers.\\

Of course, John had an enormous impact on my own research, in fact, on my whole life. He opened my eyes for a sharp and clear view of Nature paired with honesty and modesty, and for the beauty in scientific thinking. With the taste of Bell's Universe I could enjoy the many collaborations I had in my life in the field of particle physics, mathematical physics and quantum physics.\\

\begin{figure}
\begin{center}
\includegraphics[width=0.75\textwidth]{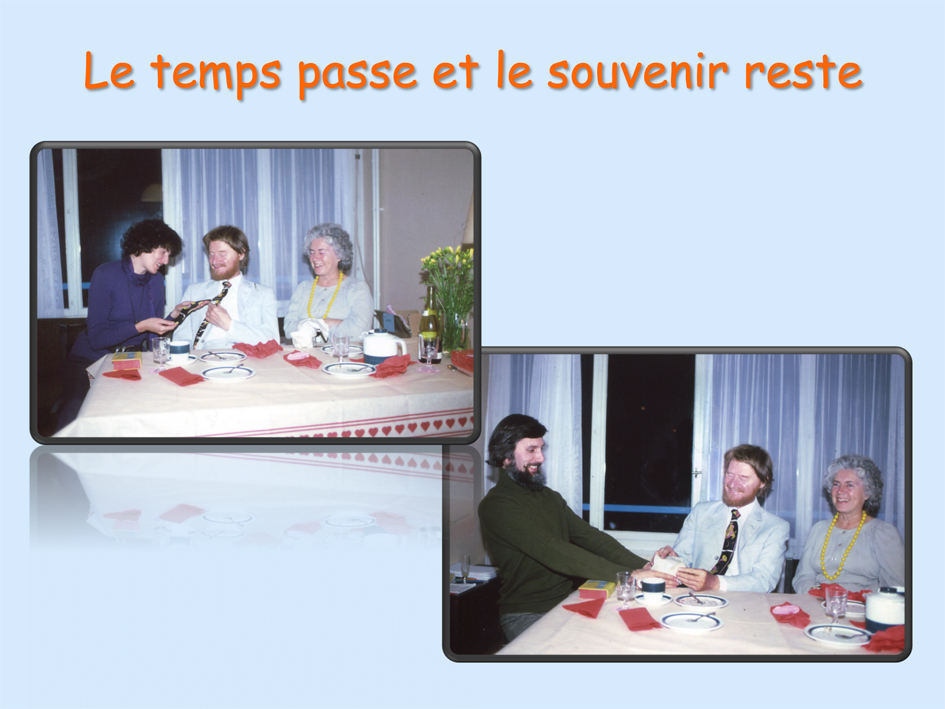}
\normalsize{
\caption{Renate (left picture) and Reinhold Bertlmann (right picture) having fun during a dinner with John and Mary Bell in Bertlmann's apartment in Geneva in 1982. Fotos: \copyright Renate Bertlmann.}
\label{fig:LeTempsPasse}}
\end{center}
\end{figure}

My essay would not be complete without reporting on Mary Bell, John's wife, who was a committed physicist as well. In my memory are always both, John and Mary. Renate and myself have spent a pleasant time with the Bells and we also had great fun together, as can be seen on Fig.~\ref{fig:LeTempsPasse}.\\

I really feel privileged and thankful for the time I could spend with John and I would like to end with the French saying:

\begin{center}
\emph{``Le temps passe et le souvenir reste.''}
\end{center}

\begin{acknowledgments}

First of all, I am grateful to Mary Bell for the generosity of sharing her memories of her scientific career and her life with John. Many thanks, of course, to all my collaborators and friends, to Beatrix Hiesmayr, Nicolai Friis, Walter Grimus, Philipp K\" ohler, Heide Narnhofer, Gabriele Uchida and Anton Zeilinger, for the exciting and joyful discussions we had together. Last but not least, I would like to thank Renate Bertlmann for her lovingly company and continuous support in all that years.

\end{acknowledgments}


\end{document}